\documentclass[a4paper,10pt]{article}
\pdfoutput=1
\usepackage[utf8]{inputenc}
\usepackage{amsmath}
\usepackage{bbold}
\usepackage{graphicx}

\date{\today}
\def\lsim{\raise0.3ex\hbox{$\;<$\kern-0.75em\raise-1.1ex\hbox{$\sim\;$}}}
\def\gsim{\raise0.3ex\hbox{$\;>$\kern-0.75em\raise-1.1ex\hbox{$\sim\;$}}}

\title{NMSSM interpretations of the observed Higgs signal}
\author{Florian Domingo, Georg Weiglein}
\date{{\em Deutsches Elektronen Synchrotron (DESY),\\ Notkestraße 85, D--22607 Hamburg, Germany}}

\begin{document}

\maketitle
\vspace{-5cm}\rightline{DESY 15-179}\vspace{5cm}
\begin{abstract}
While the properties of the 
signal that was discovered in the Higgs searches at the LHC
are consistent so far with the Higgs boson of the Standard Model (SM), 
it is crucial to investigate to what
extent other interpretations that may correspond to very different
underlying physics are compatible with the current results.
%experimental results on the
%observed signal, the limits from Higgs searches and other existing
%constraints. 
We use the Next-to-Minimal 
Supersymmetric Standard Model (NMSSM) as a well-motivated theoretical
framework with a sufficiently rich Higgs phenomenology to address this
question, making use of the public tools \verb|HiggsBounds| and
\verb|HiggsSignals| in order to take into account comprehensive
experimental information on both the observed signal and on the existing
limits from Higgs searches at LEP, the TeVatron and the LHC. 
We find that besides the decoupling limit resulting in a single light
state with SM-like properties, several other configurations 
involving states lighter or quasi-degenerate with the
one at about 125~GeV turn out to give a competitive fit to the Higgs
data and other existing
constraints. We discuss the phenomenology and possible future
experimental tests of those scenarios, and compare the features of
specific scenarios chosen as examples with those arising from a more
global fit.
%While the Higgs-like signal measured at LHC for a mass of
%$\sim125.6$~GeV seems roughly consistent with a standard Higgs boson so
%far, it may also hide richer Higgs phenomenologies. The Next-to-Minimal
%Supersymmetric Standard Model (NMSSM) is a well-motivated extension of
%the Standard Model which involves a populous Higgs sector. It may lead
%to several unconventional configurations and / or allow for specific
%mechanisms. We study several such scenarii with the help of the public
%codes \verb|NMSSMTools|, \verb|HiggsBounds| and \verb|HiggsSignals|. A
%number of setups, involving states lighter or quasi-degenerate with the
%one at $\sim125.6$~GeV, prove to give a competitive fit to the data and
%we discuss the associated phenomenology.
\end{abstract}

\section{Introduction}
After the discovery of a signal with a mass of about 125~GeV
%$\sim125.6$~GeV 
in the Higgs searches at the LHC~\cite{Aad:2012tfa,Chatrchyan:2012ufa}, the prime goal is now to identify the 
underlying nature of the new state and to determine the mechanism of electroweak symmetry breaking.
While the properties of the observed state 
%determined so far 
are compatible with the ones predicted for
the Higgs boson of the Standard Model (SM) 
at the current level of precision, also 
a wide range of alternative interpretations could be possible, 
%which could correspond to quite 
corresponding to very
different underlying physics. In particular, in models with an extended Higgs sector the observed state
would be accompanied by several other Higgs bosons, in contrast to the minimal formulation of the SM
where a single $SU(2)_L$-doublet is responsible for electroweak-symmetry breaking.

Supersymmetry (SUSY)~\cite{SUSY} is commonly regarded as the most appealing extension of the SM, 
since it provides a solution for stabilising the huge hierarchy between the Planck scale and the
weak scale~\cite{HierProb} and offers further attractive features such as unification of the gauge
couplings and a natural candidate for cold dark matter in the Universe.
A crucial prediction of supersymmetric
extensions of the SM are their extended Higgs sectors: 
the holomorphicity of the superpotential 
(as well as the cancellation of gauge anomalies) implies that at least two $SU(2)_L$ doublets with 
opposite hypercharge have to be present, so as to 
generate mass terms for both up- and down-quarks (in a Type II 2-Higgs-Doublet-Model fashion).
The minimal supersymmetric extension of the SM (MSSM)~\cite{MSSM} is based on the minimal 
Higgs sector of this kind comprising two Higgs doublets, whereas the Higgs sector of the next-to-minimal 
supersymmetric extension of the SM (NMSSM)~\cite{NMSSM,NMSSM1} contains an 
additional (complex) gauge-singlet. It has long been recognised that the NMSSM provides an elegant
solution~\cite{fayet} to the ``$\mu$-problem''~\cite{Kim:1983dt} of the MSSM. In the context of the
relatively high mass value of about 125~GeV of the observed state, this model has received particular
attention lately since the mass of the light doublet-like state receives an additional contribution at
lowest order as compared to the MSSM, which dominates at low values of $\tan\beta$ (the ratio of the
vacuum expectation values, v.e.v.'s, of the two Higgs doublets). In this case significantly smaller higher-order
corrections are required to obtain a Higgs-boson mass in the appropriate range~\cite{largelamb}
as compared to the MSSM,
where the lowest-order prediction for the mass of the light CP-even Higgs boson is bounded from above by
the mass of the Z~boson, $M_Z$. Furthermore, also the singlet--doublet mixing can give rise to an uplift
of the mass of the doublet state, provided that the CP-even singlet state is lighter than the doublet
state. It has been argued in this context that the relaxed requirement on the size of the higher-order
corrections as compared to the MSSM makes it possible to obtain a Higgs-mass prediction of about 125~GeV
in a ``more natural'' way~\cite{natsusy}.

% stands at the center of the NMSSM interpretation of the $\mu$-problem \cite{fayet}. 

In the following we will focus on the NMSSM as a theoretically
well-motivated alternative to the SM with a potentially rich
phenomenology in the Higgs sector. For simplicity, we will restrict to
the CP-conserving case, for which the spectrum of physical Higgs states
of the NMSSM consists of three CP-even, two CP-odd and a pair of charged
Higgs states (while we do not explicitly consider CP-violating effects
giving rise to a mixture between the five neutral states, it should
be noted that cases where a CP-even and a CP-odd state are nearly
mass-degenerate essentially mimick a scenario where a single state is an
admixture of CP-even and CP-odd components). Furthermore, while several
versions of the NMSSM can be formulated, depending on the form of the singlet
and singlet-doublet interaction terms in the superpotential, we will focus on the 
$\mathbb{Z}_3$-conserving version only, where the solution to the ``$\mu$-problem''
is more immediate (on the other hand, this simple model could lead to a domain
wall problem \cite{Abel:1995wk} but we will not address this question here).  
The Higgs sector of this version of the
NMSSM is characterised by six parameters (at tree-level), in contrast to the two
parameters of the MSSM. While we shall borrow most of our notations from
\cite{NMSSM}, we recall the Higgs terms entering the superpotential of the model:
\begin{equation}
 W_{\mbox{\small NMSSM}}\ni\lambda S H_u\cdot H_d+\frac{\kappa}{3}S^3
\end{equation}
where $S$ denotes the singlet (super)field, $H_{u}$ and $H_d$ the doublets, while $\cdot$ stands 
for the $SU(2)$ product.

%Three CP-even, two CP-odd and one pair of charged Higgs states are thus expected in this model, with
%the prospects of a rich associated phenomenology. Six parameters (against two in the MSSM) are now present in the Higgs sector, resulting in several
%new mechanisms and evasion of MSSM correlations. 

%Among the recorded effects, we can spend some attention to the benefit of the mass of the light
%doublet-like state: this quantity receives an additional contribution at tree level $\propto \lambda^2 v^2\sin^2{2\beta}$, dominating at low 
%values of $\tan\beta$ and offering an alternative \cite{largelamb} to the large-radiative-correction mechanism, which the MSSM requires so as 
%to generate a Higgs mass in the appropriate range. Alternatively, one may exploit the singlet-doublet mixing, provided the CP-even singlet state 
%is lighter, in order to uplift the mass of the doublet state. Such contraptions arguably alleviate the question of the Little-Hierarchy 
%in the NMSSM, making a Higgs mass at $\sim125.6$~GeV more natural \cite{natsusy}.

When confronting the predictions of an extended Higgs sector with the
observed signal and the limits from the Higgs searches at LEP, the
TeVatron and the LHC, the most obvious interpretation of the signal at
about 125~GeV is to associate it with the lightest CP-even Higgs boson
of the considered model. The case where all other Higgs (as well as all new physics) states of a
supersymmetric extension of the SM (and the same is true for various
other extended Higgs sectors) are significantly heavier corresponds to
the ``decoupling region'' of the model under consideration, where the
couplings of the light Higgs boson to gauge bosons and SM fermions are
very close to the ones of the SM. Revealing deviations of those
couplings from their SM counterparts in such a case will require
high-precision measurements, where in many cases the expected deviations
do not exceed the level of a few per cent. An additional source of
possible deviations from the SM could be decays of the SM-like state
into new-physics particles. Such a decay could in particular occur into
a pair of dark matter particles, if the mass of the latter is less than
half of the mass of the Higgs state, i.e.\ below about 60~GeV. 
This would give rise to an
invisible decay mode of the observed state, providing a strong
motivation for searches of decays of the observed signal into
invisible final states.

Besides the interpretation of the observed state as the lightest CP-even
Higgs boson of an extended Higgs sector, it is also possible, at least
in principle, to identify the observed signal with the second-lightest
state of an extended Higgs sector. This interpretation would have the
immediate consequence that there should be an (or more generally at
least one) additional Higgs state
{\em below\/} the one observed at about 125~GeV. The phenomenology of
such a scenario is very different from the case of the decoupling limit
discussed above, because of the presence of at least one more light
state in the spectrum. Within the MSSM this interpretation is in
principle possible~\cite{MSSMheavyH,MSSMfit}, but gives rise to a rather exotic
Higgs sector where in fact all additional Higgs bosons are light, i.e.\
in the vicinity of the state at about 125~GeV or below. It is remarkable
that a global fit within the MSSM within this interpretation has
resulted in an acceptable fit probability~\cite{MSSMfit}, but lately
this interpretation, which implies in particular a light charged Higgs
boson below the mass of the top quark (see in particular
Ref.~\cite{MSSMnewbench}), has come under increased pressure from the
limits obtained in the charged Higgs searches by
ATLAS~\cite{ATLASchargedHiggs} and, more recently, CMS~\cite{CMSchargedHiggs}.

The NMSSM provides a well-suited and theoretically well motivated
framework for investigating to what extent
%example of a realistic and well motivated model 
%with a quite rich phenomenology in the Higgs
%sector. It is therefore of interest to investigate to which extent
%within this model 
interpretations that go beyond the obvious case of a 
single light state in the decoupling limit are compatible with the
latest experimental results both with respect to the properties of the observed state
and to the limits obtained from Higgs searches (as well as other
existing constraints). It is the purpose of the present paper to
perform such an analysis.
It is obvious that compatibility with the observed signal requires much
more than just a Higgs state (or possibly more than one) in the spectrum 
with a mass of about 125~GeV.
In order to properly take into account the latest experimental results
from Higgs search limits and from measurements of the
properties of the observed state, we make use of the public tools
{\tt HiggsBounds}~\cite{HiggsBounds} and {\tt
HiggsSignals}~\cite{HiggsSignals}, which incorporate a comprehensive set
of results from
ATLAS~\cite{ATLAS}, CMS~\cite{CMS} and the TeVatron~\cite{TeVatron}.
%Tools (state of the art): {\tt HiggsBounds}, {\tt HiggsSignals}, \ldots
We do not explicitly impose limits from the direct searches for
SUSY particles at the LHC. While some of the scenarios discussed in this paper could
be affected by constraints from SUSY particle searches, we have checked
that the qualitative features of the Higgs phenomenology of those scenarios 
are maintained also for somewhat heavier SUSY particle spectra. Note however
that most of the SUSY spectra that we employ (especially for coloured particles) are beyond the mass-range tested in the Run-I of the LHC.

%Note that LHC limits
%on supersymmetric particles are not included within our analysis and some spectra may seem over-optimistic in that respect. We checked however that
%the scenarios under consideration also apply to heavier supersymmetric spectra without spoiling, qualitatively, the good agreement with the Higgs
%measurement data. 

%Generating a Higgs mass in the appropriate range, although crucial, is not sufficient however to claim good agreement with the Higgs measurement
%data: the success of the searches in the conventional channels entails requirements on the nature and the couplings of the observed state, but also
%on the rest of the spectrum. A pure singlet state at $\sim125.6$~GeV for instance is unlikely to provide the expected behaviour. Neither is a SM-like
%state that would decay predominantly to invisible states. On the other hand,
%moving to the full SM-limit of the NMSSM would represent an extreme posture, unlikely to prove necessary from the standpoint of the experimental
%data and jeopardizing much of the benefits which one could gain with a model as phenomenologically rich as the NMSSM. For this reason, a deeper
%understanding of the requirements inflicted by the ``Higgs measurement'' signals from TeVatron \cite{TeVatron}, ATLAS \cite{ATLAS} and CMS \cite{CMS} 
%upon NMSSM Higgs scenarii seems  indicated. Here, we aim at highlighting a few unconventional setups which however seem to 
%provide a satisfactory agreement with the measurements reported by TeVatron and LHC.

As mentioned above, 
already since the very early hints of a signal at about 125~GeV
the Higgs sector of the NMSSM
%$\sim125.5$~GeV, much attention has been drawn to the Higgs sector of the 
%NMSSM, first in connection
has found a lot of attention in this context. Besides the mass prediction 
in comparison with the MSSM case~\cite{natsusy}, the 
%with the naturalness of a Higgs state at this mass \cite{natsusy} and the
possibility of modified rates has been discussed, particularly in the 
diphoton channel~\cite{diphoton}. 
%The question was also considered from the point of view of
Furthermore, the case of
universal (or semi-universal) SUSY-breaking conditions at the GUT 
scale~\cite{CNMSSM}, gauge-mediation~\cite{gamed} and other related 
scenarios~\cite{othcon} have been considered in this context. The CP-violating version
of the NMSSM also received some attention \cite{CPviol}. \cite{Z3viol} analysed the fine-tuning
in a $\mathbb{Z}_3$-violating version of the NMSSM and variants.
Other groups confronted the presence of a Higgs state at
this mass with direct searches for SUSY particles at the LHC or Dark-Matter
constraints~\cite{othcon2}. Scenarios with a 
%Much interest was raised by the possible presence of a 
light singlet-like state around $\sim100$~GeV 
%which may betray its cover in nearby searches or may appear with 
%unexpected rates
have found considerable interest~\cite{singsearch}. 
Another possibility involves a singlet and a doublet that are almost
mass-degenerate at about 125~GeV and may mix with each other, see 
Ref.~\cite{2CPeven} (and
%possibly mixing, almost degenerate at $\sim125.5$~GeV 
the last reference of \cite{diphoton}). Several studies also suggested to
exploit pair production processes at the LHC in order to
distinguish the SM from the NMSSM and/or to look for a
light singlet in this fashion~\cite{pairprod}. 
%Finally, the tentative
%existence of a very light CP-odd (or CP-even) Higgs has not escaped
%attention \cite{lightCPodd}: it was addressed with several search proposals
Scenarios with a very light CP-odd (or CP-even) Higgs boson were
addressed with several search proposals
in direct production, unconventional light charged-Higgs decays, or cascade
decays from SM-like / light singlet states; large Higgs-to-Higgs decays were
also considered from the point of view of the SM-compatible nature of the
observed state~\cite{lightCPodd}. Recent studies of the properties
of a light pseudoscalar in the NMSSM~\cite{Bomark:2014gya} have emphasized the 
relevance of indirect production modes for the investigation of this scenario
at the LHC. In a different direction, 
the authors of \cite{Christensen:2013dra} focussed on NMSSM Higgs scenarios 
with a low-scale doublet sector. Furthermore, \cite{King:2014xwa}, and more recently 
\cite{Guchait:2015owa}, studied 
the discovery prospects of NMSSM Higgs states in the LHC run at $13$~TeV.
In our analysis we go beyond the previous work in several respects:
while many of the afore-mentioned analyses discussed
scenarios which are compatible with existing limits, our inclusion of
a fitting tool allows us to highlight the quality of the various
scenarios in view of the available data. Furthermore, we aim at a comprehensive
discussion from the point of view of the NMSSM Higgs phenomenology, hence do
not confine to a specific scenario (within our assumptions on the model,
perturbativity constraints and choices of simplicity with regards to the SUSY spectrum). We also focus on Higgs physics and thus
try, without spoiling the physical content, to avoid emphasis on questions of 
secondary importance with respect to this topic (e.g.\ the details of the 
supersymmetric spectrum). 
Finally, much experimental data has become available in the last few years,
narrowing the possibilities in the Higgs sector, and most of the recent
developments are included within the tools on which our discussion is based.

The paper is organised as follows: in Sect.~2 we describe the framework
that we use for the analyses in this paper, in particular the
statistical approach used in our fits, the treatment
of external constraints and the tools that we apply. As a first step of
our analysis, in Sect.~3 we briefly consider the SM case and the corresponding
decoupling limits of the MSSM and the NMSSM. The SM result is used for
comparison with $\chi^2$ analyses in different NMSSM scenarios, which we
perform in Sects.~\ref{lsing}--\ref{ldoub}. In Sect.~\ref{fp},
we focus on specific points of the NMSSM parameter space and discuss
in more details the Higgs phenomenology and the consequences for future
searches, should the corresponding spectrum be realised in nature.
In Sect.~\ref{gs} a more global scan is carried out, and
the features observed in the global scan are discussed in view of the 
results obtained for the specific NMSSM scenarios that we have
considered before. Sect.~\ref{Conc} contains our conclusions.

%We shall first review briefly the tools which we will trust in our investigation. Then we will focus on specific scenarios before we summarize with
%a global scan. 

%\section{Computing Tools}
\section{Framework of the analysis: treatment of external constraints and applied tools}
The NMSSM parameter space is explored with the help of the spectrum generator \verb|NMSSMTools_4.4.0| \cite{NMSSMTools}, computing the Higgs masses 
up to leading two-loop double-log order (we will be using the default mode only), in an effective potential approach. This code 
considers a certain number of phenomenological limits, several among which are kept within our analysis. The first class of such tests are consistency 
requirements and are (necessarily) included as hard cuts:
\begin{itemize}
 \item stability of the EWSB-vacuum: positivity of the scalar squared-masses, absence of deeper minimum;
 \item absence of Landau poles below the GUT scale: while this requirement is sometimes omitted in order to probe effects associated 
with large values of the parameter $\lambda$ and under the assumption that new-physics or specific properties of the non-perturbative regime would
smoothen the theoretical difficulty of the Landau poles, we choose to keep this theoretical limit;
 \item requirement for Higgs soft squared-masses at the TeV scale: the potential-minimization procedure in NMSSMTools trades
these masses for the Higgs v.e.v.'s, so that the naturalness requirement that soft masses intervene at the TeV scale must be checked explicitly;
 \item requirement for a neutralino LSP (the impact of which, however, is of secondary importance in our discussion).
\end{itemize}
Another type of constraints are supersymmetric searches at LEP. Given that we are chiefly interested in the Higgs sector, we also keep these
limits under the form of a hard cut:
\begin{itemize}
 \item $Z\to\mbox{inv.}$ decay ($<1.71\cdot10^{-3}$)
 \item mass lower limits on squarks ($m_{\tilde{t}}>93.2$~GeV, $m_{\tilde{q}_{1,2}}>100$~GeV), gluino ($m_{\tilde{g}}>180$~GeV), sleptons 
($m_{\tilde{l}}>99.9$~GeV), charginos ($m_{\tilde{\chi}^{\pm}}>103.5$~GeV);
 \item limits on $\tilde{t}\to bl\tilde{N}$, $\tilde{t}\to c\chi^0$, $\tilde{b}\to b\chi^0$.
\end{itemize}
We remind the reader that LHC limits on SUSY searches are not considered in our analysis. However, the SUSY spectra that we employ are 
typically beyond the mass-range of the searches in the Run-I. In this
context, the inclusion of LEP limits as mentioned above has only a minor
impact.

NMSSMTools also computes several low-energy observables:
\begin{itemize}
 \item limits from the bottomonium sector: non-observation of a signal in $BR(\Upsilon\to\gamma A\to \gamma l^+l^-)$, excessive contribution to the 
$\eta_b(1S)-A$ mixing \cite{bottomonium}. Only light CP-odd Higgs below $\sim10$~GeV are concerned by these constraints and the limits are kept
as a $95\%$ C.L. cut\footnote{Note that the points excluded by such limits are stored while scanning, however.}.
 \item limits from $B$-factories (under a strong Minimal Flavour Violation hypothesis, i.e.\ neglecting all possible tree-level flavour-changing 
neutral currents): $BR(B\to X_s\gamma)$, $BR(B^+\to\tau\nu_{\tau})$, $BR(\bar{B}_s\to\mu^+\mu^-)$, $BR(B\to X_s\mu^+\mu^-)$, $\Delta M_{d,s}$
\cite{Bphys}. Instead of treating the limits as a hard cut, we combine them in a $\chi^2$ function relying on the central value and error bars 
computed in NMSSMTools:
\begin{equation}
 \chi^2=\sum_i\frac{(O_i^{\mbox{\tiny NMSSM}}-O_i^{\mbox{\tiny exp.}})^2}{\sigma_i^{2\,\mbox{\tiny theo.}}+\sigma_i^{2\,\mbox{\tiny exp.}}}
\end{equation}
the corresponding experimental central values $O_i^{\mbox{\tiny exp.}}$ and standard deviation $\sigma_i^{\mbox{\tiny exp.}}$ are summarized in
Table~\ref{expBlim}. The theoretical error estimate is the result of an involved calculation: errors relative to SM-like contributions are taken 
from the corresponding SM estimate; the uncertainty on new-physics contributions is estimated to $30\%$ (if only leading-order effects are included) / 
$10\%$ (if next-to-leading $\alpha_S$ corrections are present) of the total corresponding contributions and are added linearly to the 
SM error; additional error sources are mostly CKM matrix elements (taken from tree-level measurements exclusively) and hadronic parameters 
(decay constants, taken from lattice calculations); to obtain the final theoretical uncertainty range, both higher-order and parametric 
uncertainties are varied within these previously-discussed limits.
 \item $(g-2)_{\mu}$ \cite{g-2}: similarly to $B$-observables, we add a contribution to the $\chi^2$ with experimental-SM input shown in
Table~\ref{expBlim}, where the errors have been added in quadrature. The theoretical uncertainty associated to new-physics contributions 
and higher orders is calculated as the sum of a fixed error $2.8\cdot10^{-10}$, a $2\%$ error estimate on 1-loop contributions (which
do not involve coloured particles) and a $30\%$ error estimate on 2-loop effects (involving coloured particles).
\end{itemize}
Additionally, given that a candidate for the interpretation of the signal observed at the LHC seems necessary, we require that the spectrum produces one 
CP-even Higgs state in the mass-range $[120,131]$~GeV.

Other limits are deliberately ignored, at least as implemented within NMSSMTools:
\begin{itemize}
 \item dark matter searches: relic-density (via MicrOMEGAs), XENON 100 \cite{DM}. The reasons for not taking such limits into account 
come from the observation that they are strongly dependent on the SUSY spectrum, while we want to focus on the Higgs sector: confining to collider 
constraints allows us to handle simple supersymmetric spectra, which play a secondary part in our analysis, while these would likely have to be 
finely adjusted if one were to include, e.g., the relic-density bounds. We note also that the dark matter phenomenology may involve mechanisms
(e.g.\ light gravitino LSP) which may alter the conclusions in the dark sector, while all such considerations are not the focus of our discussion.
 \item LEP Higgs searches: $e^+e^-\to Zh$, $h\to\mbox{inv.},2\mbox{jets},2\gamma,\{2A\to4b,4\tau,2b+2\tau,\mbox{light pairs}\}$, $e^+e^-\to hA\to4b,4\tau,2b+2\tau,
3A\to\{6b,6\tau\}$, $Z\to hA$ (Z width);
 \item TeVatron limits on $t\to bH^+$, $H^+\to cs,\tau\nu_{\tau},W^+\{A_1\to2\tau\}$;
 \item LHC Higgs limits: $t\to bH^+$, $m_h\in[122,129]$~GeV, effective $h\gamma\gamma$, $hbb$, $hZZ$ couplings excessive \cite{Belanger:2013xza}.
\end{itemize}

\begin{table*}
\begin{tabular}{|c||c|c|c|c|c}
\hline
$O_i$ & $BR(B\to X_s\gamma)$ & $BR(B^+\to\tau\nu_{\tau})$ & $BR(\bar{B}_s\to\mu^+\mu^-)$ & $BR(B\to X_sl^+l^-)$&\ldots\\
 & & & &$|_{1~{\mbox{\tiny GeV}^2<s_{l^+l^-}<6~\mbox{\tiny GeV}^2}}$&\ldots\\\hline 
$O_i^{\mbox{\tiny exp.}}$ & $3.43\cdot10^{-4}$ & $1.14\cdot10^{-4}$ & $3.2\cdot10^{-9}$ & $1.6\cdot10^{-6}$ &\ldots\\\hline
$\sigma_i^{\mbox{\tiny exp.}}$ & $0.22\cdot10^{-4}$ & $0.22\cdot10^{-4}$ & $0.7\cdot10^{-9}$ & $0.5\cdot10^{-6}$ &\ldots\\\hline
%$X_{\mbox{\tiny min}}^{\mbox{\tiny exp}}$ & $3.04\cdot10^{-4}$ &  $0.89\cdot10^{-4}$ & $2.0\cdot10^{-9}$ & $0.6\cdot10^{-6}$ &\ldots\\\hline
%$X_{\mbox{\tiny Max}}^{\mbox{\tiny exp}}$ & $4.06\cdot10^{-4}$ &  $2.45\cdot10^{-4}$ & $4.7\cdot10^{-9}$ & $2.6\cdot10^{-6}$ &\ldots\\ \hline
\end{tabular}

\null\hspace{4.9cm}\begin{tabular}{c|c|c|c||c|}
\hline
\ldots & $BR(B\to X_sl^+l^-)$ & $\Delta M_{d}$ & $\Delta M_{s}$ & $\Delta a_{\mu}^{\mbox{\tiny exp-SM}}$ \\
\ldots & $|_{14.4~{\mbox{\tiny GeV}^2<s_{l^+l^-}}}$ & & & \\\hline
\ldots & $4.4\cdot10^{-7}$ & $0.507~\mbox{ps}^{-1}$ & $17.719~\mbox{ps}^{-1}$ & $27.4\cdot10^{-10}$ \\\hline
\ldots & $1.2\cdot10^{-7}$ & $0.004~\mbox{ps}^{-1}$ & $0.043~\mbox{ps}^{-1}$ & $9.3\cdot10^{-10}$ \\\hline
%\ldots & $2.0\cdot10^{-7}$ & $0.499~\mbox{ps}^{-1}$ & $17.633~\mbox{ps}^{-1}$ & $8.77\cdot10^{-10}$ \\\hline
%\ldots & $6.8\cdot10^{-7}$ & $0.515~\mbox{ps}^{-1}$ & $17.805~\mbox{ps}^{-1}$ & $46.11\cdot10^{-10}$ \\\hline
\end{tabular}
\caption{Experimental central values and uncertainties for $B$ physics observables and $(g-2)_{\mu}$ as implemented in NMSSMTool\_4.4.0.}
\label{expBlim}
\end{table*}

LEP, TeVatron and LHC limits on Higgs searches are checked through the code \verb|HiggsBounds_4.2.0| \cite{HiggsBounds}, 
%\mla
%{\bf [Explain that these are pre-ICHEP results and that we comment on
%new input from ICHEP where it is relevant for the present analyses.]}
which we interfaced to 
NMSSMTools via a subroutine. HiggsBounds is used with the 
default settings, using hard cuts on the allowed regions and an individual test for all Higgs states (`full' method). Note that the more sophisticated 
implementation of limits on the Higgs sector within HiggsBounds leads to divergences with the channel-after-channel checks implemented within NMSSMTools, in 
particular in the LEP $e^+e^-\to hA$ channels: as a consequence, lighter Higgs (doublet-like) states are accessible, while the corresponding points 
would be rejected by NMSSMTools. The version of HiggsBounds which we use includes all the released LHC limits till the end of 2014.

Another remark concerns the distinction between `hard-cut constraints' and `observables to include in our fit': it is bound to carry some
arbitrariness. One could object, for instance, that implementing LEP limits as `hard cuts', one loses the sensitivity to small deviations 
in the LEP data so that no benefit is associated in the fit. On the other hand, those may prove to be only statistical fluctuations. In practice, 
we treat all search limits as hard cuts while observables kept in the fit have been actually measured.

\verb|HiggsSignals_1.3.1| \cite{HiggsSignals} performs the comparison to the TeVatron+LHC-observed Higgs data, delivering a $\chi^2$ fit to the 
Higgs-measurement observables. The version we employ collects all released experimental material till the end of 2014. Here follow 
a few remarks concerning the setting of the options that are offered.
\begin{itemize}
 \item {\em Statistical test method - }Two statistical methods, `peak-centered' (comparing theoretical and experimental signals at 
masses determined by the experimental signals) and `mass-centered' (comparing theoretical and experimental signals at the masses that are defined 
by the theoretical input) are available in HiggsSignals. Only the `peak-centered' test shall be performed. $81$ channels are tested within the analysis.
 \item {\em Higgs pdf - }The probability density function assumed for the Higgs boson mass is modelled as a gaussian.
 \item {\em Theoretical mass uncertainties - }We allow for a $\pm3$~GeV uncertainty on the theoretical Higgs mass predictions delivered by NMSSMTools. 
This ensures that the phenomenology that we are discussing applies to the NMSSM and not to an arbitrary spectrum. Given also that our study is limited
by the density of the scans that we perform, the mass uncertainty smoothens the impact of phenomenological limits. On the other hand, one could argue 
that, in view of the Higgs-like signals at LHC, the interesting part of the phenomenological study concerns the workings of a given spectrum, 
irrespectively to the model hiding behind. Moreover, considering that the NMSSM has a large number of parameters in the Higgs sector, it may be
possible to absorb a substantial part of the higher-order corrections within a shift in parameter space. In such a context, one could be concerned that large
mass uncertainties might blur the phenomenology and provide unlikely spectra with undue attention. Yet this feature does not happen and the scenarios
that we propose would appear with qualitatively comparable fit values even though the mass uncertainty was set to be smaller.
\end{itemize}
%\end{tabular}
\begin{figure}[!htb]
    \centering
    \includegraphics[width=15cm]{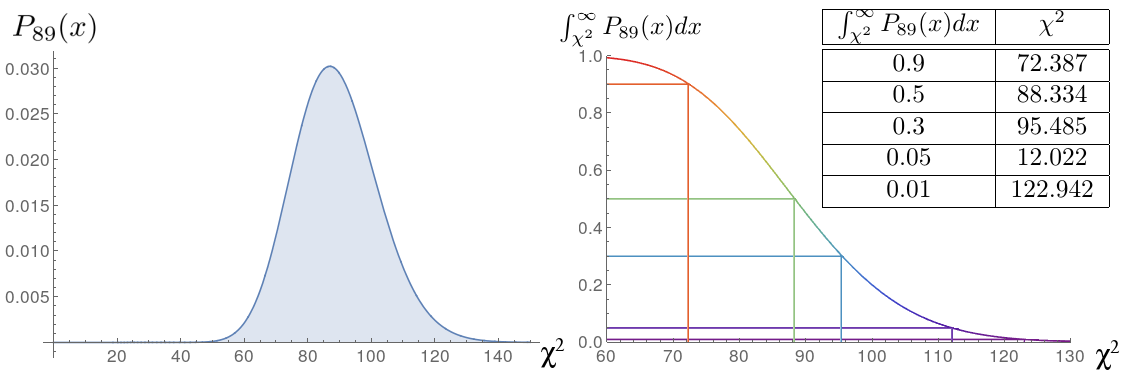}
  \caption{$\chi^2$ distribution and a few values for the integral of its tail (P-values).}
  \label{chi2}
\end{figure}
This $\chi^2$ test to the Higgs-measurement data, delivered by HiggsSignals, is added to the corresponding tests from $B$-physics and $(g-2)_{\mu}$.
The resulting quantity, $\chi^2_{(total)}$, will be at the center of our discussion in the following sections. Let us comment briefly on its 
interpretation. The $\chi^2$ is the sum of squared deviations between experimental measurements and theoretical inputs, weighed by the 
corresponding uncertainties. Assuming the corresponding quantities are random and independent (gaussian) variables, the $\chi^2$ would follow a 
probability law given by the $\chi^2$-distribution of $N^{\mbox{\tiny th}}$ degree $P_{N}$, where $N$ is the number of variables in the sum: in 
our case, $N=81+7+1=89$, since there are $81$ channels involved in the HiggsSignals test, $7$ $B$-observables and $(g-2)_{\mu}$ finally. 
Correspondingly, one may define the compatibility of the $\chi^2$ test (`P-value') as the probability to fall farther than the obtained $\chi^2$ 
value (i.e.\ the probability that the generated deviations be larger). With this definition, the compatibility of a $\chi^2$ value with the test 
involving $N$ degrees of freedom is obtained as: $C(\chi^2)\equiv\int_{\chi^2}^{\infty}{P_{N}(x)dx}$. In this context, for the $\chi^2$ 
distribution of $89^{\mbox{\scriptsize th}}$ degree, the compatibility of the spectrum with the data reaches $90\%$ for $\chi^2\simeq72$, $50\%$ for 
$\chi^2\simeq88$, $30\%$ for $\chi^2\simeq95$, $5\%$ for $\chi^2\simeq112$ and $1\%$ for $\chi^2\simeq123$ (see Fig.~\ref{chi2}). (Note that
HiggsSignals directly provides an approximate P-value for the Higgs-fit.) This test would be the statistically relevant confrontation of one isolated 
point to the experimental data. This approach raises a few issues, however, as the corresponding statistics is then critically dependent on the list of
tested channels and the precise definition of these. Moreover, we will not be considering isolated points but scan over various portions of the NMSSM 
parameter space, thus introducing degrees of freedom which would have to be substracted from the statistical test. Absolute $\chi^2$ values are thus
difficult to interpret. Consequently, we will base our phenomenological discussion on relative $\chi^2$-differences with respect to a best-fit point,
which prove to be a more robust interpretation in the given context.

\section{SM and Decoupling limits}\label{decoup}
\begin{figure}[!htb]
    \centering
    \includegraphics[width=10cm]{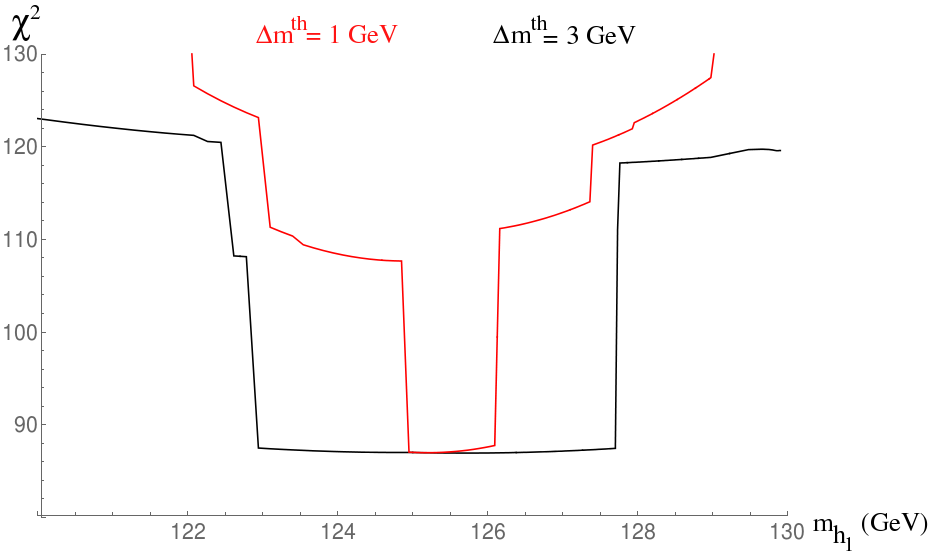}
  \caption{$\chi^2$ in the SM-limit of the (N)MSSM, as a function of the SM-like mass: $\tan\beta\in[1,50]$, $2M_1=M_2=\mu=1$~TeV, $M_3=3$~TeV, 
$m_{\tilde{Q}}=2$~TeV, $m_{\tilde{L}}=1$~TeV, $A_t=-4$~TeV, $A_{b,\tau}=-1.5$~TeV, $M_A=2$~TeV, $A_{\kappa}=-1$~TeV, $\lambda=\kappa=1\cdot10^{-5}$.
While the uncertainty on the Higgs masses, $\Delta m^{\mbox{\scriptsize th}}$, is set to $3$~GeV in the rest of the paper, we also include a plot for
$\Delta m^{\mbox{\scriptsize th}}=1$~GeV here, in order to illustrate the impact of this quantity on the fit.}
  \label{SMlim}
\end{figure}
The Standard-Model limit of the NMSSM is obtained when decoupling the singlet sector -- MSSM limit: $\lambda\sim\kappa\to0$; the singlet
states then have vanishing couplings to their doublet counterparts, while an effective $\mu$-term, $\mu\equiv\lambda \left<S\right>$ is generated from the 
singlet v.e.v. -- and pushing the masses of the MSSM non-standard states to very large values. The scale of the heavy doublet sector is controlled at 
tree-level by the parameter $M_A$ -- the doublet diagonal entry in the CP-odd Higgs mass matrix at tree-level -- which can be used directly as an input 
(instead of $A_{\lambda}$) within NMSSMTools: the decoupling condition reads $M_A\gg M_Z$. Similarly, the following scales enter the supersymmetric 
spectrum: the sfermion, $m_{\tilde{f}}$, gaugino, $M_{1,2,3}$, and higgsino, $\mu$, masses, which can be chosen far from $M_Z$. At low energy, one is 
then left with an effective SM, whose Higgs boson, the remaining light doublet Higgs state, has indeed SM-like couplings and a mass falling within the 
appropriate range ($\sim125$~GeV), provided soft stop terms $A_t$ and moderate-to-large $\tan\beta\gsim10$ are chosen accordingly. New physics effects
are then suppressed in accordance with the high scales that they involve or to the vanishing couplings of new states to the SM ones, so that this limit is virtually
undiscernible (at low energy) from a genuine SM\footnote{Note however that the higher the new-physics scales, the weaker becomes the case of supersymmetry 
as a solution to the hierarchy problem.}. We consider this trivial scenario in order to `calibrate' our fit -- i.e.\ set a point of comparison with 
other NMSSM scenarii -- and display our results in Fig.~\ref{SMlim}: the $\chi^2$-value is plotted as a function of the mass of the SM-like Higgs for a multi-TeV 
heavy supersymmetric and second-Higgs doublet spectrum. The best-fit points receive a $\chi^2$ of about $\sim87$, which statistically places the SM-limit within 
$1\sigma$ compatibility with the considered observables. This fact is not surprising, since the measurements of the Higgs signal at the LHC are grossly
consistent with a SM interpretation (within $1\sigma$). Similarly, no tension develops in the $B$-sector, where the considered observables are also 
compatible with the SM. On the other hand, the SM-limit is difficult to reconcile with the anomalous magnetic moment of the muon, which generates a $\chi^2$-pull of 
$\sim8-9$ (depending on the chosen scale for the sleptons, $m_{\tilde{l}}$): a limited departure from the strict SM limit -- e.g.\ lowering sleptons and 
gaugino masses -- would remedy this discrepancy. Finally, we wish to comment on the general aspect of the curves in Fig.~\ref{SMlim}: the minimum appears 
there as a broad step, extending from $\sim123$~GeV to $\sim128$~GeV. This appearance is driven by the treatment of the theoretical uncertainty on the 
Higgs masses, $\Delta m^{\mbox{\scriptsize th}}$, within HiggsSignals, as appears clearly when comparing both cases $\Delta m^{\mbox{\scriptsize th}}=1$~GeV and
$\Delta m^{\mbox{\scriptsize th}}=3$~GeV.

\begin{figure}[!htb]
    \centering
    \includegraphics[width=16.cm]{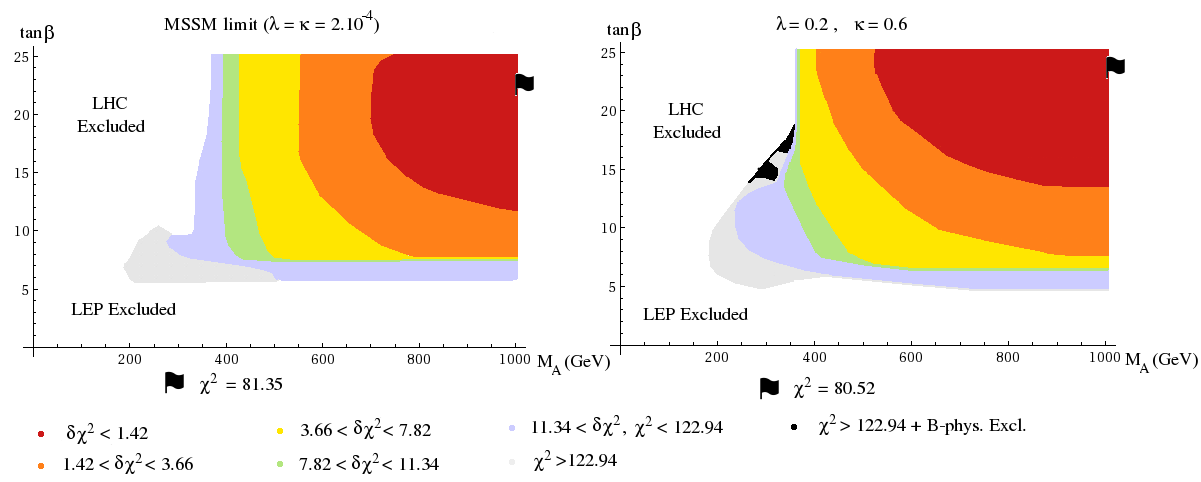}
  \caption{Scan in the $\{M_A,\tan\beta\}$-plane: $\tan\beta\in[1,25]$, $M_A\in[50,1000]$~GeV, $A_{\kappa}\in[-1.5,0]$~TeV, $\mu=200$~GeV, $2M_1=M_2=500$~GeV, $M_3=1.5$~TeV, 
$m_{\tilde{Q}_{1,2}}=1.5$~TeV, $m_{\tilde{Q}_3}=1.1$~TeV, $A_t=-2.3$~TeV, $A_{b,\tau}=-1.5$~TeV. }
  \label{MATBscan}
\end{figure}

Without turning to the full SM limit, the most frequent interpretation of the Higgs signal at $\sim125$~GeV within supersymmetric models consists in 
identifying it with the lightest Higgs state, this prejudice being motivated by the current absence of conclusive experimental signals for a Higgs 
boson at lower mass values. The corresponding configuration is most naturally achieved in the decoupling limit, that we define as $M_A\gg M_Z$, 
without necessarily requiring that the supersymmetric or the singlet spectra are much heavier: the light doublet state is then largely SM-like, at least
at tree-level. The possible presence of new-physics particles at neighbouring scales might then affect the couplings of this light Higgs state at 
the radiative level, which would allow for tests in precision physics -- unless the induced effect is negligible; note that the current LHC results 
allow for a relatively broad range of coupling strength in the vicinity of the SM values. The major concern in this configuration of the Higgs spectrum actually 
lies in generating a mass for the light doublet state in the appropriate range in order to identify it with the measured signals. In the MSSM limit 
($\lambda\sim\kappa\to0$), this can be achieved by saturating the tree-level contribution to the mass $\propto M_Z\cos2\beta$ -- therefore turning to 
$\tan\beta\gsim10$ -- and relying on substantial loop corrections (heavy stops / large trilinear couplings). This solution can be extended to the NMSSM 
(i.e.\ departing from $\lambda,\kappa\ll1$), although specific NMSSM effects can also be employed, as we will show in the following sections. Here, we just
illustrate the decoupling limit in Fig.~\ref{MATBscan}, where we display the results of a scan in the plane $\{M_A,\tan\beta\}$
both for the MSSM limit and a case with non-vanishing $\lambda$ and $\kappa$. Both configurations lead to a fit result where the lowest $\chi^2$ values 
are obtained in the range of large $M_A$ ($O(\mbox{TeV})$) and significant $\tan\beta$, i.e.\ in the decoupling limit. The best-fit point is indicated
in the plots by a flag. Note that the preference for $\tan\beta\gsim15$ is driven only
partially by the requirement of a Higgs mass close to $\sim125$~GeV. The discrepancy of the SM with the anomalous magnetic moment of the muon can indeed be 
cured by supersymmetric contributions, which then favour sizable $\tan\beta$: this is the main pull in the $\tan\beta$ direction, otherwise the $\chi^2$ 
distribution is mostly flat as soon as $\tan\beta\gsim10$.
It should be noted that, while $(g-2)_{\mu}$ in general 
favours large values of $\tan\beta$ for a relatively heavy SUSY spectrum, 
the preferred range does of course depend on the details of the SUSY spectrum, and especially the masses of the sleptons and charginos / neutralinos. We also wish to comment that the main limiting factor at low $\tan\beta$, in Fig.~\ref{MATBscan}, rests 
with the requirement of a Higgs state close to $125$~GeV. As such this (LHC) constraint supersedes the LEP lower bound of $\sim114$~GeV for the mass of 
a SM-like state. Note also that the SUSY spectrum (and especially the masses and mixings in the stop sector) plays a crucial part in the resulting lowest 
value accessible for $\tan\beta$ ($\sim5-6$ in Fig.~\ref{MATBscan}), as it controls the magnitude of the radiative corrections to the mass of the light Higgs doublet.

\section{Light CP-even singlet}\label{lsing}

A quite natural NMSSM scenario, already noted for its admissible application to a (speculative) enhancement of the $h[\sim125~\mbox{\small GeV}]\to2
\gamma$ decay rate \cite{diphoton}, is that of a light CP-even singlet state with mass under $\sim125$~GeV. This setup offers several interesting 
phenomenological features: 
the presence of a Higgs state around $\sim100$~GeV could account for a small excess observed in the $h\to b\bar{b}$ LEP data \cite{Barate:2003sz}; 
furthermore, the mixing between singlet and doublet (for non-vanishing $\lambda$) offers a lifting mechanism for the mass of the state identified with 
the LHC-observed signal; finally, the interplay with the singlet allows for an enhanced flexibility in the composition of the state at $\sim125$~GeV 
-- the mixing matrix in the CP-even sector has now three mixing angles, instead of only one in the pure-doublet case -- so that small deviations from SM-like 
couplings might be interpreted in this fashion. In contrast to the prejudice according to which light 
states should already have left tracks in experimental searches, the presence of a light CP-even singlet proves phenomenologically viable, as the 
large singlet component entails a suppressed production cross-section of this state -- via a suppressed coupling to SM-particles, e.g.\ gauge bosons or 
fermions --, at colliders. Moreover, the singlet induces no major perturbation in the SM fermion and gauge-boson sectors. This scenario can be studied 
e.g.\ in the limit of a heavy decoupling $SU(2)$ Higgs doublet.
\begin{figure}[!htb]
    \centering
    \includegraphics[width=15.3cm]{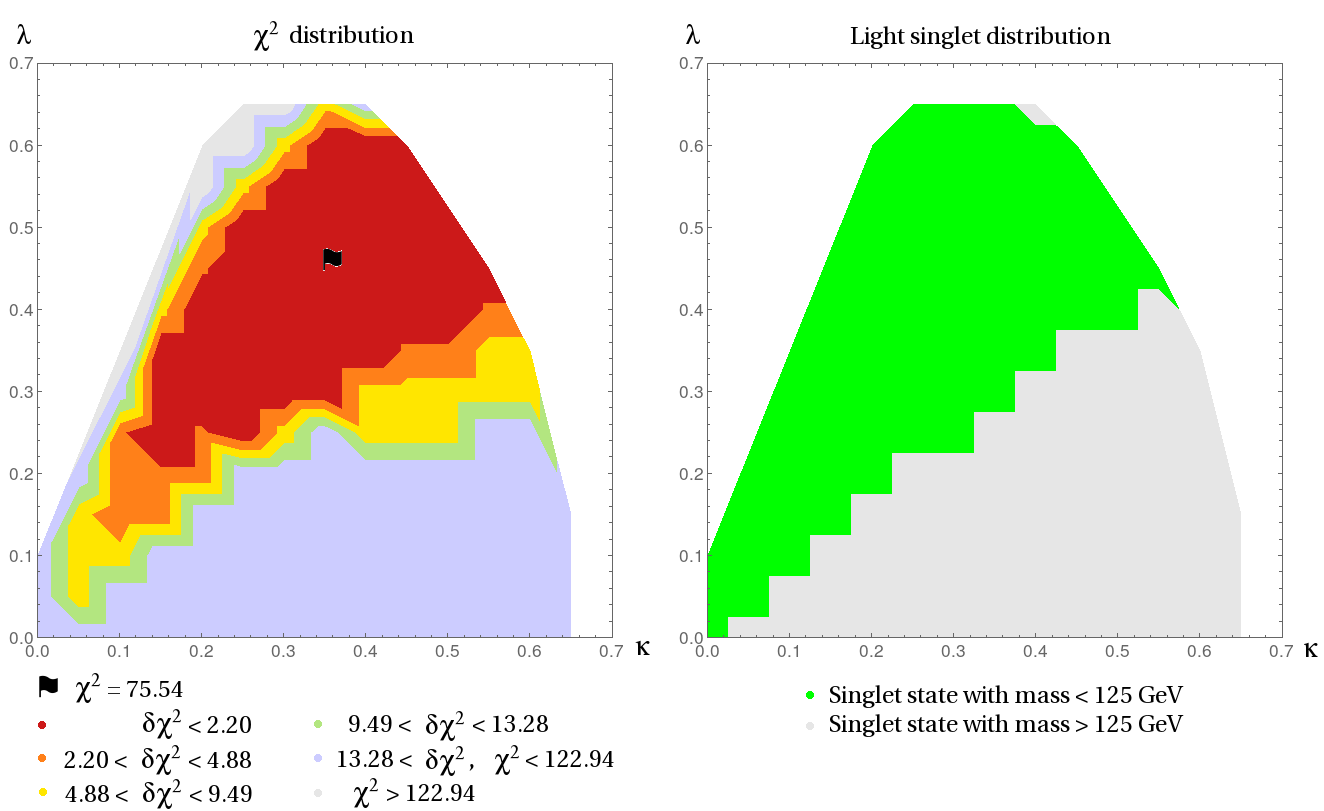}
  \caption{Scan in the $\{\kappa,\lambda\}$-plane for a heavy doublet sector: $\tan\beta=8$, $M_A=1$~TeV, $A_{\kappa}\in[-2,0]$~TeV, $\mu\in[120,2000]$~GeV, 
$2M_1=M_2=500$~GeV, $M_3=1.5$~TeV, $m_{\tilde{Q}_3}=1$~TeV, $m_{\tilde{Q}_{1,2}}=1.5$~TeV, $A_t=-2$~TeV, $A_{b,\tau}=-1.5$~TeV. The plot on the left-hand
side shows the $\chi^2$ distribution in the plane while the one on the right identifies the region with light singlet states.\label{heavydoub}}
\end{figure}

In Fig.~\ref{heavydoub}, we show such a region of the NMSSM parameter space, involving a heavy doublet sector ($M_A=1$~TeV) and $\tan\beta=8$: the points
are distributed in the $\{\kappa,\lambda\}$ plane. Points with $\lambda^2+\kappa^2\gsim0.7^2$ are discarded by the scan as they would lead to Landau 
Poles below the GUT scale. Moreover, the region with `large' $\lambda$ and moderate $\kappa$ tends to lead to unstable electroweak symmetry-breaking, 
as negative Higgs mass-squared are produced via the large singlet-doublet mixing (as soon as $\tan\beta\gsim5$). The best-fit points, with $\chi^2$ 
down to $\sim75$, involve a light singlet state: this fact is made evident when comparing the plots on both sides of Fig.~\ref{heavydoub}, as the region
including the best-fitting points (left-hand plot) largely coincides with that involving light singlets (right-hand plot). A determining factor for this
correlation rests with the uplift of the mass of the light Higgs-doublet via the mixing effect (of only $\sim1-2$~GeV in the particular configuration of
Fig.~\ref{heavydoub}). Note that varying $\tan\beta$ (or the squark spectrum) displaces the favoured region in the $\{\kappa,\lambda\}$ plane: indeed the 
magnitude of the mass contribution, which originates from the mixing among Higgs-states and shifts the mass of the light doublet state to a value closer
to the center of the LHC signals, changes accordingly. Another reason for the improved fit values in the presence of a light singlet is associated with small 
deviations (at the percent level) from the standard values in the couplings of the light doublet to SM particles: the mixing with the singlet results in
an increased flexibility of the doublet-composition of the state, which in turn allows for a possibly improved match with the measured signals. 

\begin{figure}[!htb]
    \centering
    \includegraphics[width=15.3cm]{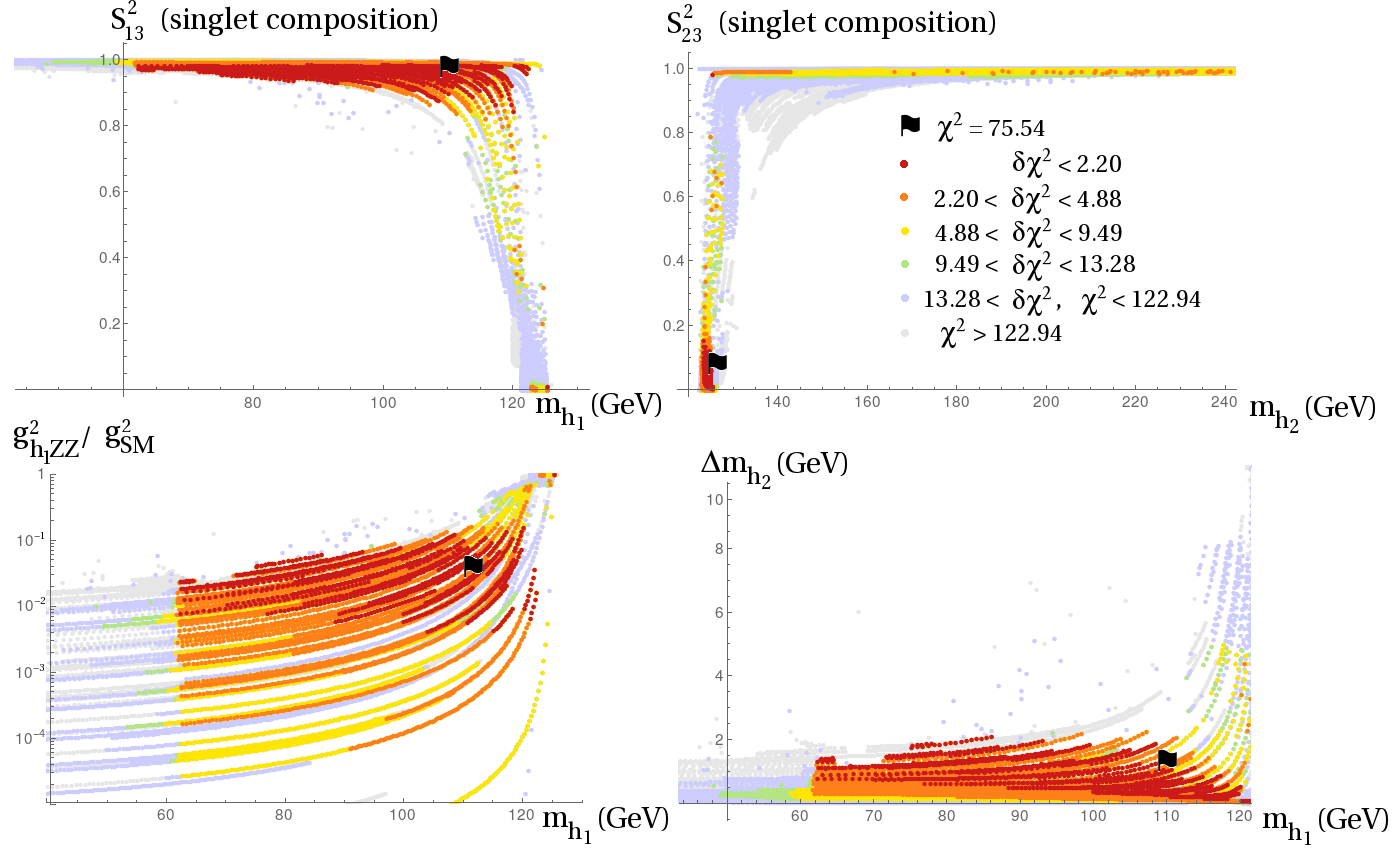}
  \caption{Same scan as in Fig.~\ref{heavydoub} but showing the characteristics of the CP-even states (mass, singlet-composition, relative squared coupling $h_1ZZ$, 
mass-shift of the doublet-like $h_2$).\label{lightsing}}
\end{figure}
The composition of the two lightest CP-even states in the scan of Fig.~\ref{heavydoub} is displayed in the upper part of Fig.~\ref{lightsing}: $S_{ij}$ 
denotes the orthogonal matrix rotating the CP-even Higgs sector from the gauge eigenstates -- second index `$j$'; $j=3$ stands for the singlet component -- to 
the mass eigenbase -- first index `$i$'; the mass states are ordered with increasing mass. One observes that significant singlet-doublet mixing up to 
$\sim20\%$ can be reached in the vicinity of $m_{h_1^0}\sim100$~GeV, although best-fitting points show a mixing under $\sim5\%$. This latter fact is 
related to the size of the mass-shift bringing the mass of the doublet-like state $m_{h_2^0}$ in agreement with the window of the LHC signal (larger 
mixing would lead to $m_{h_2^0}$ beyond the desirable $\sim125$~GeV range in the present configuration).

This mass shift of the doublet state via its mixing with the light singlet, $\Delta m_{h_2^0}$, is defined in the following fashion: regarding the heavy 
doublet sector as essentially decoupled, the squared-mass matrix of the singlet and light-doublet CP-even Higgs states may be approximated as the 
decoupled block\footnote{Note that we derive here an approximate formula, under the assumption that the heavy doublet state has negligible effect. This
approach is qualitatively justified, at least at the level of the mass shift, provided $M_A\gg M_Z$. A more exact expression, accounting for the heavy
doublet state though losing somewhat in clarity, may be derived in a similar fashion however.}:
\begin{equation}
 \begin{pmatrix}
  m^2_{h_S^0} & m^2_{h_S^0h_D^0}\\m^2_{h_S^0h_D^0} & m^2_{h_D^0}
 \end{pmatrix}=\begin{pmatrix}
  \cos\theta_S & -\sin\theta_S\\\sin\theta_S & \cos\theta_S
 \end{pmatrix}\begin{pmatrix}
  m^2_{h^0_1} & 0\\0 & m^2_{h^0_2}
 \end{pmatrix}\begin{pmatrix}
  \cos\theta_S & \sin\theta_S\\-\sin\theta_S & \cos\theta_S
 \end{pmatrix}
\end{equation}
Up to a sign, one can identify $\cos\theta_S\simeq S_{13}$, which determines the singlet-doublet mixing angle. In the presence of a lighter singlet-like 
state, the upward shift of the doublet state is defined as $\Delta m_{h_2^0}\equiv m_{h_2^0}-\sqrt{m^2_{h_D^0}}\simeq m_{h_2^0}-
\sqrt{m_{h_1^0}^2+S_{13}^2(m_{h_2^0}^2-m_{h_1^0}^2)}$. This quantity, still for the scan of Fig.~\ref{heavydoub}, is shown in the lower right-hand 
portion of Fig.~\ref{lightsing}: while the uplift in mass may reach up to $\sim8$~GeV, shifts of only $1-2$~GeV are favoured by the fit in this 
particular scan. Note that the formula that we have just derived only makes sense if $h_2^0$ is indeed the doublet-like state: for this 
reason, when displaying $\Delta m_{h_2^0}$ as a function of $m_{h_1^0}$, we cut the plot at $m_{h_1^0}<120$~GeV, since, for $m_{h_1^0}\gsim120$~GeV, the 
doublet-like state is likely to become $h_1^0$ in order to match the LHC signals at $\sim125$~GeV.

The plot on the lower left-hand side of Fig.~\ref{lightsing} shows the squared-coupling of the singlet-like state $h_1^0$ to $Z$ bosons -- controlling the 
production cross-section at LEP; it essentially coincides with $1-S_{13}^2$ here: at $m_{h_1^0}\sim100$~GeV, this quantity reaches $\sim5\%$ (for best fits) up 
to $\sim20\%$ of its SM value: for memory, the $\sim2.3\sigma$ LEP excess\footnote{Local significance without taking into account the `look-elsewhere'-effect.} in $H\to b\bar{b}$ observed in this mass-range would be compatible with a 
Higgs-like state, the squared coupling strength of which is reduced to $\sim10\%$ of its SM value. 

\begin{figure}[!htb]
    \centering
    \includegraphics[width=15cm]{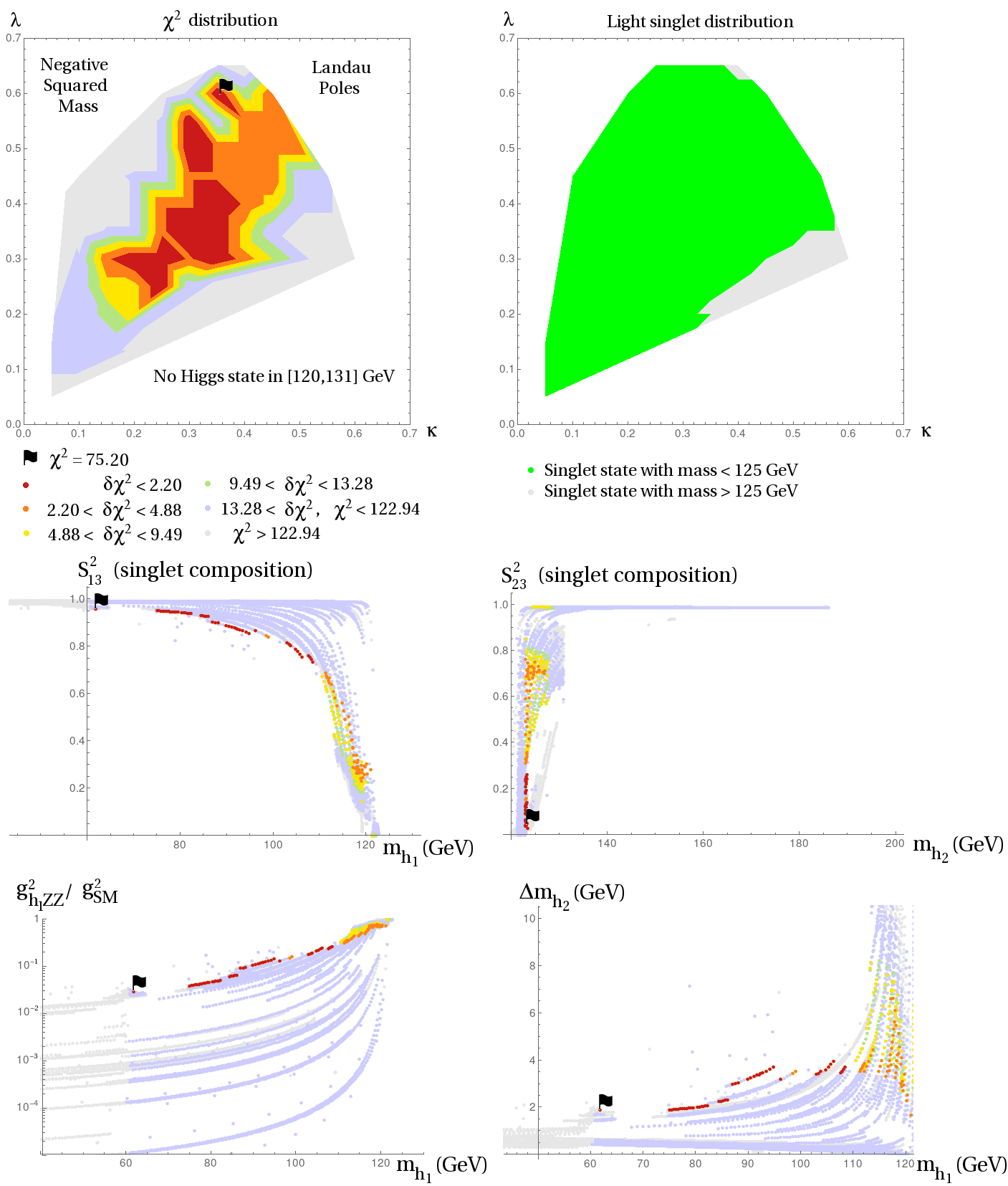}
  \caption{Scan in the $\{\kappa,\lambda\}$-plane for a heavy doublet: $\tan\beta=8$, $M_A=1$~TeV, $A_{\kappa}\in[-2,0]$~TeV, $\mu\in[120,2000]$~GeV, 
$2M_1=M_2=500$~GeV, $M_3=1.5$~TeV, $m_{\tilde{Q}_3}=1$~TeV, $m_{\tilde{Q}_{1,2}}=1.5$~TeV, $A_{t,b,\tau}=-1.5$~TeV.\label{largemix}}
\end{figure}
One can observe that (in this particular scan) the case where the lightest state is a doublet -- represented by the limit 
$S_{13}^2\to0$, $m_{h_1^0}\to125$~GeV -- yields a slightly worse fit than the scenario with a lighter singlet: two factors are at work here. The first 
one is related to the value of the mass characterising the doublet (`would-be-observed') state in this limit: it typically reaches $\sim121-123$~GeV only 
-- which lies on the margin of the uncertainty-allowed window. Note in particular that the mixing-effect tends to push the mass of the `visible' state (now $h_1^0$) 
into the `wrong direction' (to lower it) when the singlet is heavier. Yet, for some of the points under consideration, $m_{h_1^0}$ reaches $\sim125$~GeV, 
hence evades this first argument: in this case, the main penalty with respect to the points involving a lighter singlet originates from the details of 
the couplings of the `observed' state to SM particles, hence of its production and decay rates at the LHC. While both configurations -- with a lighter 
singlet or a lighter doublet -- provide (doublet) couplings within a few percent of each other, and of those values that a SM Higgs boson at this mass 
would take, small deviations can provide a closer match to the LHC data. In our particular scan, for instance, the $\gamma\gamma$ rate is slightly 
enhanced when the lighter state is dominantly singlet, resulting in an improved agreement with the ATLAS measurement.

More generally, the effects that 
the presence of a light singlet state may have on the couplings of the light doublet are related to the increased flexibility inherent in the $3\times3$
Higgs-mixing matrix $S_{ij}$ when compared to the case of a pure doublet $2\times2$ matrix. While one degree 
of freedom controls the singlet-composition of the Higgs state at $\sim125$~GeV -- i.e.\ its `invisible', for phenomenological reasons subdominant, 
component --, the other two modulate the relative proportions amongst the two doublet components $H_u$ and $H_d$ -- which are fixed in the case of pure 
doublets: in the limit $M_A\gg M_Z$, the corresponding $H_d/H_u$ ratio would be $\sim\tan^{-1}\beta$ --, therefore granting room for small deviations (or 
not, if the relative proportions are left unchanged) at the level of the couplings, with respect to the naive SM-like case. While this mechanism would 
offer an interpretation for slightly non-SM couplings, should this case be motivated by precision measurements of the Higgs properties, note that 
similar effects can also be obtained e.g.\ via loop-effects involving the supersymmetric spectrum. We shall come back to the question of non-standard 
couplings of the `observed' state in the following section (\ref{ltb}).

We complete this discussion with Fig.~\ref{largemix}, whose scan differs from the previous one only by a lower value of the trilinear stop coupling 
$|A_t|$ -- which thus tends to decrease the magnitude of the corrections to the mass of the light doublet originating in radiative effects. The 
situation is essentially comparable to the previous case, except for the fact that 
larger uplifts of 
the mass of the doublet-like $h_2^0$ -- $\Delta m_{h_2^0}\sim2-4$~GeV -- are now favoured. Larger singlet-doublet mixings, hence larger squared 
couplings of the light singlet to $Z$ bosons, $1-S_{13}^2\simeq15-20\%$, are correspondingly preferred, in the vicinity of $m_{h_1^0}\sim100$~GeV. With 
slightly heavier singlets $m_{h_1^0}\simeq110-115$~GeV, we observe that large mixings, up to $\sim25\%$ may appear. Note that the best-fit point
lies in an isolated region with mass close to $\sim60$~GeV: this isolated position results both from the limited scan density and from the 
marginal situation with respect to the LEP limits.

Finally, we note that, in the plots of Fig.~\ref{lightsing} and \ref{largemix}, the mass of the singlet may reach values as low as $\sim62$~GeV without spoiling 
the quality of the fit. The case of states under $\sim62$~GeV opens the possibility of $h_2^0\to2h_1^0$ decays and will be treated in a separate 
section (\ref{la1}).

\begin{figure}[!htb]
    \centering
    \includegraphics[width=15.3cm]{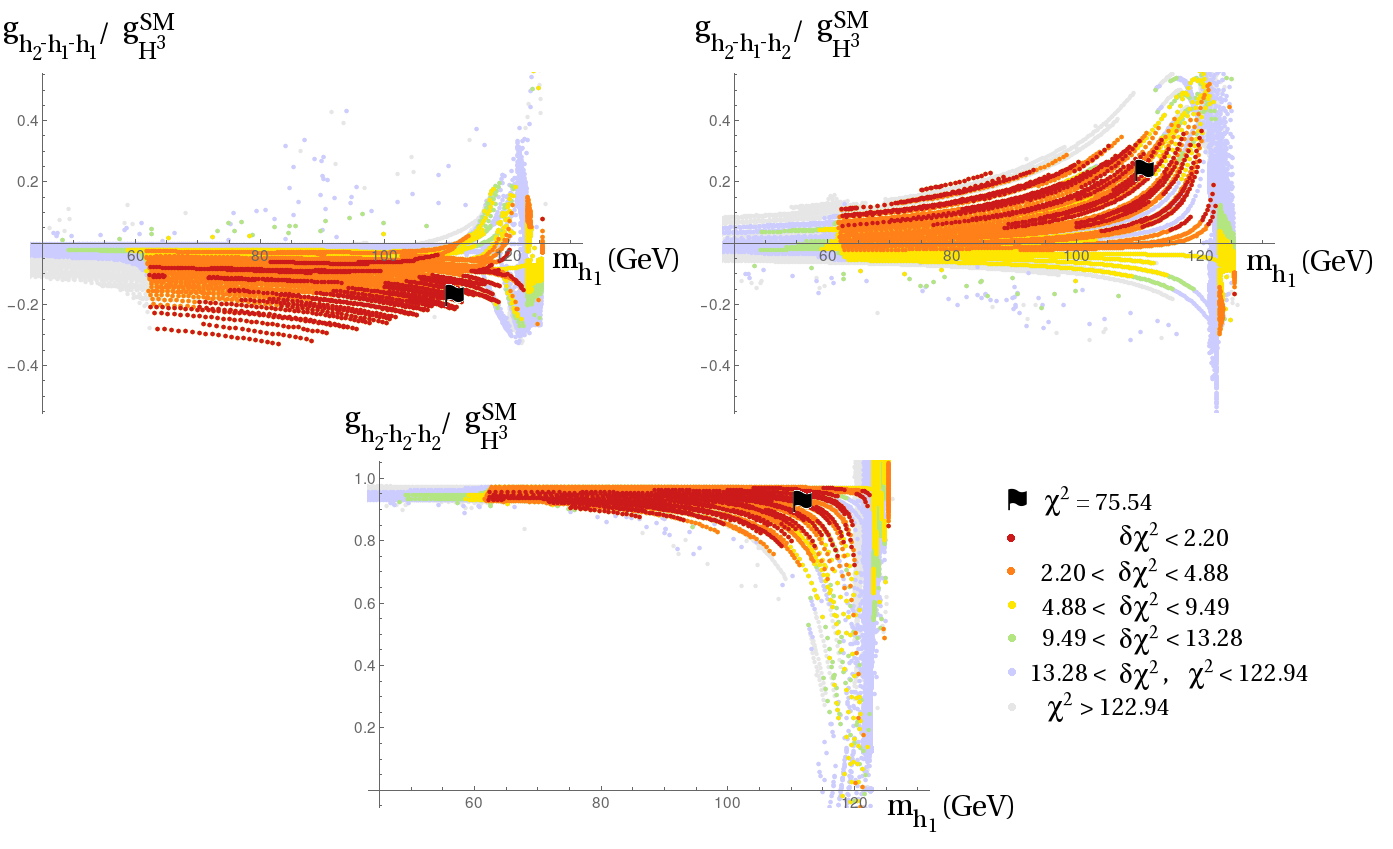}
  \caption{Same scan as in Fig.~\ref{heavydoub}, now displaying the strength of the Higgs-to-Higgs couplings. The latter were normalized 
by the SM triple-Higgs coupling $g_{H^3}^{\mbox{\scriptsize SM}}\simeq192$~GeV (see text).\label{lightpair}}
\end{figure}
While precision tests at the level of the couplings of the Higgs state at $\sim125$~GeV could provide arguments in favour of this scenario involving a
light singlet, as we already mentioned, the most convincing evidence would lie in the detection of the light singlet itself. The latter is likely to 
appear as a `miniature' Higgs boson, i.e.\ with decay rates grossly comparable to those of a SM Higgs boson at the same mass but a reduced production 
cross-section (and a smaller width) -- indeed the singlet component decouples from SM-particles and production is thus only achieved via a small doublet component. The 
observability of the singlet state thus critically rests with the magnitude of its doublet component. In the discussion above, we stressed that the doublet 
composition ($1-S_{13}^2$) of the light CP-even singlet may reach $O(10\%-20\%)$ and it is likely that the corresponding signal would be large enough to 
allow detection -- at least in the form of a local excess, while discovery at the $5\sigma$ level could remain challenging, provided the LHC searches are 
extrapolated to the low-mass region. Yet, apart from LHCb searches in the $\tau\tau$ channel \cite{Aaij:2013nba}, which are still quite far from the 
necessary sensitivity in order to probe such a scenario, only the recent ATLAS results in the $\gamma\gamma$ channel \cite{Aad:2014ioa} consider 
masses below $\sim110$~GeV. (Note that, in the scans of Fig.~\ref{lightsing} and \ref{largemix}, the typical cross-section at $8$~TeV in the diphoton 
channel lies below $1$~fb; see also the discussion of Fig.~\ref{hgamgam} below.) On the other hand, smaller singlet-doublet mixings, at the percent level or below, would also fit the picture 
adequately, allowing for a mass-uplift of the doublet and / or small variations in its couplings to SM particles -- such are actually the best-fit points 
of the scan in Fig.~\ref{heavydoub}, which can be found in the appendix. The visibility of the light singlet in direct production should become 
increasingly difficult as its doublet component becomes small. 

Alternative search strategies have been suggested, as the Higgs-to-Higgs couplings need not follow the same pattern as couplings to SM particles: 
singlet-doublet Higgs couplings could in principle allow for singlet production from e.g.\ the observed state, via Higgs-pair productions. It was 
stressed, however, that even such channels did not ensure the visibility of the light singlet, as the presence of this particle does not necessarily 
entail significantly larger inclusive rates than those of a single doublet state \cite{pairprod} (although an enhancement by a factor up to $2-3$ has 
been reported for certain points). Let us highlight the fact that trilinear singlet-doublet couplings, while possibly as large as SM 
Higgs-to-Higgs couplings, $g_{H^3}^{\mbox{\scriptsize SM}}\equiv\frac{3m_{H}^{2\,\mbox{\tiny SM}}}{\sqrt{2}v}\simeq192$~GeV, where we took 
$m_{H}^{\mbox{\scriptsize SM}}=125.6$~GeV, $v\simeq174$~GeV, may also remain much smaller without contradicting the light-singlet scenario. 
In Fig.~\ref{lightpair}, we display the strength of the triple Higgs couplings involving the light singlet and the light doublet in the scan of 
Fig.~\ref{heavydoub}: while the trilinear coupling of the state $h_2$ with mass $\sim125$~GeV remains SM-like, the couplings involving both singlet and 
doublet-like states reach only up to $\sim30\%$ of the SM-strength. Assuming that the cross section for pair production follows a similar pattern 
-- which is only justified at high center of mass energy --, we see that a discovery in such channels would be challenging experimentally. 
On the other hand, pair production close to threshold is very sensitive to a small imbalance among triangle and box contributions so that the 
presence of a light singlet may affect this observable. An estimate of such effects goes beyond the scope of this paper, and we refer the reader to the 
discussions in \cite{pairprod}. As a summary, let us stress that, if Higgs-pair production is a viable search channel for light singlet states, should 
this state be present in nature, it does not automatically ensure the discovery of the singlet at the LHC. Production associated to Higgs-gauge 
couplings was also discussed in \cite{King:2014xwa} but again depends critically on the magnitude of the doublet component of the light singlet-like state. 

Other production modes would involve the supersymmetric spectrum or the heavy Higgs states. In particular, the decays 
$h_3^0\to h_1^0h_1^0,\ h_1^0h_2^0,\ h_2^0h_2^0$ might be discovered for a resonant production of $h_3^0$. In this respect, NMSSM effects may intervene 
at several levels:
\begin{itemize}
 \item The doublet-to-doublet Higgs couplings differ from their MSSM equivalent due to NMSSM-specific terms in the Higgs potential ($\propto\lambda,\ 
\kappa$) so that the associated width may differ significantly.
 \item Singlet-doublet couplings induce a decay of the heavy Higgs into the singlet-like state.
 \item Kinematically accessible additional final states (e.g.\ singlinos) also affect the branching ratios of $h_3^0$. More generally, the branching ratios are
strongly dependent on the details of the supersymmetric spectrum.
 \item The decay rates (into e.g.\ $b\bar{b}$, $\gamma\gamma$) of the decay products $h_1^0$, $h_2^0$ may vary with respect to the 
naive standard rates. 
\end{itemize}
Unless the decay products $h_1^0h_1^0,\ h_1^0h_2^0,\ h_2^0h_2^0$ can be observed separately (using kinematical cuts), it would thus be difficult 
to infer the presence of a singlet from the inclusive rates, as several factors could explain a deviation from the predictions of a type II 2HDM.

\begin{figure}[!htb]
    \centering
    \includegraphics[width=15cm]{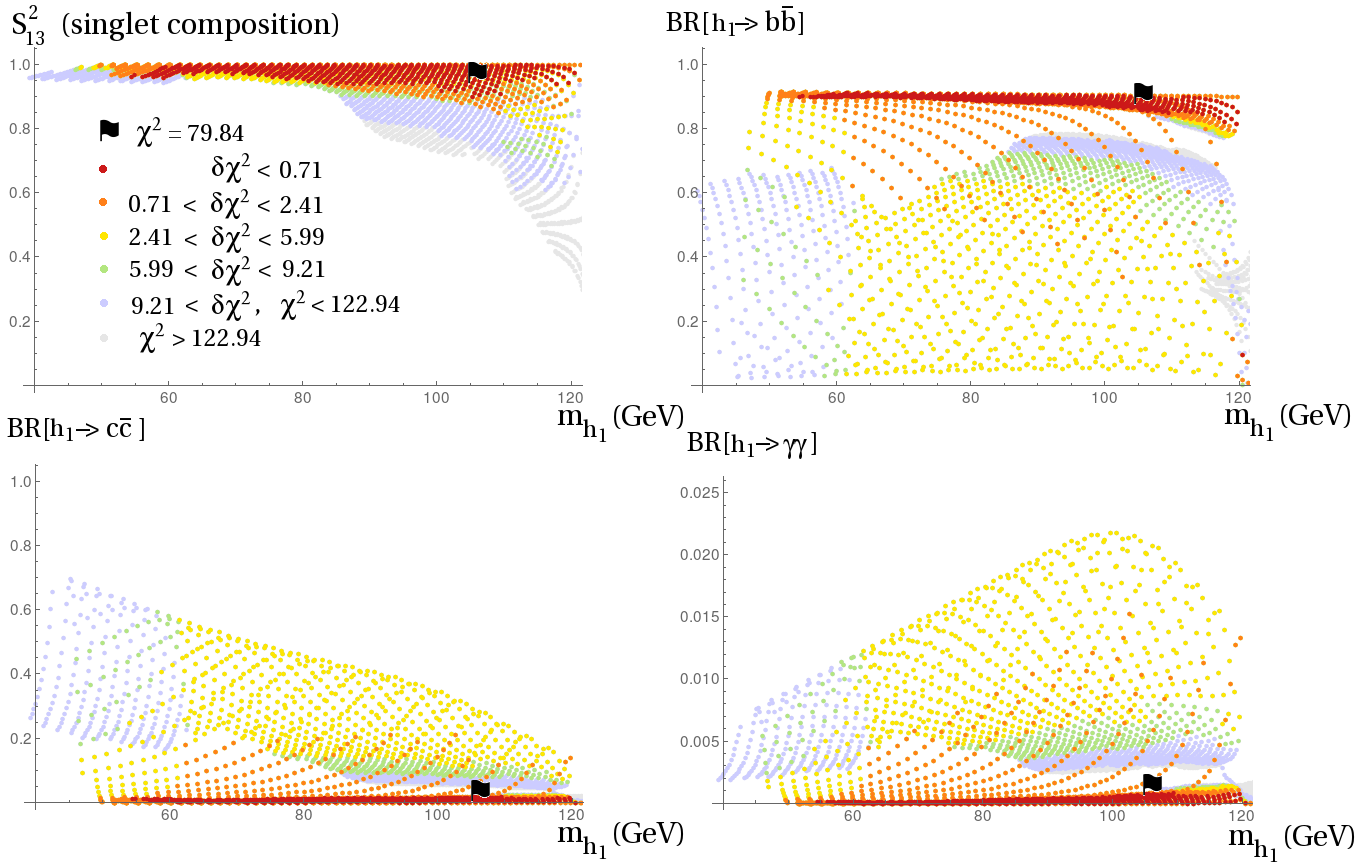}
  \caption{Modified rates of the light singlet: $\lambda=0.1$, $\kappa=0.05$, $\tan\beta=12$, $M_A\in[0,2]$~TeV, $A_{\kappa}\in[-2,0]$~TeV, $\mu=125$~GeV, 
$2M_1=M_2=500$~GeV, $M_3=1.5$~TeV, $m_{\tilde{Q}_3}=1$~TeV, $m_{\tilde{Q}_{1,2}}=1.5$~TeV, $A_{t}=-2$~TeV, $A_{b,\tau}=-1.5$~TeV.\label{singrates}}
\end{figure}
Another remark on the decay rates of the light singlet is in order: as mentioned above, one may naively expect them to coincide 
(coarsely) with those of a SM Higgs boson at the same mass. Yet, it has also been stressed that these rates could show unconventional behaviours in
specific cases, at low \cite{Ellwanger:2010nf} or large $\tan\beta$ \cite{singsearch}: couplings to down-type quarks could be suppressed indeed, 
which would lead to an apparent enhancement of decay channels such as $c\bar{c}$ or $\gamma\gamma$. Extreme cases with an up-to-seven-times enhanced 
diphoton branching fraction of the singlet, allowing for cross-sections at the level of their SM equivalent -- despite the reduced production 
cross-section --, have received much attention in view of their remarkable consequences on the possible discovery of such a singlet. We illustrate this 
possibility of non-conventional singlet rates with Fig.~\ref{singrates}: with an intermediate value of $\tan\beta=12$, we observe that the $b\bar{b}$ 
branching fraction may be strongly suppressed, while the other rates (here $c\bar{c}$ and $\gamma\gamma$) are enhanced, together with acceptable fit 
values -- note however that the best-fit points lie in a region of more SM-like behaviour.

Note that points involving light singlets are quite common in the NMSSM parameter space. The only difficulty consists in stabilizing the low singlet
mass and keeping the singlet-doublet mixing under control: too strong a mixing would push the squared mass of the lightest state towards negative values. 
The typical scale entering the singlet mass is $\frac{\kappa}{\lambda}\mu$, so that light singlets favour low ratios $\kappa/\lambda$. As 
$\tan\beta$ increases, however, the balance among terms entering the mixing of the light doublet and singlet CP-even states is disturbed,
such that the region with large $\lambda$ and low $\kappa$ becomes increasingly unstable. For larger values of $\kappa/\lambda$, one observes that 
$\mu$ tends to be driven to low values $\sim100$~GeV -- in order to keep the singlet mass at the electroweak scale without relying too much on 
accidental cancellations. This issue becomes even more severe when $\lambda\sim\kappa$ becomes large: one then relies exclusively on
the accidental cancellation in the singlet diagonal and the singlet-doublet mixing mass-matrix entries. It is therefore most natural to 
consider the scenario involving a light singlet in the low $\tan\beta$ regime, allowing for small $\kappa/\lambda$: this is the focus of the next section.

\section{\boldmath Low $\tan\beta$ and large $\lambda$}\label{ltb}
The region with low $\tan\beta$ ($\sim2$) and large $\lambda$ ($\sim0.6-0.7$) is particularly interesting in the NMSSM parameter space: as we just
mentioned, light CP-even singlets under $\sim125$~GeV appear most naturally there; furthermore, the squared mass of the light-doublet Higgs state 
receives an F-term contribution of the form $\lambda^2v^2\sin^2{2\beta}$ at tree-level (see e.g. Eq.(2.23) of \cite{NMSSM}), which is maximized in this regime, 
so that significantly smaller radiative corrections than in the MSSM case are needed to reach the experimental value of the observed signal; finally, 
the region of low $\tan\beta$ is `specific' to the NMSSM, in the sense that current phenomenological requirements on the Higgs sector forbid it in the 
MSSM. One may typically consider this setup in the Peccei-Quinn (PQ) limit $\kappa/\lambda\ll1$ -- note that, in view of the Landau Pole constraints, 
turning to relatively large values of $\lambda$ automatically implies moderate values of $\kappa$ anyway.

\begin{figure}[htb!]
    \centering
    \includegraphics[width=15cm]{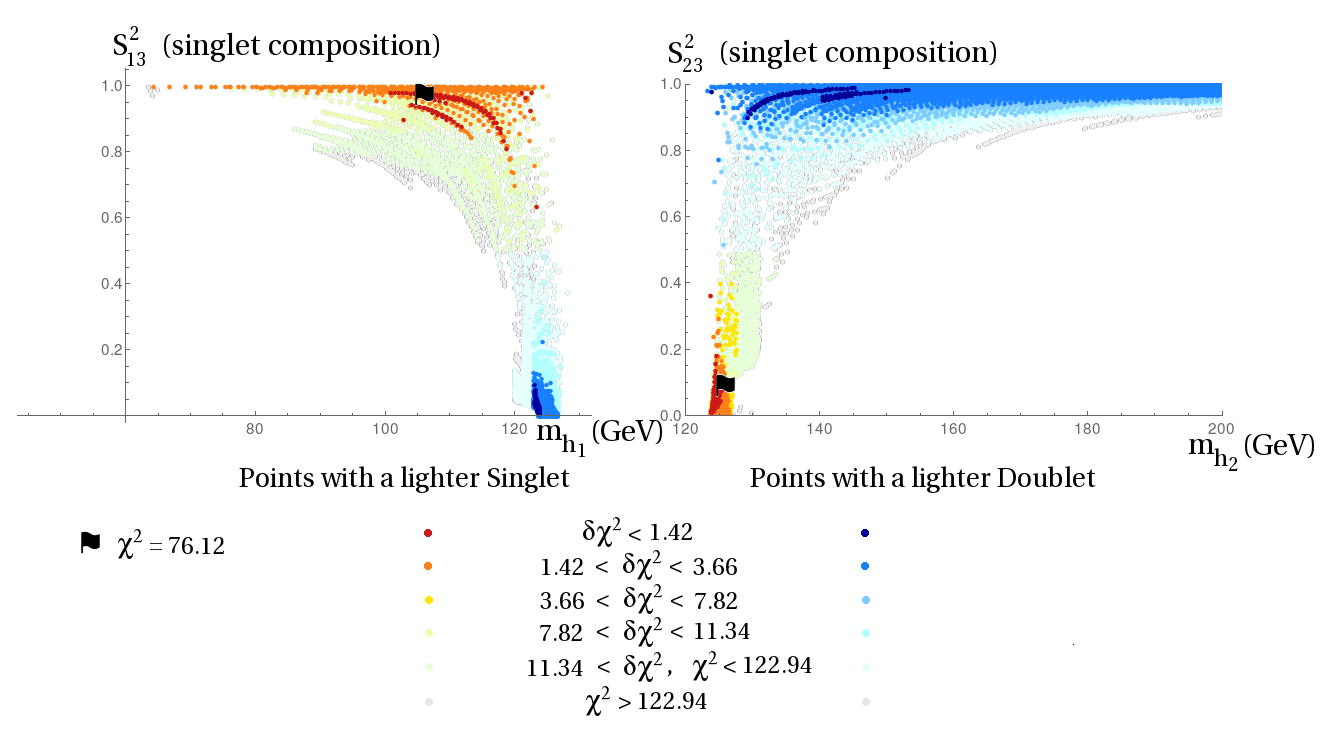}
  \caption{$\chi^2$ in the PQ-limit of the NMSSM, with a very-light supersymmetric sector: $\tan\beta=2$, $\lambda=0.7$, $\kappa=0.1$, 
$\mu\in[120,2000]$~GeV, $M_A\in[0,3]$~TeV, $A_{\kappa}\in[-500,500]$~GeV, $2M_1=M_2=150$~GeV, 
$M_3=1.5$~TeV, $m_{\tilde{Q}_3}=500$~GeV, $m_{\tilde{Q}_{1,2}}=1.5$~TeV, $m_{\tilde{L}}=110$~GeV,$A_{t,b,\tau}=-100$~GeV.}
  \label{PQlim}
\end{figure}
Given that, in this context, one does not have to rely on large corrections from the stop sector to uplift the mass of the light doublet Higgs state 
into the mass range $\sim125$~GeV where it can be identified with the LHC-observed signal, we will consider moderate stop masses ($\sim0.5$~TeV) and 
trilinear couplings ($\sim0.1$~TeV). While we do not check the compatibility of this choice with LHC limits on supersymmetric searches, note that this 
is not a binding requirement but purely an illustration of the fact that radiative corrections to the Higgs mass are of lesser importance in this region of the 
parameter space: the specific tree-level contribution and / or the mixing with a light singlet are mechanisms enough to generate the mass of the light 
doublet Higgs state in the $\sim125$~GeV range. In fact,
if in this regime the SUSY sector is such that it generates large radiative
corrections to the mass of the light Higgs doublet state (e.g.\ via a sizable 
mixing in the stop sector), this would lead to predictions for the mass
of the light doublet state in the region of low $\tan\beta$ and large $\lambda$ 
that tend to be higher than the mass value detected at the LHC
-- up to $\sim140$~GeV if it is the lightest CP-even state 
and beyond if it is the second lightest: see e.g.\ \cite{largelamb}. Yet, a 
Higgs mass that is compatible with the observed value can still be
obtained even in the presence of large contributions
from both the NMSSM tree-level and SUSY radiative corrections through the 
effect of the mixing between the doublet and the singlet. This mixing
effect lowers the mass of the light
CP-even doublet, provided the singlet is heavier. Note finally that $B$-physics is of almost no concern in this scenario since both charged-Higgs and 
supersymmetric effects remain small, due to a heavy $H^{\pm}$ (decoupling limit) and low $\tan\beta$ (absence of an enhancing-factor in radiative 
effects). On the other hand, low values of $\tan\beta$ tend to suppress supersymmetric contributions to $(g-2)_{\mu}$, translating into a typical pull 
of a few units in $\chi^2$. The presence of light sleptons / charginos / neutralinos (with mass close to $100$~GeV) could balance this effect, however, 
so that we will assume low masses for these states in the following.

\begin{figure}[htb!]
    \centering
    \includegraphics[width=15cm]{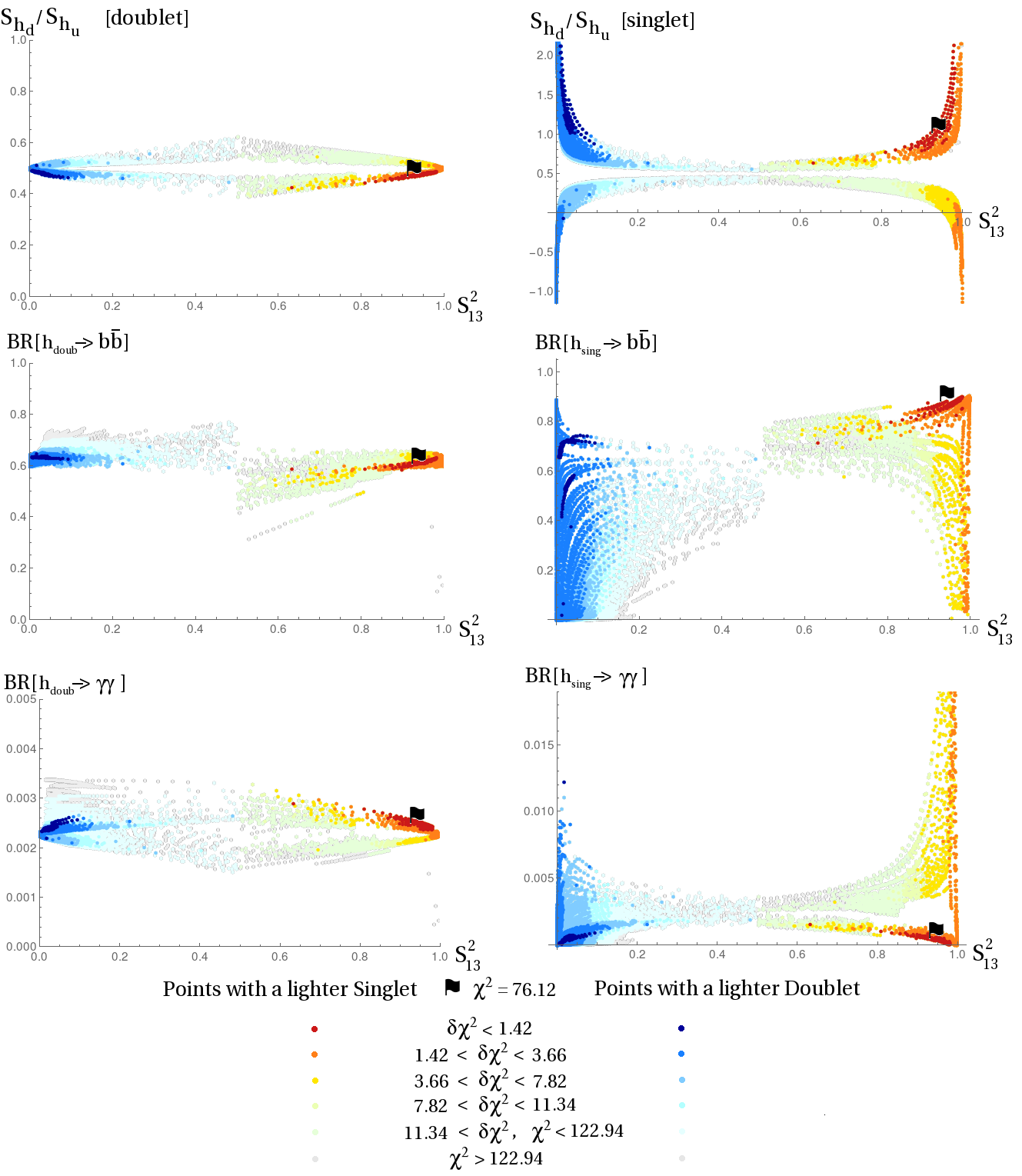}
  \caption{Same scan as in Fig.~\ref{PQlim}: consequences of the three-state mixing on the couplings of the light states are shown. The plots
on the left concern the mostly-doublet state (identified with the observed signal at $\sim125$~GeV); those on the right give information concerning
the lighter (yellow-red points) or heavier (bluish points) singlet. The first pair of plots displays the proportion of $H_d/H_u$ components (in 
comparison to the value $\tan^{-1}\beta=0.5$ expected in the decoupling limit). The branching ratios into $b\bar{b}$ and $\gamma\gamma$ are provided 
in the lower part of the figure.}
  \label{cpsing}
\end{figure}
Fig.~\ref{PQlim} shows the $\chi^2$ distribution in a scan where $\tan\beta=2$, $\lambda=0.7$, $\kappa=0.1$, and the super\-symmetric spectrum is 
relatively light (at least for the third generation of sfermions, with small trilinear couplings). The plots illustrate how the fit distributes
in terms of the mass and singlet composition -- i.e.\ singlet component squared $S_{i3}^2$ -- of the first and second lightest CP-even Higgs states. 
Points where the lightest Higgs state has a dominantly singlet nature ($S_{13}^2>0.5$) are shown in yellow-red shades. Bluish tones correspond to 
points where the lightest state is dominantly doublet ($S_{13}^2<0.5$). Both configurations give an excellent agreement with the measurements 
reported by the LHC and the TeVatron, and their fit values improve somewhat on the SM limit. One observes that the best fit points (with 
$\chi^2\sim76-79$) tend to cluster in the vicinity of $S_{i3}^2=0$ or $1$, that is for moderate singlet-doublet mixing. This seems reasonable since one 
expects a `full' doublet state at $\sim125$~GeV: the corresponding experimental signals would have been suppressed in proportion to the singlet 
composition otherwise. Still, fairly large values of the mixing ($S_{i3}^2\sim0.5$) turn out to be a possible scenario, giving acceptable $\chi^2\sim77-80$: 
this situation occurs only when both states are almost degenerate and within a few GeV of $\sim125$~GeV. This last configuration will be studied in 
more detail in section \ref{deg}.
\begin{figure}[htb!]
    \centering
    \includegraphics[width=15cm]{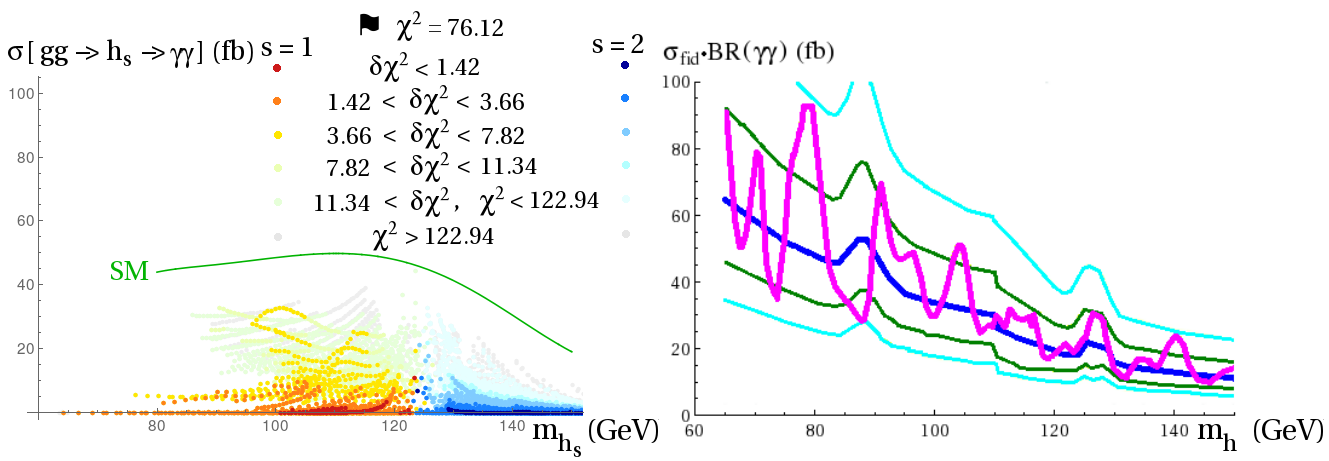}
  \caption{On the left: gluon-gluon-fusion cross-section for the mostly-singlet state, then decaying into a pair of photons, for a center of mass energy
of $8$~TeV, resulting from the scan of Fig.~\ref{PQlim}; the corresponding value for a SM Higgs boson is given by the green curve. On the right, a 
reproduction of the ATLAS limit on the fiducial cross-section for a light Higgs state (in the presence of the $\sim125$~GeV one) decaying into photons.}
  \label{hgamgam}
\end{figure}
\begin{figure}[htb!]
    \centering
    \includegraphics[width=16cm]{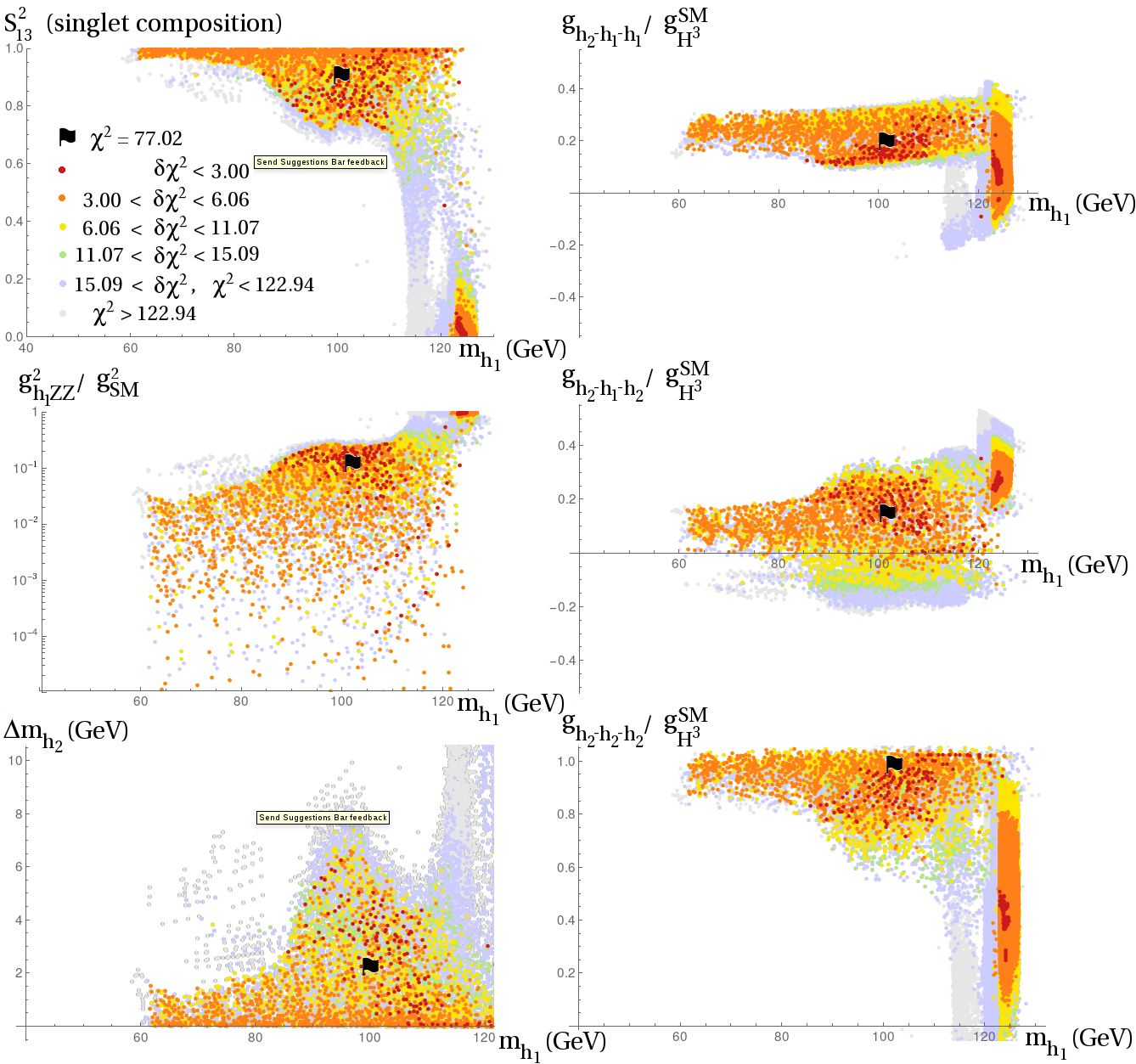}
  \caption{Similar to Fig.~\ref{PQlim}, but with an additional scan on $\tan\beta\in[1,4]$ and $\lambda\in[0.6,0.7]$ in order to probe possible 
singlet-doublet mixing. On the left-hand side, from top to bottom: singlet composition of the light CP-even Higgs state, squared coupling of the light 
Higgs state to $Z$-bosons relative to the SM, size of the mass-uplift for the doublet state (as defined in the previous section). On the right-hand 
side: magnitude of the triple Higgs couplings (relative to $g_{H^3}^{\mbox{\scriptsize SM}}$).}
  \label{PQlimmix}
\end{figure}

In Fig.~\ref{cpsing}, we display some information concerning the coupling properties of the light states to SM particles obtained with the scan of 
Fig.~\ref{PQlim}: the plots on the left-hand side illustrate the possible effects that the three-state mixing could have on the couplings of the 
doublet-like state -- the one which is identified with the LHC-observed signal. The plot on the upper part shows the proportion of $H_d/H_u$ components 
in this state as a function of the 
mixing: deviations of up to $\sim25\%$ appear relative to the naive ratio $0.5=\tan^{-1}\beta$ (expected in the decoupling limit), and the maximal 
effects are obtained for maximal mixing $S_{13}^2\sim0.5$. Deviations at the level of the branching ratios reach $\sim25\%$ for $b\bar{b}$ and 
$\sim50\%$ for $\gamma\gamma$. Note that best fit regions favour more moderate variations however. Note also that, even for large mixings, the 
deviations in doublet proportions or in branching ratio may remain negligible.

The plots on the right-hand side of Fig.~\ref{cpsing} provide similar information but concerning the singlet-like state. We observe that the corresponding
coupling properties may depart very significantly from the naive behaviour when the state is almost purely singlet -- that is, in the vicinity of $S_{13}^2\sim0$ 
and $\sim1$ -- and that the corresponding points offer excellent fit to the Higgs data. The $b\bar{b}$ channel may then become subdominant while the 
diphoton channel is enhanced by a factor of up to $\sim7$. In such a spectacular case, the light singlet could be more easily observed in direct 
production at the LHC. On the other hand, the fit tends to associate the large diphoton branching fraction tightly with the limit of a pure singlet 
state, that is with vanishing production cross-sections. Note also that the naive scenario of a singlet-like state with dominant decays towards 
down-type fermions is also represented and actually provides the best-fit points of the scan. Unconventional decay rates also appear as a possibility 
when the singlets are beyond $\sim125$~GeV (blue points), even though the maximal diphoton rates remain below $\sim1\%$.

In Fig.~\ref{hgamgam}, we study how the Higgs production cross-section at $8$~TeV compares to the ATLAS limits on the fiducial 
cross-section for the diphoton decay channel \cite{Aad:2014ioa}. We estimated the cross-section for the light Higgs states of the scan of Fig.~\ref{PQlim}
in the following way: we multiplied the SM gluon-gluon-fusion cross-section delivered by {\sc SusHi} \cite{sushi} by the squared effective coupling of 
$h_1^0$ to gluons, relative to its SM value at the same mass, and the diphoton branching ratio of $h_1^0$. We observe that the cross-section may almost reach 
the order of magnitude probed experimentally, both when the singlet is heavier or lighter than $125$~GeV (note that in the immediate vicinity of $125$~GeV,
comparing the cross-section of the mostly-singlet state with the ATLAS limit has limited sense, due to the possibly large mixing between singlet and doublet
states), although the best-fitting points tend to cluster around much smaller values -- at or below the $1$~fb range. Further searches in the low-mass region, 
in the diphoton but also in the fermionic channels, would be an interesting probe and place limits on the light-singlet scenario.

In Fig.~\ref{PQlimmix}, we vary $\tan\beta$ and $\lambda$ somewhat so as to modulate the strength of the F-term contribution to the tree-level 
doublet Higgs mass. As a result, larger singlet-doublet mixings are favoured: the two-state mixing uplift can indeed compensate the decreased tree-level
contribution and thus help maintain the mass of the light doublet state in the vicinity of $\sim125$~GeV. In agreement with our discussion in section 
\ref{lsing}, we observe that large singlet-doublet mixing, up to $\sim25\%$, may be achieved for a singlet mass in the range $[90-100]$~GeV, with 
excellent fit-values to the Higgs measurement data. Therefore, this low $\tan\beta$ regime also motivates the search for a light singlet state, possibly 
responsible for the $\sim2.3\,\sigma$ (local) excess in 
the LEP $e^+e^-\to Z h\to b\bar{b}$ channel. The magnitude of the mass uplift for the doublet state in this 
region may again reach up to $6-8$~GeV, as we observe on the plot on the bottom left-hand side of Fig.~\ref{PQlimmix}.

Concerning the prospects of discovery of the light state in pair production, the Higgs-to-Higgs couplings in the scan of Fig.~\ref{PQlimmix} are 
displayed on the right-hand side of this figure. The typical magnitude is about $10-40\%$ of $g_{H^3}^{\mbox{\scriptsize SM}}$ for 
$h_2-h_1-h_1$, $0-30\%$, for $h_2-h_2-h_1$, and $85-100\%$, for $h_2-h_2-h_2$ (in the region where the lightest state is a singlet). The impact of the
singlet-doublet couplings on the apparent Higgs pair production cannot be simply estimated as the latter depends on several interfering diagrams. We
see however that the typical couplings reach $\sim30\%$ of the pure-doublet value.

Although all these observations are essentially similar to our discussion in section \ref{lsing}, the crucial point rests upon the fact that such a
Higgs phenomenology is also achievable in this low $\tan\beta$ / large $\lambda$ regime, {\em without} relying on large radiative corrections to the 
Higgs masses. This provides a motivation for relatively-light supersymmetric spectra (at least, as far as the third generation is concerned). In the 
case where the search for stops at the LHC would provide experimental support for such a configuration, deviations of the Higgs couplings from the SM 
expectations could be generated at the loop level and be addressed in precision tests.

\section{\boldmath Light CP-odd (even) Higgs states under $m_{h[125]}/2$}\label{la1}
A durably-considered NMSSM scenario (see for instance \cite{cascade}) is that involving light neutral states, e.g.\ a light CP-even singlet or a light 
CP-odd state, allowing for unconventional Higgs decays ($h\to2A_1/2h_1$). As the width of a SM-like Higgs is quite narrow -- $\sim4$~MeV at $\sim125$~GeV --, 
Higgs-to-Higgs decays can easily dominate the width of a Higgs state as soon as they are kinematically allowed. This configuration could have 
explained the invisibility of a doublet Higgs state $h$ at LEP -- provided $m_{h}\gsim85$~GeV (i.e.\ above the limit from the decay-mode independent 
search \cite{Abbiendi:2002qp}) and $A_1\to\tau^+\tau^-$ would represent the dominant decay channel of the pseudoscalar. Furthermore, this mechanism 
suggested an interpretation of the $2.3\,\sigma$ (local) excess in $e^+e^-\to Z h\to b\bar{b}$ originating from suppressed branching ratios into SM 
particles of a CP-even state at $\sim100$~GeV, instead of a suppressed production as in the case of a light 
singlet, and would have led to a reduced visibility of the conventional channels at the LHC -- even if the light CP-even doublet had been heavier than 
$100$~GeV. Considering the success of these conventional searches and the approximately SM-like behaviour of the LHC-observed state, Higgs-to-Higgs 
decays of the SM-like state no longer appear as a favoured option, unless one would correspondingly enhance the SM-like modes, which is not easily 
achieved in the context of the NMSSM. Yet, we already noted in the previous sections that the existence of light states -- under the 
kinematic threshold at $\sim62$~GeV -- was still possible (see e.g.\ Fig.~\ref{lightsing}); we will discuss here under what conditions. In brief, if the 
invisible decay channel is not kinematically closed, compatibility with the measured rates of the observed signal demands the decoupling of the light 
state -- suggesting a singlet-like nature with reduced Higgs-to-Higgs couplings \cite{lightCPodd}. 

\begin{figure}[!htb]
    \centering
    \includegraphics[width=15cm]{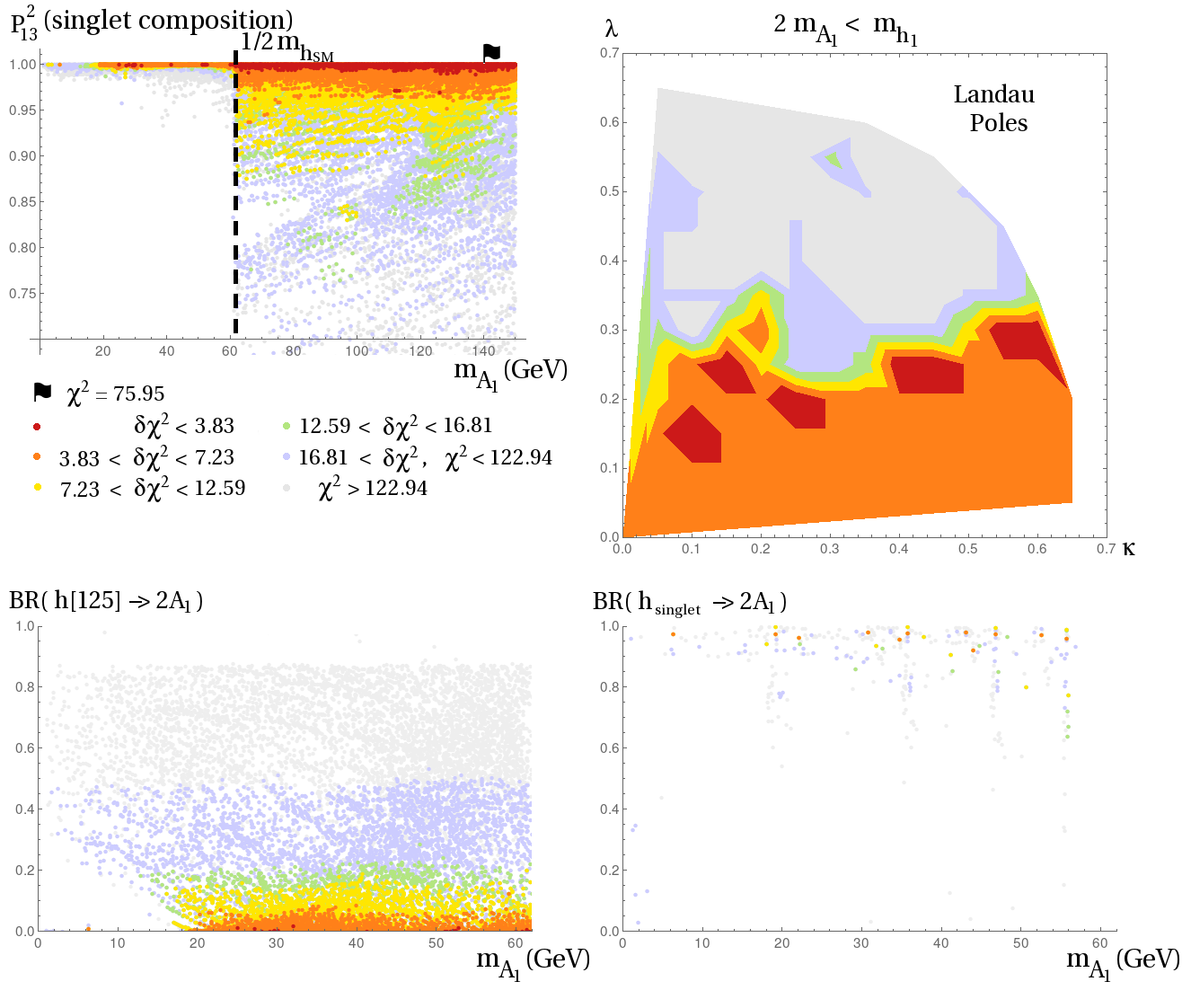}
  \caption{$\chi^2$ in the presence of a light CP-odd Higgs: $\tan\beta\in[2,22]$, $M_A\in[200,1200]$~GeV, $\mu\in[120,1600]$~GeV, $A_{\kappa}\in[-400,400]$~GeV, 
$\lambda\in[2\cdot10^{-4},0.65]$, $\kappa\in[2\cdot10^{-4},0.65]$, $2M_1=M_2=500$~GeV, $M_3=1.5$~TeV, $m_{\tilde{Q}_{1,2}}=1.5$~TeV, 
$m_{\tilde{Q}_3}=1.2$~TeV, $A_t=-2.5$~TeV, $A_{b,\tau}=-1.5$~TeV. The top-left plot shows the distribution of $\chi^2$ as a function of the light CP-odd mass 
and the singlet composition of this state ($P^2_{13}$); on the right, distribution of the points with $m_{A_1}<62$~GeV in the $\{\kappa,\lambda\}$ plane.
The figures at the bottom give the branching ratios $h[\sim125~\mbox{\small GeV}]\to2A_1$ on the left and $h_{\mbox{\tiny singlet}}\to2A_1$ (in
the presence of a singlet as the lightest CP-even state) on the right.}
  \label{lightA}
\end{figure}
Fig.~\ref{lightA} shows how the fit distributes in the presence of light CP-odd Higgs states. Note that the SUSY sector, in particular the 
mixing in the stop sector, is set to provide radiative corrections to
the mass of the light doublet Higgs of the appropriate magnitude in view of the 
signal at $\sim125$~GeV, as already discussed in section \ref{decoup}. While $\tan\beta$ is left free, larger values $\gsim10$ are accordingly favoured
by the requirement of a SM-like Higgs state at $\sim125$~GeV, but also by $(g-2)_{\mu}$ (for which the exact prefered range is determined by our
choice of spectrum in the slepton, neutralino and chargino sectors). One 
observes in the plot on the top left-hand quadrant that, when 
the CP-odd Higgs is lighter than half of the mass of the `observed' Higgs, its doublet composition\footnote{Similarly to the CP-even case, we denote the 
mixing matrix in the CP-odd sector as $P_{ij}$, with $i$ the mass-state index and $j$ the flavour index; $j=3$ stands for the singlet component.} must 
fall under $\sim1\%$ to provide an acceptable fit to the data, while fairly larger singlet-doublet mixings are allowed beyond this mass. This condition 
can be interpreted as follows: if the channel $h[\sim125~\mbox{\small GeV}]\to2A_1$ is not kinematically forbidden and $A_1$ has a significant doublet
component, doublet Higgs couplings lead to a large (dominant) decay of the state $h[\sim125~\mbox{\small GeV}]$ -- which is identified with the LHC Higgs 
discovery -- into a pair of pseudoscalars; yet, this is in contradiction with the roughly standard behaviour of the signal observed at the LHC, hence a
disfavoured possibility; it thus follows that the light pseudoscalar with mass under the threshold must be essentially singlet in nature, for its 
presence not to spoil the fit to LHC Higgs data. Note, on the other hand, that for masses beyond $62$~GeV, the doublet composition of the light CP-odd 
state can reach up to $\sim100\%$, as we will discuss in the light-doublet section (\ref{ldoub}). However, apart from this rather isolated possibility, 
LHC limits -- in particular searches in the $\tau\tau$ channel -- do not leave much room for heavy Higgs doublets under $\sim400$~GeV (see e.g.\ 
Fig.~\ref{MATBscan}): as a result, the doublet composition of the light CP-odd state (beyond $62$~GeV), though possibly larger than for masses below the 
threshold, does not take large intermediate values (or only for disfavoured fits).

\begin{figure}[!htb]
    \centering
    \includegraphics[width=8cm]{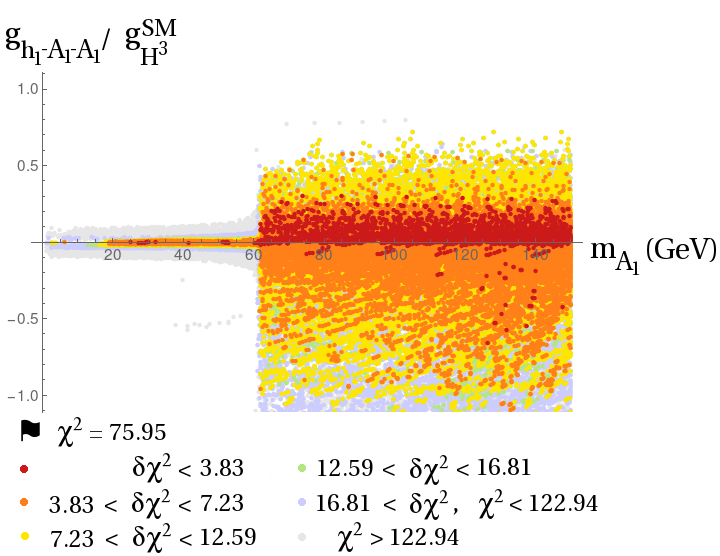}
  \caption{Triple Higgs coupling $h_1-A_1-A_1$ in the scan of Fig.~\ref{lightA}, relative to the SM value $g_{H^3}^{\mbox{\scriptsize SM}}$.}
  \label{h1A1A1}
\end{figure}

Coming back to CP-odd states under the kinematic threshold of $h[\sim125~\mbox{\small GeV}]\to2A_1$, one would naively expect favoured values of the
parameter $\lambda$ to remain moderate, in order to suppress singlet-doublet couplings which may still allow for a significant decay width of the 
SM-like state into singlet pseudoscalars. Yet, when one considers the distribution of the points with $m_{A_1}$ under $62$~GeV in the 
$\{\kappa,\lambda\}$ plane (right-hand side of Fig.~\ref{lightA}), it turns out that values of $\lambda$ up to $0.3$ still give an excellent fit to the 
LHC / TeVatron data. Two factors should be considered in order to understand this fact. The first one is related to the observation that the decay
$h[\sim125~\mbox{\small GeV}]\to2A_1$ receives a kinematic suppression in the immediate vicinity of the kinematic threshold: this effect leaves some 
manoeuvering space for moderate $h[\sim125~\mbox{\small GeV}]-A_1-A_1$ couplings. However, the second, and main, reason for the compatibility of rather 
large values of $\lambda$ with the Higgs measurement data in the presence of a light $A_1$ originates from accidental cancellations within the 
$h[\sim125~\mbox{\small GeV}]-A_1-A_1$ coupling. The latter indeed involves several terms (which are pondered by mixing angles and numerical 
coefficients):
\begin{enumerate}
 \item doublet-doublet interactions $\propto g^2v$: their effect on the $h[\sim125~\mbox{\small GeV}]-A_1-A_1$ coupling is suppressed when the 
pseudoscalars have a mostly singlet nature, as we mentioned before;
 \item singlet-singlet interactions $\propto \kappa A_{\kappa}$ and $\kappa^2s$: the state at $\sim125$~GeV being essentially doublet (to ensure a 
SM-like behaviour), such terms would also contribute little to the $h[\sim125~\mbox{\small GeV}]-A_1-A_1$ coupling; some effect can develop in 
proportion to the singlet component of the CP-even doublet, however;
 \item singlet-doublet interactions $\propto \lambda^2v$, $\propto \lambda\mu$, $\propto\lambda M_A^2/\mu$, $\propto \lambda\kappa v$ and 
$\propto \kappa \mu$: they naively dominate the couplings of the light CP-even doublet and the singlet pseudoscalar.
\end{enumerate}
The interplay of these various terms can thus give rise to very small $h[\sim125~\mbox{\small GeV}]-A_1-A_1$ couplings for certain points in the NMSSM 
parameter space. Note however that radiative corrections are likely to spoil the cancellation at tree-level -- leading corrections to the Higgs couplings 
are included within NMSSMTools as well --, but their main effect simply consists in shifting in the parameter space the 
regions with accidentally vanishing $h[\sim125~\mbox{\small GeV}]-A_1-A_1$ coupling: the corresponding case is of course a peculiar region of parameter space
where different contributions conspire, albeit possible. We observe, in Fig.~\ref{h1A1A1}, that best-fitting points in the scan of Fig.~\ref{lightA} indeed 
involve negligible values of the coupling $h[\sim125~\mbox{\small GeV}]-A_1-A_1$.

In view of the early LHC results, the maximal branching ratio into lighter Higgs states of the state at $\sim125$~GeV which remains compatible with
the observed signals had been estimated at $\sim20\%$ \cite{lightCPodd,Belanger:2013kya}. In the bottom left-hand quadrant of Fig.~\ref{lightA}, the 
unconventional branching ratio $BR(h[\sim125~\mbox{\small GeV}]\to2A_1)$ is displayed as a function of the Higgs mass. Best-fit (yellow-to-red) points
cluster within $BR(h[\sim125~\mbox{\small GeV}]\to2A_1)<20\%$ indeed. Interestingly, a light CP-odd Higgs may coexist with a light CP-even singlet-like 
state at reduced $\kappa/\lambda$: the CP-even singlet may then decay dominantly into $2A_1$, as shown in the plot at the bottom right-hand corner
of Fig.~\ref{lightA}. In this configuration, the observability of such a CP-even singlet state is hindered both by its reduced production cross-section and 
its unconventional decay. However, this setup typically requires that $BR(h[\sim125~\mbox{\small GeV}]\to2A_1)$ be suppressed through an accidental 
cancellation of the $h[\sim125~\mbox{\small GeV}]-A_1-A_1$ coupling.

\begin{figure}[!htb]
    \centering
    \includegraphics[width=15cm]{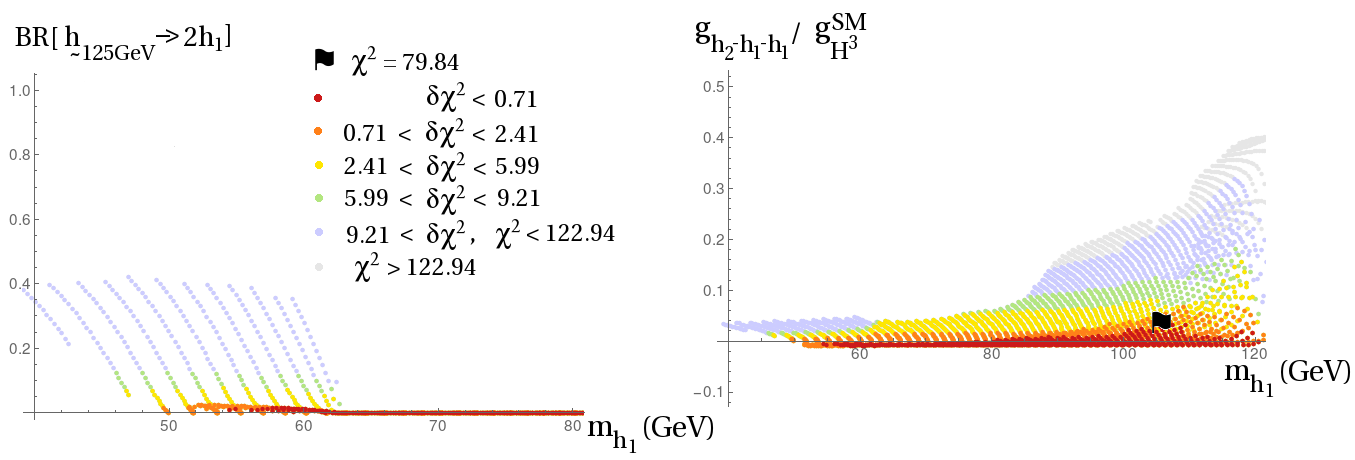}
  \caption{Scan of Fig.~\ref{singrates}. Decay rate of the state at $\sim125$~GeV into the light state and magnitude of the $h_2-h_1-h_1$ coupling
(relative to $g_{H^3}^{\mbox{\scriptsize SM}}$).\label{singH2H}}
\end{figure}
The case of light CP-even states with mass under the decay threshold of the observed state essentially follows the same principles: from 
Figs.~\ref{lightsing}, \ref{largemix}, \ref{PQlim} and \ref{PQlimmix}, we observe that light CP-even singlets under $\sim62$~GeV do not seem to offer 
particularly good fits to the Higgs-measurement data, as points with acceptable fit values drastically disappear from the low-mass tail under $\sim62$~GeV. 
Note that in the corresponding plots, $\lambda$ tended to be large, so that the decay $h_2\to2h_1$ would typically retain a significant branching 
fraction (which is disfavoured by the fit), even though $h_1$ is purely singlet: this explains why no red points persist at $S_{13}^2\simeq1$ below 
$62$~GeV in these plots, contrarily to the case of $P_{13}^2\simeq1$ in Fig.~\ref{lightA}. An example for this scenario can be found in Table 
\ref{gs_fit} (Point 9). In Fig.~\ref{singrates}, on the contrary, we observe points with mass down to $40$~GeV, some of them having an excellent fit
value. The fact that these points are dominantly singlet was already a requirement from LEP limits. However, in Fig.~\ref{singH2H}, we display the magnitude of
the branching fraction $BR(h_2^0\to2h_1^0)$ and the $h_2^0-h_1^0-h_1^0$ coupling: reduced values are evidently preferred below the threshold.

From this discussion, it follows that the success of the conventional Higgs searches prohibits\footnote{Note that this conclusion holds for the 
NMSSM only. Although this is likely to prove a common trend in most models where such light (pseudo)scalars are directly associated with the Higgs 
sector, there also exist theoretical frameworks where the light state has naturally suppressed couplings to the SM-like Higgs boson, without 
necessarily decoupling from SM-particles.} possible light Higgs states below $\sim62$~GeV to play a significant part in the phenomenology of the 
state which was discovered by the LHC, but also to intervene in connection to other standard particles: we have observed indeed that a condition for 
suppressed Higgs-to-Higgs decays laid in an almost-pure singlet nature of the light Higgs state. Therefore, the latter does not possess relevant 
couplings to SM-fermions or gauge bosons -- these develop only in proportion to the doublet components. As a consequence, direct production of 
the light state, e.g.\ via its coupling to a quark line (in associated production with tops or $b$'s), proves even less promising than searches in 
Higgs-to-Higgs processes: the corresponding cross-sections would 
indeed receive suppression factors of the form $\sim(1-S_{13}^2)$ / $(1-P^2_{13})$ for a CP-even / CP-odd state\footnote{Extreme values of $\tan\beta$ 
may partially compensate this suppression in the couplings to down-type quarks and leptons, however.}. On the other hand, the existence of such avoidant 
states remains a phenomenological possibility. The most straightforward search channel remains a possible decay (with branching ratio of a few percent) 
of the state observed at $\sim125$~GeV towards the light (pseudo)scalars, which may result in pairs of $b\bar{b}$'s, $\tau\bar{\tau}$'s or 
$\gamma\gamma$'s in the final state. Note that these remarks are consistent with the results of the recent analysis \cite{Bomark:2014gya}. Other search 
channels would involve new Higgs or supersymmetric particles, depending critically on the characteristics of these states.

\begin{figure}[!htb]
    \centering
    \includegraphics[width=14cm]{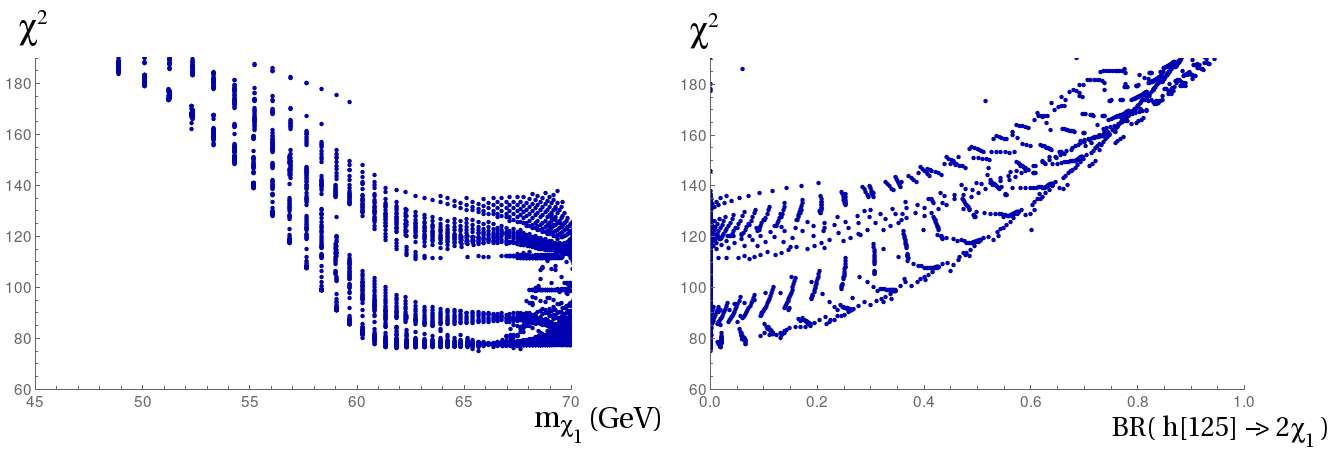}
  \caption{$\chi^2$ in the presence of a light singlino: the input corresponds to the scan of Fig.~\ref{PQlim}. The mass of the light Higgs doublet was 
constrained to fall within $[124,127]$~GeV.}
  \label{lightsingli}
\end{figure}
As a side-remark, let us mention that another type of unconventional Higgs decays would involve light supersymmetric particles, e.g.\ a light singlino.
This scenario has raised interest in dark matter studies \cite{DM}. While our framework is not suited to discuss dark matter phenomenology, we may still 
comment on the tentative presence of a light neutralino with mass under $\sim62$~GeV from the point of view of Higgs physics. Decays of the Higgs state
which is identified with the signal observed at the LHC into a pair of light neutralinos (invisible decay) would indeed lead to a suppression of the 
rates in the conventional search channels, which contradicts the LHC / TeVatron results (unless this effect would be compensated by an enhanced 
production mode). In the case of a light singlino, the corresponding width is 
related to $\lambda^2$, so that large values of $\lambda$ are again disfavoured as long as this decay channel is kinematically open. 
Fig.~\ref{lightsingli} in the low $\tan\beta$ / large $\lambda$ regime illustrates this comment: the region below the threshold for 
$h[$125$~\mbox{\small GeV}]\to2\chi^0_1$ gives rise to rather large $\chi^2$ values. A comparable limit would apply in the presence of light binos: 
the couplings to a light doublet Higgs 
would then be of electroweak magnitude, which could lead to a disfavoured decay of the Higgs state towards neutralinos. However, contrarily to the case
of light scalar states, the constraints on light new fermions are typically milder since such decays are of a similar type to the standard channels, hence
less likely to dominate the branching ratio as soon as the kinematical threshold opens.  

\section{\boldmath Two states in the vicinity of $\sim125$~GeV}\label{deg}
In \cite{2CPeven} the possible presence of two CP-even Higgs states in the vicinity of $\sim125$~GeV was highlighted, the rates of both states adding while the 
experimental resolution remained too broad to distinguish between them. Note that, while individual channels -- e.g.\ searches in the 
diphoton final state -- provide an excellent precision on the mass where the Higgs signals are centered, it will be challenging for ATLAS and CMS,
because of the limited detector resolution, to resolve two 
Higgs signals separated by less than $2-3$~GeV. On the theoretical side, a typical separation scale is the SM Higgs width of $4$~MeV. The aim in 
\cite{2CPeven} originally consisted in exploiting the presence of two states in the signal region, so as to enhance the diphoton and $ZZ$ rates in the 
context of a NUHM version of the NMSSM -- thence explaining tentative deviations of the apparent Higgs rates from the standard values via this approximate 
degeneracy of Higgs states. In this section, we shall discuss such configurations in more detail.

\begin{figure}[!htb]
    \centering
    \includegraphics[width=15cm]{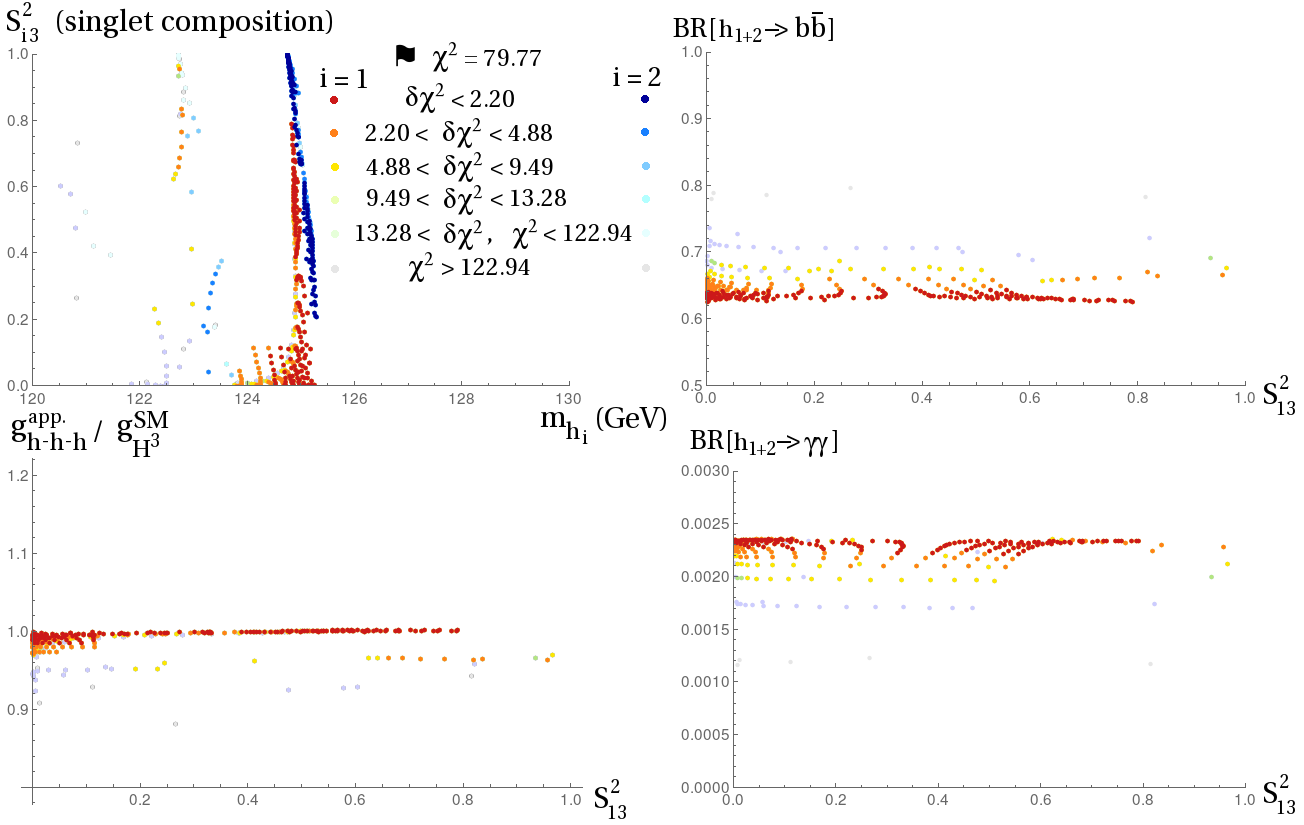}
  \caption{Quasi-degenerate CP-even states in the MSSM limit: $\lambda=\kappa=1\cdot10^{-3}$, $\tan\beta\in[3,22]$, $M_A\in[100,2000]$~GeV, 
$A_{\kappa}\in[-2,0]$~TeV, $\mu\in[120,2000]$~GeV, $2M_1=M_2=500$~GeV, $M_3=1.5$~TeV, 
$m_{\tilde{Q}_{1,2}}=1.5$~TeV, $m_{\tilde{L}}=300$~GeV, $m_{\tilde{Q}_3}=1.2$~TeV, $A_t=-2.5$~TeV, $A_{b,\tau}=-1.5$~TeV. $BR(h_{1+2}\to\ldots)$
denote the apparent `global' decay rates related to the two light, almost degenerate, CP-even Higgs states. $g_{h-h-h}^{\mbox{\tiny app}}$ represents
the apparent trilinear Higgs coupling accounting for both states. Only points with $m_{h_2^0}-m_{h_1^0}<1$~GeV have been retained in the scan.}
  \label{2CPeven}
\end{figure}
While the CP-even singlet and the light doublet may be close in mass, requiring them to lie within a few $100$~MeV / $1$~GeV demands that the mixing entry in
the mass matrix approximately vanishes. This is a tuning requirement, in general, unless one moves to the MSSM limit of the NMSSM (and it is then still 
necessary to adjust the diagonal entries of the mass matrix so that they are approximately equal). As a consequence of the small size of this 
off-diagonal mass entry, both states can be quasi degenerate while their composition, hence their coupling properties to SM particles, may vary from 
that of pure singlet / doublet to strong admixtures. Note that the typical widths involved ($\lsim4$~MeV) are still very small compared to the mass
differences which we are considering, so that interference effects such as those discussed in \cite{Fuchs:2014ola} should remain small. Therefore, 
as long as the two single states are not resolved, the overall behaviour in interactions with SM 
particles would naively coincide with that of the doublet-like state taken alone. Only the apparent Higgs-to-Higgs couplings may show deviations -- to 
which even a HL-LHC is unlikely to be sensitive to, due to the limited experimental precision, and which would still have to be disentangled from other 
sources of non-standard Higgs-to-Higgs couplings for the state at $\sim125$~GeV. On the other hand, the 
composition of the apparent light doublet may differ slightly from that of a pure doublet state due to the three-state mixing effects (similarly to what 
we discussed in sections \ref{lsing} and \ref{ltb}): here again, this effect could account for small deviations from the standard rates. On the side of 
the singlet, the simplest case consists in having all particles to which it directly couples (CP-odd singlet, higgsinos) heavier than half its mass: 
then, the only possible decay products are SM particles, the corresponding widths depending exclusively on the doublet component. In this context, the 
singlet does not contribute to the total width. If, on the contrary, significant decays of the singlet towards other new-physics states are allowed, 
then we are back to a case similar to what we discussed in section \ref{la1}: indeed, the singlet-doublet mixing would then dilute the branching ratios 
into standard particles, and could hence contradict the signals observed at LHC -- large singlet-doublet mixings would then be disfavoured. 
This second configuration will not emerge from the scans.

We have already come across points involving quasi-degenerate CP-even Higgs states, e.g.\ in section \ref{ltb}: some examples have been recorded in 
Table~\ref{BestFit} below. As a first illustration of this scenario, we consider the MSSM limit in Fig.~\ref{2CPeven}: the off-diagonal mass term between 
singlet and light doublet state is then naturally suppressed. One observes, however, that significant singlet-doublet mixing, up to $50\%$ can develop 
for favourable fit values, provided the two states are almost degenerate. On the other hand, no effect at the level of the rates is expected 
in the MSSM limit (because of the decoupling of the singlet). We display the apparent branching ratio into $b\bar{b}$ and $\gamma\gamma$ as a function 
of the singlet-doublet mixing on the right-hand side of Fig.~\ref{2CPeven}, for points where $m_{h_2^0}-m_{h_1^0}<1$~GeV: while some variations are 
present, these are entirely independent of the mixing or the presence of the singlet, and their origin can actually be traced back to the scale of the 
heavy doublet sector ($M_A$) and the value of $\tan\beta$, hence appears as a pure doublet effect. Similarly, consequences on Higgs pair production are 
minimal since the singlet-doublet interactions are suppressed in this limit. While singlet-singlet interactions formally enter the relevant Higgs-to-Higgs 
couplings, the associated effects are projected onto the doublet component -- due to the coupling to SM particles in the process of the Higgs production 
-- and vanish when summed over both degenerate states. In the lower left-hand quadrant of 
Fig.~\ref{2CPeven} one observes indeed that the apparent trilinear Higgs coupling accounting for both states in the vicinity of $\sim125$~GeV remains 
SM-like. Therefore, in this scenario, only the presence of two separate peaks in the Higgs searches -- as the result of a large mixing between the
singlet and the doublet state --, with widths adding to $\sim4$~MeV, would be observable and document the quasi-degeneracy of the singlet with the light 
doublet. On the other hand, the singlet could also be almost degenerate but remain (almost) unmixed, in which case, its `peak' would remain invisible. 

\begin{figure}[!htb]
    \centering
    \includegraphics[width=15cm]{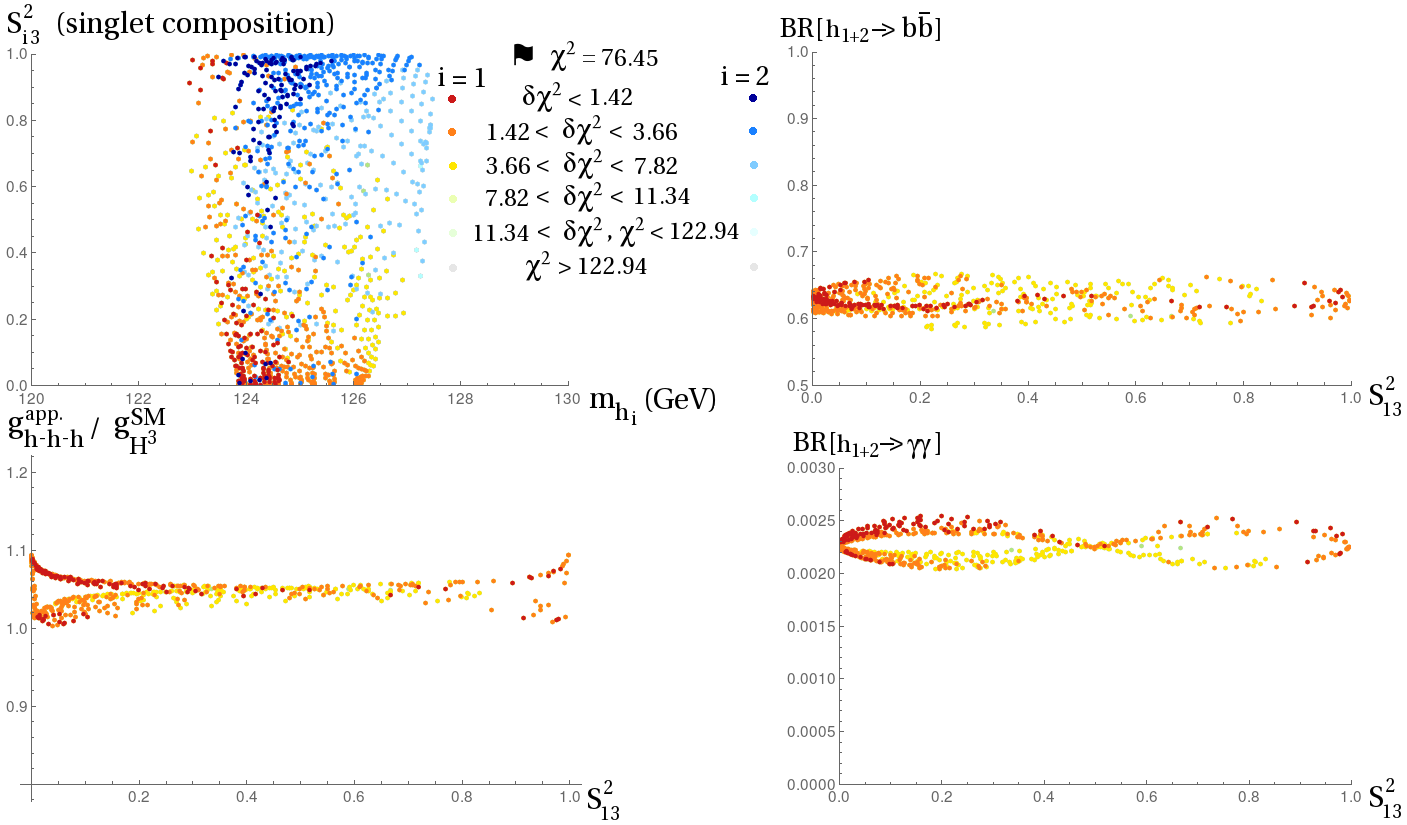}
  \caption{Quasi-degenerate CP-even states in the low $\tan\beta$ / large $\lambda$ regime: $\tan\beta=2$, $\lambda=0.7$, $\kappa=0.1$, 
$\mu\in[120,1200]$~GeV, $M_A\in[0.8,3]$~TeV, $A_{\kappa}\in[-300,0]$~GeV, $2M_1=M_2=150$~GeV, $M_3=1.5$~TeV, $m_{\tilde{Q}_3}=500$~GeV, 
$m_{\tilde{Q}_{1,2}}=1.5$~TeV, $m_{\tilde{L}}=110$~GeV,$A_{t,b,\tau}=-100$~GeV. Only points with $m_{h_2^0}-m_{h_1^0}<1$~GeV were retained 
in the scan.}
  \label{deg2}
\end{figure}
The same configuration is considered in Fig.~\ref{deg2}, but in a different regime: we turn to the low $\tan\beta$ / large $\lambda$ region again. Only 
points where the two lightest CP-even states have masses within $1$~GeV of each other are kept in this scan. Deviations at the percent level can be 
observed in the apparent rates into SM-particles. The corresponding effect is obviously associated with the three-state mixing -- this is most visible in 
the case of the $\gamma\gamma$ channel -- in contrast to the case of the MSSM limit. Note that pure doublet effects are negligible here as $M_A\gg M_Z$ 
and $\tan\beta$ is fixed. Large singlet-doublet couplings affecting the apparent triple-Higgs coupling at $\sim125$~GeV may develop as well, leading to 
a $\sim10\%$ increase (at most) of the apparent trilinear Higgs coupling (see the lower left-hand quadrant).

Note that while this kind of scenario may occur in
different parameter regimes, the existence of a doublet state 
at $\sim125$~GeV always relies on
the mechanisms described in the previous sections, i.e.\ substantial
radiative corrections driven by the SUSY spectrum and larger $\tan\beta$
or the specific
tree-level contribution associated with large $\lambda$ and low
$\tan\beta$. On the other hand, singlet-doublet mixing does not provide
a mass uplift for the doublet in the scenario under consideration.

Another possibility would consist in the presence of a CP-odd state in mass-proximity to the SM-like state. Such a scenario might even be considered as 
a CP-conserving approximation of the CP-violating case, where the observed state would appear as a superposition of CP-even and CP-odd components. 
Considering LHC limits associated with Higgs searches in the $\tau\tau$ channel as well as (indirectly) with top decays to a charged Higgs, it proves 
difficult to admit a (pure) CP-odd doublet state in the desired mass range. The possibility of a CP-odd Higgs close to $\sim125$~GeV thus rests with 
singlet-like states, which may however carry a significant doublet component. In Fig.~\ref{lightA}, one observes that a doublet composition of $5-10\%$ 
still receives an acceptable fit to the data in the vicinity of $\sim125$~GeV. The fermion rates at $\sim125$~GeV would be the observables that are
most significantly affected in 
this setup, with an apparent increase of a few percent, while also the diphoton channel may receive a small subsidiary contribution.

Note that in the CP-even as in the CP-odd case, the additional state may well be purely singlet so that it would remain undetected in direct 
production due to vanishing couplings to SM particles. Only at the level of pair productions, making use of Higgs-to-Higgs couplings, would there be
a deviation from the naive one-particle case -- which would be extremely challenging to detect with sufficient precision experimentally. Note also
that, e.g.\ in the MSSM limit, a singlet that is mass degenerate with the state at $\sim125$~GeV could even have no impact at all on the rates or on 
multi-Higgs production.

\section{Two-light-doublet scenario}\label{ldoub}
Identifying the LHC signal with a heavy doublet state -- which suggests that a lighter Higgs doublet state has eluded searches so far -- is a scenario 
which was originally considered in the MSSM \cite{MSSMheavyH,MSSMfit}. It was recently put under further pressure by the publication of charged Higgs searches 
in top decays \cite{ATLASchargedHiggs} -- see the newest reference in \cite{HiggsBounds} for a discussion in the context of the MSSM --  and 
\cite{CMSchargedHiggs}, however. In this section, we will discuss the situation in the context of the NMSSM. 
%The main advantage of the NMSSM from the perspective of this scenario, rests with the increased flexibility of the spectrum: 
%charged states receive an additional
%contribution at tree-level $\propto\lambda^2v^2$; beyond new tree-level contributions, the neutral sector may involve significant mixing among singlets
%and doublets, displacing the masses and diluting the `observable' doublet component; more generally, there are now as much as $6$ relevant parameters 
%in the Higgs sector, already at tree-level. 
%indeed the strict correlation among Higgs masses and couplings in the MSSM can be loosened somewhat, which potentially allows the mass of NMSSM Higgs 
%states to reach less-sensitive mass-ranges or even, in certain channels, values below the LHC search-range. Such an interpretation of the `Higgs' 
%signals as the signature of a heavy CP-even doublet is thus expected to prove more resilient.

The presence of a light CP-even Higgs -- with mass below $125$~GeV -- in the spectrum does not entail major phenomenological difficulties by itself.
Indeed we have already outlined two possible strategies which allow such a scenario to evade LEP limits:
\begin{itemize}
 \item the production cross-section at LEP could be suppressed by a small coupling of the light Higgs state to electroweak gauge bosons -- we have 
discussed above how a singlet Higgs naturally fulfills this requirement;
 \item the decays in the standard search channels could be blurred by large non-conventional decays -- in this case, the difficulty lies in explaining
why these unconventional decays do not affect the observed state at $\sim125$~GeV.
\end{itemize}
While the success of the LHC Higgs searches makes the second approach difficult to implement in a realistic model, the first strategy remains viable,
even in the case of a doublet state -- even though the cancellation of the couplings of the light Higgs to electroweak gauge bosons is then largely 
accidental. Turning to direct limits from the LHC, it is to be noted that searches in the $\tau\tau$ channel for an additional Higgs boson essentially 
disfavour neutral states with 
mass above $\sim110$~GeV, while the light doublet can be well below $100$~GeV. In practice, we could indeed find NMSSM parameter points where a light
doublet Higgs with vanishing couplings to electroweak gauge bosons -- typically in the $70$~GeV mass range -- escapes all direct limits.

Yet, if all the CP-even doublet states are at or below $~125$~GeV, the correlations of the masses in the doublet sector will force the other doublet 
masses -- those of the CP-odd and charged states -- to be close. Indeed, as a good approximation, we have:
\begin{align}
 &m_{H^{\pm}}^2\simeq M_{A,\mbox{\tiny eff}}^2+M_W^2-\lambda^2v^2\\
 &m_{A,\mbox{\tiny doub}}^2\simeq M_{A,\mbox{\tiny eff}}^2\nonumber\\
 &m_{h_{1,\mbox{\tiny doub}}}^2\simeq \frac{1}{2}\left[M_{A,\mbox{\tiny eff}}^2+M_{Z}^2-\sqrt{(M_{A,\mbox{\tiny eff}}^2-M_Z^2)^2\cos^22\beta+(M_{A,\mbox{\tiny eff}}^2+M_Z^2-2\lambda^2v^2)^2\sin^22\beta}\right]\nonumber\\
 &m_{h_{2,\mbox{\tiny doub}}}^2\simeq \frac{1}{2}\left[M_{A,\mbox{\tiny eff}}^2+M_{Z}^2+\sqrt{(M_{A,\mbox{\tiny eff}}^2-M_Z^2)^2\cos^22\beta+(M_{A,\mbox{\tiny eff}}^2+M_Z^2-2\lambda^2v^2)^2\sin^22\beta}\right]+\delta_{\mbox{\tiny rad}}\nonumber
\end{align}
In this list of (approximate) masses for the doublet states, $M_{A,\mbox{\tiny eff}}$ does not exactly correspond to the NMSSMTools input $M_A$, but 
represents a corrected value absorbing radiative corrections. There are, of course, additional deviations at the loop level, and 
$\delta_{\mbox{\tiny rad}}$ only denotes the bulk of the large shift due to top/stop effects -- hence associated to the $H_u$ flavour.

\begin{figure}[!htb]
    \centering
    \includegraphics[width=16cm]{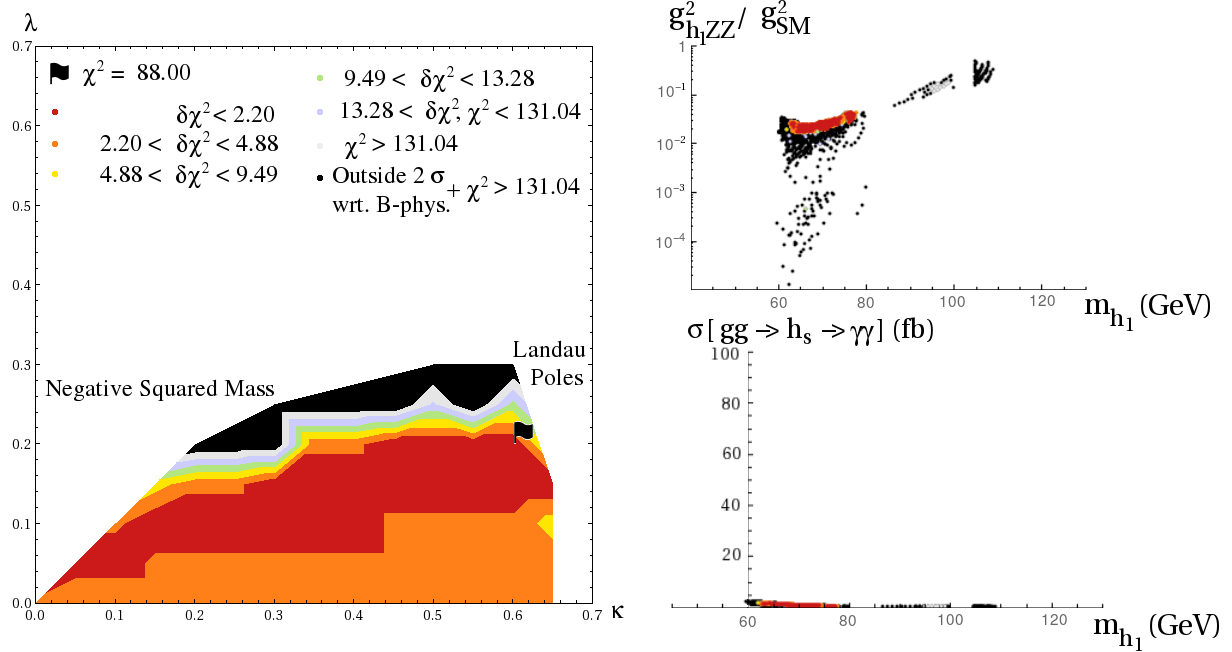}
  \caption{Left plot: Scan in the $\{\kappa,\lambda\}$-plane: $\lambda\in[2\cdot10^{-4},0.65]$, $\kappa\in[2\cdot10^{-4},0.65]$, $\tan\beta\in[3,20]$, $M_A=130$~GeV, $\mu\in[100,300]$~GeV, 
$A_{\kappa}\in[-1.5,0]$~TeV, $2M_1=M_2=500$~GeV, $M_3=1.5$~TeV, $m_{\tilde{Q}_{1,2}}=1.5$~TeV, $m_{\tilde{Q}_3}=1.1$~TeV, $A_t=-2.3$~TeV, $A_{b,\tau}=-1.5$~TeV.
On the right: coupling properties of the light CP-even doublet to $Z$-bosons (upper part) and the typical cross-section for the 
production of this state in the $8$~TeV run of the LHC in gluon-gluon fusion, with a diphoton decay (lower plot). The colour coding remains the same as in 
the plot on the left.}
  \label{lightdoublk}
\end{figure}

\begin{figure}[!htb]
%    \centering \includegraphics[width=11cm]{lightdoub_MA=130.png}\vspace{0.5cm}
    \includegraphics[width=14cm]{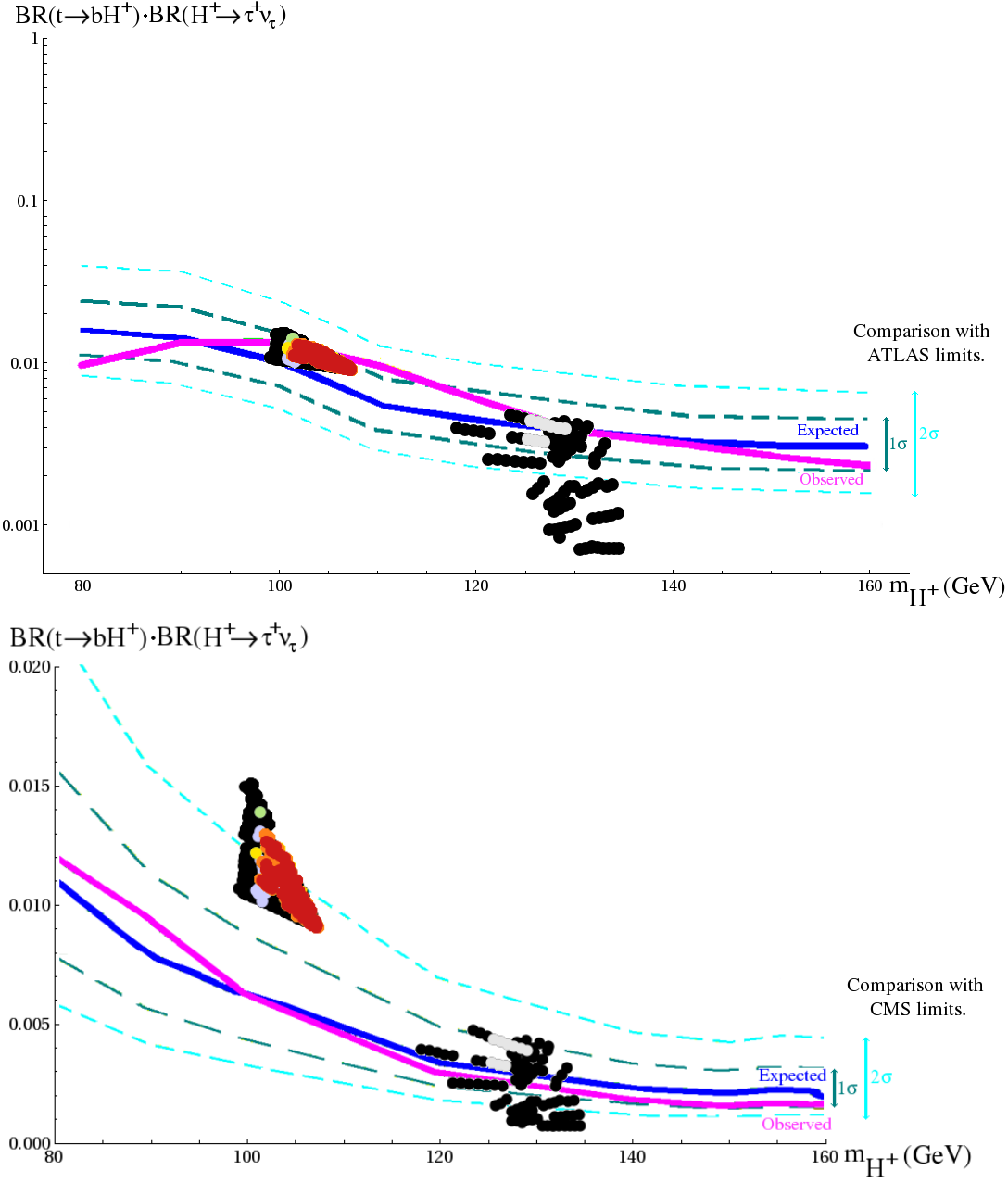}
%  \caption{Scan for fixed $M_A=130$~GeV: top $\rightarrow$ distribution in the $\{\lambda,\kappa\}$ plane; bottom $\rightarrow$ superposition with 
%the ATLAS limits on $t\to H^+b$. $\lambda\in[2\cdot10^{-4},6.5]$, $\kappa\in[2\cdot10^{-4},6.5]$, $\tan\beta\in[3,20]$, $M_A=130$~GeV, $\mu\in[100,300]$~GeV, 
%$A_{\kappa}\in[-1.5,0]$~TeV, $2M_1=M_2=500$~GeV, $M_3=1.5$~TeV, 
%$m_{\tilde{Q}_{1,2}}=1.5$~TeV, $m_{\tilde{Q}_3}=1.1$~TeV, $A_t=-2.3$~TeV, $A_{b,\tau}=-1.5$~TeV.}
\caption{$BR(t\to H^+b)\cdot BR(H^+\to\tau^+\nu_{\tau})$ in the scan of Fig.~\ref{lightdoublk}; the limits from ATLAS \cite{ATLASchargedHiggs} (upper 
plot) and CMS \cite{CMSchargedHiggs} (lower plot) on this branching ratio $BR(t\to H^+b)$ -- assuming a $100\%$ decay $H^{+}\to\tau^+\nu_{\tau}$ -- are 
reproduced for comparison.}
  \label{brtHnew}
\end{figure}

The CP-odd doublet Higgs causes limited concern as its mass typically falls under $100$~GeV and its presence was tested at LEP only via pair production 
processes (since a CP-odd Higgs has vanishing couplings to gauge bosons). Only when its mass falls below the threshold $\sim125/2$~GeV does it entail 
indirect limits from the LHC measurement -- in accordance with our discussion in section \ref{la1}.

On the other hand, the existence of a light charged Higgs is problematic in view of the existing limits. This possibility is already severely 
constrained in view of the tensions
that it produces in the $B$-sector -- e.g.\ in the channels $B\to X_s\gamma$, $B^+\to\tau\nu_{\tau}$ or $B_s\to\mu^+\mu^-$. Yet, to the price of a
$\chi^2$ pull of at least $\sim4$, these flavour limits could be satisfied. On the other hand, a light charged Higgs opens the possibility of top 
decays: these channels have been investigated by ATLAS \cite{ATLASchargedHiggs} and CMS \cite{CMSchargedHiggs}, e.g.\ in the $b\bar{\tau}\nu_{\tau}$ final state,
and the current constraints exclude all the points that were preferred by the fit performed in Fig.\ref{lightdoublk}.

%In addition to these limits originating from Higgs searches, the presence of a light charged-Higgs leads to tensions with e.g.\ $BR(B\to X_s\gamma)$: 
%yet the $H^{\pm}/t$ contribution can be balanced by supersymmetric effects intervening with the opposite sign. Nevertheless, large supersymmetric 
%effects tend to be disfavoured by $B$-observables mediated by Higgs penguins (e.g. $B_s\to\mu^+\mu^-$: remember that light neutral doublet states 
%mediating this transition are present in the spectrum; note also that the corresponding requirements on 
%the supersymmetric spectrum must be conciliated with the need for large radiative corrections to the mass of the Higgs state $h_{2\mbox{\tiny doub}}$). 
%$BR(B^+\to\tau\nu_{\tau})$ provides a tree-level limit on light charged Higgs states but suffers from large uncertainties. On the whole, if observables 
%in the $B$-sector can be kept within the $2\sigma$-range, the corresponding channels are chronically under some strain, which translates into a 
%$\chi^2$ pull of at least $\sim4$. In short, we see that an extensive tuning of the parameters accompanies this scenario whenever it proves compatible 
%with phenomenological limits.

%\begin{figure}[!htb]
%    \centering
%    \includegraphics[width=12cm]{lightdoub_MA_{lk}.png}
%  \caption{Scan in the $\{\lambda,\kappa\}$-plane: $\tan\beta=9-10$, $\mu=200$~GeV, $M_A\in[80,200]$~GeV, $A_{\kappa}\in[-2.5,0]$~TeV, $2M_1=M_2=500$~GeV, 
%$M_3=1.5$~TeV, $m_{\tilde{Q}_3}=1.1$~TeV, $m_{\tilde{Q}_{1,2}}=1.5$~TeV,$A_t=-2.3$~TeV, $A_{b,\tau}=-1.5$~TeV.}
%  \label{lightdoublk}
%\end{figure}

In order to illustrate the workings of the recent LHC constraints on $t\to bH^+$, we turn to older versions of HiggsBounds -- 4.1.3 -- and 
HiggsSignals -- 1.2.0 -- where LHC results of the summer 2014 had not yet been included. To set a reference, the best-fit point in the SM-limit then 
receives a $\chi^2$ of\footnote{The higher $\chi^2$ value in this previous version is related to a revision of the tests in the diphoton 
channel (both in the presentation of the experimental data and its implementation within HiggsSignals).} $\sim94$. We then consider the region in the 
NMSSM parameter space where $M_A=130$~GeV is set as a fixed input. The results are shown in the 
$\{\kappa,\lambda\}$ plane in Fig.~\ref{lightdoublk}. The best-fitting points are obtained for $\lambda\sim0.2$. They have a somewhat better $\chi^2$
than the SM limit -- $\chi^2\sim88.0$ -- and provide the following spectrum: $m_{h_2}\simeq125.0$~GeV, $m_{H^{\pm}}\simeq107$~GeV and light $H_d$-like 
doublet states (CP-even and odd) around $70-75$~GeV (i.e.\ $M_{A,\mbox{\tiny eff}}\sim70$~GeV); the singlet states are much heavier (beyond $1$~TeV) 
and play no role in the electroweak Higgs phenomenology. The $h_2$-rates are essentially SM-like, which explains the quality of the fit. 
%Note that this setup is accompanied by an almost pure decoupling of the $H_u$ and $H_d$ components, which allows for $m_{h_2}\sim125$~GeV via large 
%loop effects for the $H_u$ component on one side, suppressed couplings to gauge bosons for the light $H_d$-like states on the other. 
On the right part of Fig.~\ref{lightdoublk}, one observes that the squared coupling of the light CP-even doublet state to $Z$-bosons is reduced 
(of the order $10^{-2}$), while its typical production cross-section in the $8$~TeV run of the LHC, accompanied by a diphoton decay, would amount to 
about a few fb's only.
Version 4.1.3 of HiggsBounds included the ATLAS limits on top decays \cite{ATLASchargedHiggs}, and we show on the upper part of 
Fig.~\ref{brtHnew} how the points of the scan in Fig.~\ref{lightdoublk} compare with these constraints -- note that the corresponding experimental bounds 
have been obtained under the assumption of a $100\%$ $H^+\to\tau^+\nu_{\tau}$ decay, so that we rescale our points by a factor $BR(H^+\to\tau^+\nu_{\tau})$: 
while sitting on the edge of the exclusion limit, the light doublet scenario appears compatible with these constraints. 
%Note that two separate regions compatible with the current bounds appear in this scan, one in the vicinity of $m_{H^{\pm}}=105$~GeV (with $m_{h_1^0}\lsim80$~GeV), 
%the other in that of $m_{H^{\pm}}=130$~GeV (with $m_{h_1^0}\lsim105$~GeV), although only the lower interval in $m_{H^{\pm}}$ provides a competitive fit. 
%That these two regions are disjoint should not surprise the reader, considering the number of constraints to satisfy simultaneously (especially LHC 
%limits and the necessity for vanishing couplings of the light CP-even Higgs to electroweak gauge bosons).
On the other hand, the more recent CMS limits on $t\to H^{+}b$ \cite{CMSchargedHiggs} had not been included within HiggsBounds, and we display them 
in the lower part of Fig.~\ref{brtHnew}. All the best-fitting region is excluded while the few remaining points at $m_{H^{\pm}}\simeq130$~GeV -- 
disconnected from the best-fit region due to the interplay of constraints -- do not
offer interesting fit qualities. Therefore, while the light doublet scenario might still be realized in regions with reduced $BR(t\to H^+b)$, the 
current limits discard most of the associated parameter space.

\section{Highlight of specific points}\label{fp}
In this section, we focus on specific points in the parameter space, found in the vicinity of the best-fit points of the various plots
presented thus far (see Table \ref{BestFit}, in the appendix), and aim at discussing the associated Higgs phenomenology in more detail. Note again 
that, although the supersymmetric spectra have some impact on the phenomenology -- Higgs mass, $(g-2)_{\mu}$, $B$-physics, etc., they are not 
strictly tied to the respective scenarios -- i.e.\ analogous Higgs properties should be accessible with different, e.g.\ heavier SUSY spectra: the 
corresponding characteristics are thus purely illustrative and we shall not discuss the prospects of discovering such states at the LHC. Table \ref{fp1} 
provides the NMSSMTools inputs and the Higgs spectra of the considered points, while Table \ref{fp2} gives the Higgs couplings to SM particles. While 
most of the qualitative features that we discussed in the previous sections enter the characteristics of the few points presented below, we would like to
caution the reader against reducing the discussed phenomenology to those specific points. 

\begin{table*}[tbh!]
\null\vspace{0cm}\hspace{-0.1cm}\begin{tabular}{|c||c|c|c|c|c|c|c}
\hline
  & Point 1 & Point 2 & Point 3 & Point 4 & Point 5 & Point 6\\ 
NMSSMTools & Decoupling & Light & Low $\tan\beta$ & 2 CP-even & Light $A$ & Light\\ 
Parameters & limit & singlet & + light s. & $\sim125$~GeV & $H\to2A$ & doublet\\\hline\hline
$\lambda$ & $0.2$ & $0.55$ & $0.699$ & $0.7$ & $0.05$ & $0.1$\\ \hline
$\kappa$ & $0.6$ & $0.45$ & $0.1$ & $0.1$ & $0.05$ & $0.25$\\ \hline
$\tan\beta$ & $22.5$ & $8$ & $2$ & $2$ &  $19$ & $12.25$\\ \hline
$\mu_{\mbox{\tiny eff}}$ (GeV) & $200$ & $125$ & $330$ & $714$ & $125$ & $187.9$\\ \hline
$M_A$ (GeV) & $1000$ & $1000$ & $801$ & $1694$ & $1200$ & $130$\\ \hline
$A_{\kappa}$ (GeV) & $-8.5$ & $-288$ & $-122$ & $-176.9$ &  $-5$ & $-1100$\\ \hline
$M_1$ (GeV) & $250$ & $250$ & $75$ & $75$ & $250$ & $250$\\ \hline
$M_2$ (GeV) & $500$ & $500$ & $150$ & $150$ & $500$ & $500$\\ \hline
$M_3$ (TeV) & $1.5$ & $1.5$ & $1.5$ & $1.5$ & $1.5$ & $1.5$\\ \hline
$m_{\tilde{Q}_{1,2}}$ (TeV) & $1.5$ & $1.5$ & $1.5$ & $1.5$ & $1.5$ & $1.5$\\
$m_{\tilde{Q}_3}$ (TeV; if $\neq m_{\tilde{Q}_{1,2}}$) & $1.1$ & $1$ & $0.5$ & $0.5$ & $1.2$ & $1.1$\\ \hline
$m_{\tilde{L}}$ (GeV) & $300$ & $200$ & $110$ & $110$ & $300$ & $250$\\ \hline
$A_{t}$ (TeV) & $-2.5$ & $-2$ & $-0.1$ & $-0.1$ & $-2.5$ & $-2.3$\\
$A_{b,\tau}$ (TeV; if $\neq A_{t}$) & $-1.5$ & $-1.5$ & / & / & $-1.5$ & $-1.5$\\ \hline\hline
Higgs Spectrum &  &  &  &  & &   \\\hline\hline
$m_{h_1}$ (GeV) & $125.0$ D & $105.6$ S & $102.1$ S & $125.1$ D/S & $125.1$ D & $62.8$ D\\
$m_{h_2}$ (GeV) & $973$ D & $125.0$ D & $125.3$ D & $125.2$ S/D & $249$ S & $125.6$ D\\ 
$m_{h_3}$ (GeV) & $1192$ S & $986$ D & $796$ D & $1693$ D & $1174$ D & $605$ S\\ 
$m_{A_1}$ (GeV) & $109.7$ S & $307$ S & $165.4$ S & $280$ S & $43.7$ S & $63.3$ D\\ 
$m_{A_2}$ (GeV) & $976$ D & $983$ D & $800$ D & $1695$ D & $1174$ D & $1245$ S\\ 
$m_{H^{\pm}}$ (GeV) & $976$ & $980$ & $790$ & $1690$ & $1177$ & $101.5$ \\ \hline
$S_{13}^2$ & $\sim0\%$ & $97\%$ & $91\%$ & $48.4\%$ & $0.1\%$ & $\sim0\%$ \\ 
$S_{23}^2$ & $0.3\%$ & $1.6\%$ & $8\%$ & $51.4\%$ & $99.9\%$ & $\sim0\%$ \\ 
$S_{33}^2$ & $99.7\%$ & $1.0\%$ & $0.9\%$ & $0.2\%$ & $\sim0\%$ & $\sim100\%$ \\ 
$P_{13}^2$ & $99.7\%$ & $99.5\%$ & $98\%$ & $99.7\%$ & $\sim100\%$ & $\sim0\%$ \\ 
$P_{23}^2$ & $0.3\%$ & $0.5\%$ & $1.6\%$ & $0.3\%$ & $\sim0\%$ & $\sim100\%$ \\ \hline\hline
$M_W$[err]~(GeV) & $80.372[17]$ & $80.373[17]$ & $80.410[20]$ & $80.393[21]$ & $80.371[17]$ & $80.397[17]$  \\\hline\hline
$\chi^2$ (/89 obs.) & $81.2$ & $76.1$ & $76.0$ & $80.5$ & $80.2$ & $81.4$ (excl.) \\ \hline %$90.7$

\end{tabular}

\caption{Highlighted points: NMSSMTools input and Higgs spectra.
\label{fp1}}
\end{table*}

\begin{table*}[tbh!]

\begin{center}
 \null\vspace{0cm}\hspace{-0cm}\begin{tabular}{|c||c|c|c|c|c|c|}
\hline
Couplings$^2$/SM & Point 1 & Point 2 & Point 3 & Point 4 & Point 5 & Point 6 \\\hline\hline
$h_1WW$ & $1.0$ & $0.017$ & $0.082$ & $0.515$ & $0.999$ & $0.027$\\ 
$h_1ZZ$ & $1.0$ & $0.017$ & $0.082$ & $0.515$ & $0.999$ & $0.027$\\ 
$h_1gg$ & $0.968$ & $0.015$ & $0.058$ & $0.517$ & $0.980$ & $13.0$\\ 
$h_1\gamma\gamma$ & $1.002$ & $0.001$ & $0.048$ & $0.469$ & $1.002$ & $0.357$\\ 
$h_1t\bar{t}$ & $1.0$ & $0.014$ & $0.059$ & $0.494$ & $0.999$ & $0.007$\\ 
$h_1b\bar{b}$ & $1.035$ & $0.853$ & $0.215$ & $0.600$ & $1.027$ & $149$\\ 
$h_1\tau\bar{\tau}$ & $1.035$ & $0.855$ & $0.218$ & $0.604$ & $1.027$ & $150$\\ \hline
$h_2WW$ & $1\cdot10^{-7}$ & $0.983$ & $0.918$ & $0.485$ & $0.001$ & $0.973$\\ 
$h_2ZZ$ & $1\cdot10^{-7}$ & $0.983$ & $0.918$ & $0.485$ & $0.001$ & $0.973$\\ 
$h_2gg$ & $0.008$ & $0.973$ & $1.079$ & $0.563$ & $0.001$ & $1.232$\\ 
$h_2\gamma\gamma$ & $0.003$ & $1.023$ & $0.955$ & $0.493$ & $0.001$ & $0.894$\\ 
$h_2t\bar{t}$ & $0.002$ & $0.986$ & $0.945$ & $0.507$ & $0.001$ & $1.000$\\ 
$h_2b\bar{b}$ & $502$ & $0.800$ & $0.813$ & $0.405$ & $0.014$ & $1.043$\\ 
$h_2\tau\bar{\tau}$ & $505$ & $0.800$ & $0.810$ & $0.401$ & $0.014$ & $1.054$\\ \hline
$h_3WW$ & $6\cdot10^{-5}$ & $6\cdot10^{-6}$ & $3\cdot10^{-5}$ & $3\cdot10^{-7}$ & $8\cdot10^{-7}$ & $2\cdot10^{-6}$\\ 
$h_3ZZ$ & $6\cdot10^{-5}$ & $6\cdot10^{-6}$ & $3\cdot10^{-5}$ & $3\cdot10^{-7}$ & $8\cdot10^{-7}$ & $2\cdot10^{-6}$\\ 
$h_3gg$ & $8\cdot10^{-5}$ & $0.013$ & $0.437$ & $0.276$ & $0.005$ & $3\cdot10^{-6}$\\ 
$h_3\gamma\gamma$ & $0.002$ & $0.089$ & $3.06$ & $0.059$ & $0.011$ & $0.002$\\ 
$h_3t\bar{t}$ & $1\cdot10^{-4}$ & $0.016$ &  $0.246$ & $0.249$ & $0.003$ & $3\cdot10^{-6}$\\ 
$h_3b\bar{b}$ & $1.634$ & $63.2$ &  $3.80$ & $3.60$ & $360$ & $7\cdot10^{-4}$\\ 
$h_3\tau\bar{\tau}$ & $1.641$ & $63.3$ &  $3.97$ & $3.99$ & $361$ & $7\cdot10^{-4}$\\ \hline
$A_1WW$ & $0$ & $0$ & $0$ & $0$ & $0$ & $0$\\ 
$A_1ZZ$ & $0$ & $0$ & $0$ & $0$ & $0$ & $0$\\ 
$A_1gg$ & $0.028$ & $1\cdot10^{-4}$ &  $0.008$ & $0.002$ & $2\cdot10^{-4}$ & $14.4$\\ 
$A_1\gamma\gamma$ & $0.032$ & $1.455$ & $0.106$ & $0.014$ & $0.008$ & $2.45$\\ 
$A_1t\bar{t}$ & $6\cdot10^{-6}$ & $8\cdot10^{-4}$ & $0.004$ & $8\cdot10^{-3}$ & $6\cdot10^{-9}$ & $0.007$\\ 
$A_1b\bar{b}$ & $1.589$ & $0.325$ & $0.060$ & $0.012$ & $8\cdot10^{-4}$ & $149$\\ 
$A_1\tau\bar{\tau}$ & $1.596$ & $0.326$ & $0.062$ & $0.013$ & $8\cdot10^{-4}$ & $150$\\ \hline
$A_2WW$ & $0$ & $0$ & $0$ & $0$ & $0$ & $0$\\ 
$A_2ZZ$ & $0$ & $0$ & $0$ & $0$ & $0$ & $0$\\ 
$A_2gg$ & $0.019$ & $0.028$ & $0.343$ & $0.299$ & $0.016$ & $4\cdot10^{-8}$\\ 
$A_2\gamma\gamma$ & $0.054$ & $0.098$ & $0.546$ & $0.380$ & $0.038$ & $6\cdot10^{-5}$\\ 
$A_2t\bar{t}$ & $0.002$ & $0.016$ & $0.246$ & $0.249$ & $0.003$ & $2\cdot10^{-8}$\\ 
$A_2b\bar{b}$ & $502$ & $63.5$ & $3.80$ & $3.67$ & $360$ & $4\cdot10^{-4}$\\ 
$A_2\tau\bar{\tau}$ & $505$ & $63.7$ & $3.94$ & $3.99$ & $361$ & $4\cdot10^{-4}$\\ \hline

\end{tabular}
\end{center}

\caption{Highlighted points: Higgs squared couplings to SM particles.
\label{fp2}}
\end{table*}

The first point is characteristic of the decoupling limit: the heavy-doublet states are at $\sim1$~TeV, so as to
decouple from the SM-like Higgs, while $\tan\beta\gg1$ to maximize the tree-level contribution to the mass of this light state 
(we stress that the rather large value $\tan\beta=22.5$ is actually driven by the anomalous magnetic moment of the muon). While $\lambda$
and $\kappa$ do not vanish, which means that one is far from the MSSM limit, the impact of the singlet states on the light doublet Higgs
is negligible, and the CP-even singlet is actually quite heavy ($\sim1200$~GeV). Correspondingly the light doublet state, at $125.0$~GeV, is 
SM-like, with couplings to SM-particles within a few percent of their values at the same mass for a genuine SM Higgs boson: the corresponding rates 
at the LHC are thus comparable (still within a few percent). Distinguishing this point from the SM via the properties
of the `observed Higgs' would thus require a high precision on the couplings, achievable most likely at a Linear Collider only -- note also that one may
get even closer to a SM Higgs scenario via further decoupling of the heavy states. The heavy doublet states are essentially $H_d$ in
nature, hence give very large couplings (enhanced by the large value of $\tan\beta$) to bottoms and taus (for the neutral states). All other 
couplings to SM particles are suppressed (which is characteristic of the large $\tan\beta$ regime), so that decays into bottoms and taus dominate
the branching ratio of the heavy neutral states. Typical production modes would be gluon-gluon fusion or production in association with $b$'s.
Note that, even though LHC searches at $13/14$~TeV could likely probe such heavy states, e.g.\ in the $\tau^+\tau^-$ channel, even larger masses 
of the heavy doublet states would be easily possible in this scenario. Moreover, should the heavy
doublet states be discovered, the question of distinguishing such a point, in the Higgs sector, from a MSSM scenario remains open: the heavy 
CP-even singlet, lightly coupled to SM-particles, is likely to evade detection. As an additional feature for this point -- additional but
not binding: light CP-odd singlets are not a generic feature of this decoupling scenario -- however, one observes that the CP-odd singlet is quite light,
$\sim110$~GeV. Although its large singlet component of $\sim99.7\%$ leads to a significant suppression of the couplings of this state to SM particles
(hence of its production cross-section), the couplings to down-type fermions are again enhanced by the large value of $\tan\beta$, so that one may look 
for a signal in the $\tau^+\tau^-$ and $b\bar{b}$ channels -- in associated production with $b$'s. The signal in the $\gamma\gamma$ channel would be 
strongly suppressed so that the current ATLAS limits \cite{Aad:2014ioa} have no impact for this point. Remember also from Fig.~\ref{lightA} that CP-odd 
singlets between $63$ and $\sim150$~GeV with slightly larger doublet components are also possible, which would improve their observability, although this feature would typically be associated with a 
lowered mass for the heavy Higgs-doublet states. Another possibility lies in exploiting triple-Higgs couplings for a production of $A_1$ in pairs. 
Furthermore, the associated production of $A_1$ with a $Z$-boson only proceeds via the tiny doublet component of $A_1$, hence is also suppressed. 

The second point is representative of the light-singlet scenario of the NMSSM. Much that has been said in connection with the previous point, especially
concerning the heavy doublet states, remains valid -- note however that the lower value of $\tan\beta$ decreases the importance of the branching ratios of the 
heavy doublet states into down-type fermions in favour of cascade decays, via light Higgs or SUSY states, which hence affects the search strategy . 
The light CP-even doublet again shows SM-like couplings (within a few percent), except for the couplings to down-type fermions,
which are slightly reduced: this results from the perturbation among $H_u$ and $H_d$ components, which is itself related to the triple-state mixing 
in the presence of a singlet component. As a consequence, LHC rates into gauge bosons are slightly enhanced (while the $b\bar{b}$ and $\tau^+\tau^-$ channels
are slightly suppressed): increased precision on the Higgs decays would thus prove interesting in such a configuration. The central test of this
scenario however rests with the observability of the light CP-even singlet at $\sim106$~GeV: its reduced doublet component $\leq3\%$ indeed entails
suppressed rates (due to the reduced production cross-section), at the percent level of those of a SM Higgs boson at the same mass. The corresponding 
cross-section at $8$~TeV for the diphoton final state is at the level of $10^{-3}$~fb. Direct
searches thus carry little chance of success on short timescales. Again, one could then search the light state in Higgs pair production.

The third point is also an example of a light singlet Higgs, although it differs from the previous one in two respects: it involves very low 
$\tan\beta=2$, and the light singlet has a significant doublet component $\sim9\%$. The first feature, low $\tan\beta$, plays a key role, together with 
$\lambda\sim0.7$, in increasing the mass of the `observed' state in the presence of a light squark spectrum. We remind the reader that this effect is not 
bound to the presence of a light CP-even singlet (see section \ref{ltb}). The second aspect, $S_{13}^2\sim9\%$, has consequences on the -- however still 
challenging -- observability of the light singlet at $\sim102$~GeV, as apparent LHC rates into $b\bar{b}$ and $\tau^+\tau^-$ would now reach a somewhat 
larger fraction ($\sim6-8\%$) of the corresponding values for a SM Higgs boson at this mass -- remember from sections \ref{lsing}, \ref{ltb} that the 
doublet components of the light singlet may take phenomenologically viable values as large as $\sim20\%$. The corresponding signal would also fit the LEP 
excess in the $b\bar{b}$ channel adequately enough. Note, though, that the $h_1\to\gamma\gamma$ rate is not particularly large ($\sim25\%$ of its SM 
value at this mass), and the corresponding cross-section at $8$~TeV is under $1$~fb for this particular example. On the other hand, as for the previous 
point, the rates of the SM-like state would give rise to slight excesses in the vector channels / a slight 
suppression in the down-type fermion channels.

Several (a priori separable) features are present for the fourth point. First, as for the previous point, the low $\tan\beta=2$ and large $\lambda=0.7$ 
ensure that the mass of the light doublet state is generated in the correct range $\sim125$~GeV, despite a light sfermion spectrum and negligible 
trilinear couplings. Furthermore, the heavy doublet sector is essentially decoupled again. The most remarkable aspect of this point however
rests with the fact that the light CP-even doublet and singlet states are almost mass-degenerate (within $\sim0.1$~GeV). Moreover, both quasi-degenerate states
are mixed at almost $\sim50\%$, which means that they carry `half a SM-like Higgs boson' each (in their contribution to the LHC rates). Ideally,
one should try and resolve experimentally the two states (their width is about $2$~MeV). Another strategy consists in looking for deviations
in the Higgs pair production rate, as singlet-doublet contributions would modify the apparent trilinear Higgs couplings at $\sim125$~GeV. The various
couplings read (in units of $g_{H^3}^{\mbox{\scriptsize SM}}$): $\sim0.83$ for $h_1-h_1-h_1$, $\sim-0.16$ for $h_1-h_1-h_2$, $\sim0.35$ for $h_1-h_2-h_2$ 
and $-0.53$ for $h_2-h_2-h_2$. The apparent triple-Higgs coupling accounting for both states (in connection to external SM particles) gives $\sim205$~GeV, 
which represents a $\sim7\%$ increase with respect to the SM value (very difficult to resolve).

The fifth point is another example of the decoupling limit. Its aspect that we wish to discuss, however, is the presence of a CP-odd state at 
$\sim 44$~GeV, that is below the $h_1\to2A_1$ threshold. Yet, the corresponding width is strongly suppressed, in accordance with the LHC data, 
resulting in a branching ratio of only $BR(h_1\to2A_1)\sim10^{-4}$. The couplings of the CP-even state at $\sim125.1$~GeV and its rates at the LHC are 
thus essentially SM-like. On the other hand, the light CP-odd Higgs is almost a pure singlet so that it will evade detection in direct production.
Production of $A_1$ from a heavier state is also problematic, as only the CP-even singlet $h_2$ has a sizable decay into this state. Therefore,
the most serious hope for the observation of such a $A_1$ would still lie in increased precision in the characteristics of $h_1$ at future colliders.
This point illustrates the extreme possibility of a light CP-odd state devoid of any phenomenological impact on the standard sector. As we stressed in 
section \ref{la1}, however, it still makes sense to test a $BR(h_1\to2A_1)$ in the few percent range.

In the case of the sixth point, the whole doublet sector lies below $\sim 125$~GeV, with very light neutral states at $\sim63$~GeV while the charged 
Higgs has a mass of $101.5$~GeV. Expectedly, the couplings of $h_1$ to SM gauge bosons are suppressed, which accounts for its invisibility at LEP. Its
diphoton cross-section at $8$~TeV is at the level of $1$~fb, as that of the light $A_1$ state.
The couplings of the heavy doublet state, at $\sim125.6$~GeV, are reasonably close to their SM value, resulting in SM-like cross-sections for
the corresponding state (within a few percent): it thus offers an acceptable interpretation of the observed signals as the competitive $\chi^2$ proves. 
Yet this point is discarded by Higgsbounds: indeed, $BR(t\to H^+b)\sim1.12\cdot10^{-2}$ (together with $BR(H^{-}\to\tau\nu_{\tau})
\simeq0.97$) is beyond the recent limits established by CMS. Much tension is also present in $B$-physics observables where $BR(B\to\tau\nu_{\tau})$ 
and $BR(\bar{B}_s\to\mu^+\mu^-)$ lie beyond $2$ standard deviations.

While electroweak precision observables were not included within our fit, predictions for $M_W$ are displayed in Table \ref{fp1}, using the tools 
presented in \cite{Domingo:2011uf}; the values given in brackets (in MeV) are estimates of the theoretical uncertainty. Comparing these
values with the current experimental measurement ($M_W^{\mbox{\tiny exp.}}=80.385\pm0.015$~GeV), we observe that all points remain within $1\sigma$ 
deviation of the experimental average, with possible tensions at low $\tan\beta$ or in the presence of light doublet states.

\section{Attempt at a `global' scan}\label{gs}
Performing a global scan over the NMSSM parameter space proves to be both a challenging and potentially even misleading task. The NMSSM Higgs sector 
involves $6$ degrees of freedom ($\lambda$, $\kappa$, $\tan\beta$, $M_A$, $\mu$, $A_{\kappa}$) at lowest order, to which one should also add the stop
masses and trilinear coupling ($A_t$), which have an important impact via loop corrections. The rest of the supersymmetric spectrum cannot be altogether 
left out from a scan 
if one keeps e.g.\ $(g-2)_{\mu}$ and the requirement for a neutralino LSP in the process. Moreover, some scenarii that we described in the 
previous sections and which provide a reasonable or even an excellent fit to LHC/TeVatron data involve specific parameter configurations (so that one 
obtains the appropriate spectrum/couplings), hence disappear or become uncompetitive if the scanning density is too loose. On the other hand, keeping
a tight scanning density over so many variables would require both long scan durations and huge storage capacities.

\begin{figure}[!htb]
    \centering
    \includegraphics[width=15cm]{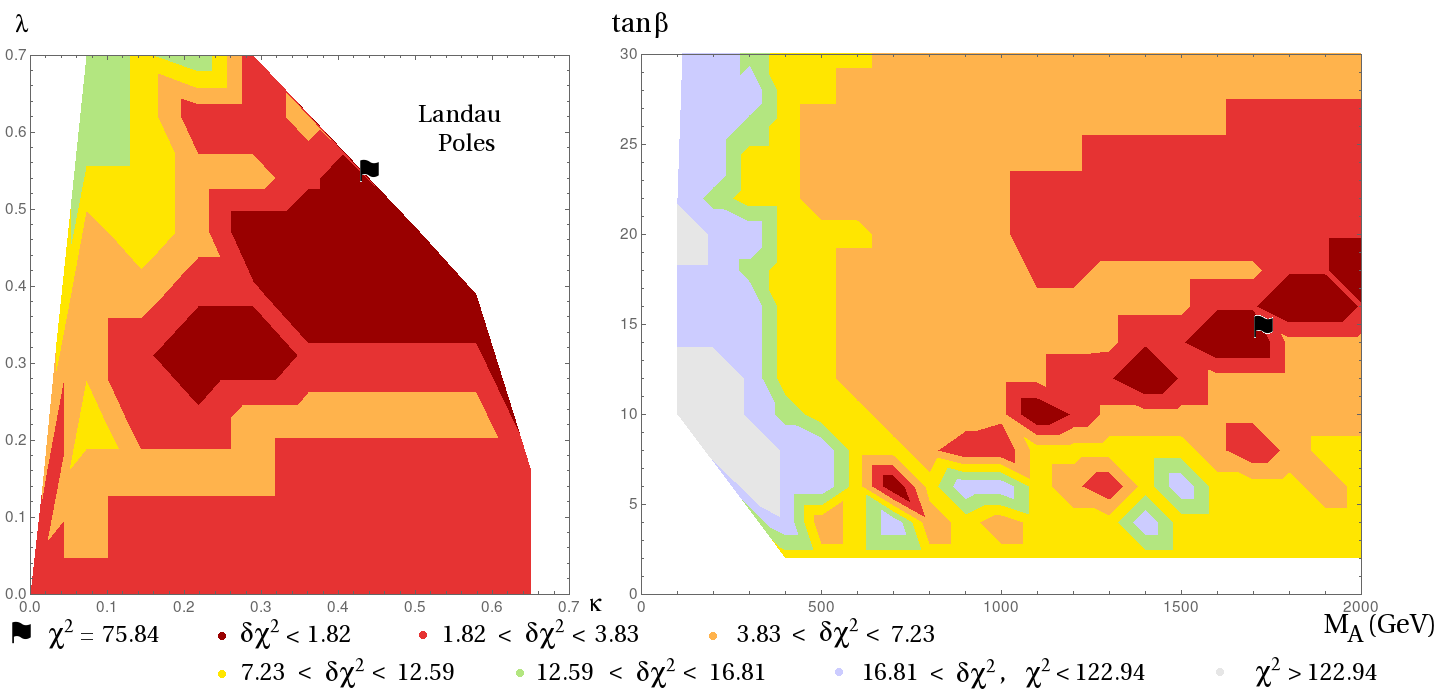}
  \caption{Parameter space in the `global' scan: $\tan\beta\in[2,30]$, $\mu\in[120,2000]$~GeV, $M_A\in[100,2000]$~GeV, $A_{\kappa}\in[-2000,200]$~GeV, 
$\lambda\in[2\cdot10^{-4},0.7]$, $\kappa\in[2\cdot10^{-4},0.65]$, $2M_1=M_2=500$~GeV, $M_3=1.5$~TeV, $m_{\tilde{Q}_3}=1$~TeV, $m_{\tilde{Q}_{1,2}}=1.5$~TeV, 
$A_t=-2.5$~TeV, $A_{b,\tau}=-1.5$~TeV, $m_{\tilde{L}}=250$~GeV. The distribution of the $\chi^2$ fit, relative to the best-fit point is shown in crimson ($\delta\chi^2<2.20$), 
red ($\delta\chi^2<3.83$), orange ($\delta\chi^2<7.23$), yellow ($\delta\chi^2<12.59$), green ($\delta\chi^2<16.81$) and blue ($\chi^2<122.94$);
grey dots represent points with $\chi^2>122.94$, black dots give $B$-physics or $(g-2)_{\mu}$ outside $2\sigma$ (and $\chi^2>122.94$).}
  \label{gs_parspace}
\end{figure}
Under these conditions, we settled for the following procedure. Since new NMSSM effects, with respect to the MSSM, essentially involve the 
tree-level parameters, we decided to freeze the supersymmetric input to a `good' MSSM value (large $A_t$ and relatively massive squarks,
along with light sleptons) and scanned over the $6$ tree-level Higgs input quantities: $\lambda\in[2\cdot10^{-4},0.7]$ --$10$ values, $\kappa
\in[2\cdot10^{-4},0.65]$ --$10$ values, $\tan\beta\in[2,30]$ --$15$ values, $M_A\in[100,2000]$~GeV --$20$ values, $\mu\in[120,2000]$~GeV
--$20$ values, $A_{\kappa}\in[-2000,200]$~GeV --$20$ values, for a total of about $12\cdot10^{6}$ points (with about $15\%$ of them not being
excluded and representing a data storage of about $0.6$~Gbytes). Note that, under these conditions, a large one-loop contribution to the SM-like 
CP-even Higgs mass is generically present, allowing for a good-fit in the decoupling limit; the benefit of large $\lambda$'s (although not 
maximal) at low $\tan\beta$ is not spoilt, however (the MSSM tree-level mass would still be too small to provide a phenomenologically viable
Higgs boson at low $\tan\beta$).

\begin{figure}[!htb]
    \centering
    \includegraphics[width=8cm]{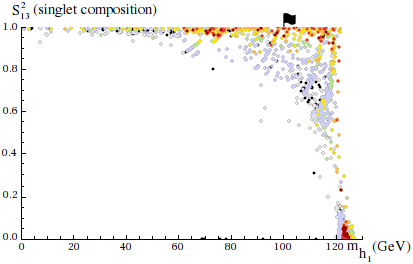}
  \caption{Singlet composition of the light Higgs state in the global
scan. (Colour code as in Fig.~\ref{gs_parspace}.)}
  \label{gs_singcomp}
\end{figure}

We show the outcome of this scan in the $\{\kappa,\lambda\}$ and the $\{M_A,\tan\beta\}$ planes in Fig.~\ref{gs_parspace}: one observes that the 
best-fitting points ($\chi^2\lsim77.5$) lay in the $\lambda\sim\kappa\sim0.3-0.5$ region and draw a band from $(M_A\sim500~\mbox{GeV},\tan\beta\sim5)$ 
to $(M_A\sim2000~\mbox{GeV},\tan\beta\sim15)$: they involve a light CP-even singlet. Next follow points fitting very well ($\chi^2\sim77.5-79$) in the 
decoupling limit with large $\tan\beta$ ($M_A\gsim1~\mbox{TeV}$, $\tan\beta\gsim20-25$). Points in the low $\tan\beta$ limit do not give competitive 
fits as they suffer from a large pull from $(g-2)_{\mu}$ (they still receive $\chi^2\sim85$, improving on the SM-limit). Fig.~\ref{gs_singcomp} shows 
the singlet composition of the lightest Higgs state. In addition to the best-fitting scenarios that we just discussed -- and which are easily recognizable 
here: light singlets at $S_{13}^2\sim1$ and mass in the range $[63,120]$~GeV, decoupling limit for $S_{13}^2\sim0$ and 
$m_{h_1^0}^2\to125$~GeV --, one observes that light singlet states under $\sim62$~GeV may come with acceptable fit values, although their doublet 
components have to be strongly suppressed. Points involving a large singlet-doublet mixing are also found in the vicinity of $m_{h_1^0}=125$~GeV.
Furthermore, one observes a few points close to the x-axis for $m_{h_1}\lsim100$~GeV: they correspond to the light doublet scenario. While they result in
a poor fit to the observed Higgs data $\chi^2\gsim135$, they escape exclusion from the unobserved top decay to a charged Higgs (with mass $\sim110$~GeV 
here) due to a suppressed branching ratio at large values of $\tan\beta$ ($\gsim20$).
Finally, note that the light CP-odd Higgs scenario is only represented by light CP-odd doublets (hence offers a poor fit to the Higgs data due to suppressed 
conventional channels for the state at $\sim125$~GeV) in our scan: light CP-odd singlets are barely represented as the scan density allows for very few 
points with low $|A_{\kappa}|$.

\begin{figure}[!htb]
    \centering
    \includegraphics[width=15cm]{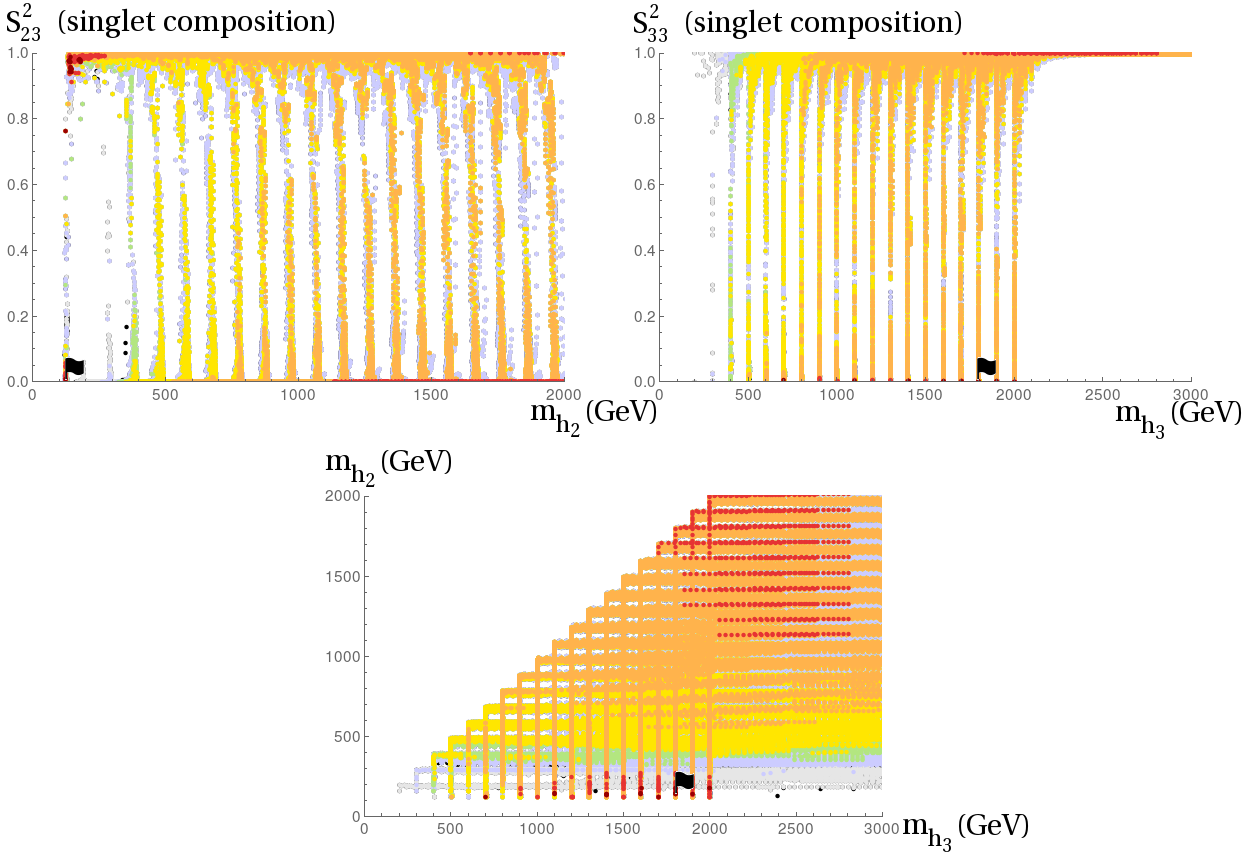}
  \caption{Characteristics of the heavier CP-even Higgs
states in the global scan. (Colour code as in Fig.~\ref{gs_parspace}.) In
the first row, we show the singlet composition
  of the second (left) and third (right) CP-even Higgs state as a
function of their masses. The plot below shows the distribution of the fit
in the plane defined by the masses of the 
  two heavier Higgs states. Note that the `ladder' and `grid' structures
in this plot are artefacts of the scan.}
  \label{gs_heavy}
\end{figure}

Fig.~\ref{gs_heavy} shows the properties of the heavier Higgs states. The plot on the top left-hand corner shows that the points which give the best fit tend to come with a light second-lightest Higgs state: 
this can be either the SM-like doublet state at $\sim125$~GeV or a
slightly heavier singlet. We also find a good fit in a region with
mass values of $\gsim1$~TeV: there, the second-lightest Higgs can be either singlet or heavy-doublet
in nature. Interestingly, sizable mixings of the singlet with the heavy
doublet state lead to fit values of secondary quality (orange vertical lines,
where the `line' shape is an artefact of the scan using discrete values
of $M_A$). 
In the top right-hand corner, the plot shows the composition of the
heaviest CP-even state. Unsurprisingly, this state 
is the heavy doublet state for the best fit points -- which involve 
light singlet states, as follows from our discussion above.
However, a good fit is also obtained for the case where the 
singlet is very heavy, as indicated by the red points forming a
horizontal line at $S_{23}^2=1$ (note that the apparent interruption
in red points at $2.9$~TeV is an artefact of the scan and that the
corresponding `red line' actually continues for heavier masses, not
displayed in the figure). The plot in the lower row shows the $\chi^2$
distribution on the plane
defined by the masses of the two heaviest CP-even Higgs states. Two
favoured regions appear: one, for $m_{h_2}\sim125$~GeV and
$m_{h_3}\gsim0.7$~TeV, corresponds to the scenario with a light singlet
state; on the contrary, in the other preferred region both 
$m_{h_2}$ and $m_{h_3}$ are heavier than 1~TeV, corresponding to the
scenario where one doublet state is light while both the singlet state and 
the other doublet state are very heavy. 
Note that both these preferred regions share
the common feature to involve a heavy doublet state with a mass close to
or above 1~TeV: this is 
characteristic of the decoupling limit and gives rise to the fact
that the light doublet state has SM-like properties.

\begin{table*}[t]
\null\begin{tabular}{|c||c|c|c|c|c|c}
\hline
 & Point A & Point B & Point C & Point D & Point E & \ldots\\ 
Parameters & (light sing.) & (light doub.) & (low $\tan\beta$) & (low $\tan\beta$) & (Dec.\ limit) & \ldots\\\hline\hline
$\lambda$ & $0.54$ & $0.16$ & $0.54$ & $0.63$ & $0.39$ & \ldots\\ \hline
$\kappa$ & $0.43$ & $0.29$ & $0.22$ & $0.36$ & $0.58$ & \ldots\\ \hline
$\tan\beta$ & $14$ & $20$ & $2$ & $2$ & $18$ & \ldots\\ \hline
$\mu_{\mbox{\tiny eff}}$ (GeV) & $120$ & $1505$ & $219$ & $318$ & $120$ & \ldots\\ \hline
$M_A$ (GeV) & $1700$ & $200$ & $500$ & $600$ & $1900$ & \ldots\\ \hline
$A_{\kappa}$ (GeV) & $-263$ & $-2000$ & $-263$ & $-147$ & $-495$ & \ldots\\ \hline\hline
Higgs Spectrum &  &  &  &  &  &  \\\hline\hline
$m_{h_1}$ (GeV) & $100.5$ S & $69.9$ D & $106.6$ S & $124.5$ D & $123.1$ D  & \ldots\\ \hline
$m_{h_2}$ (GeV) & $125.2$ D & $122.8$ D & $126.3$ D & $333$ S & $203$ S & \ldots\\ \hline
$m_{A_1}$ (GeV) & $285$ S & $61.6$ D & $279$ S & $309$ S & $512$ S & \ldots\\  \hline
$m_{H^{\pm}}$ (GeV) & $1671$ & $110.7$ & $491$ & $589$ & $1874$ & \ldots\\\hline\hline
$\chi^2$ (/89 obs.) & $75.8$ & $137.7$ & $87.6$ & $85.4$ & $78.6$ & \ldots\\ \hline
\end{tabular}
\vspace{0.5cm}

\null\hspace{1cm}\begin{tabular}{c|c|c|c|c|c|}
\hline
\ldots & Point F &  Point G & Point H & Point I & Point J \\ 
\ldots & (MSSM limit) & (2 CP-even) &  (close CP odd) & ($m_{h_1}<\frac{m_{h_{SM}}}{2}$) & ($m_{A_1}<\frac{m_{h_{SM}}}{2}$) \\\hline\hline
\ldots & $2\cdot10^{-4}$ & $0.23$ &  $0.39$ & $0.23$ & $0.08$ \\ \hline
\ldots & $2\cdot10^{-4}$ & $0.22$ &  $0.36$ & $0.14$ & $0.36$ \\ \hline
\ldots & $20$ & $6$ &  $8$ & $16$ & $26$ \\ \hline
\ldots & $1406$ & $120$ &  $120$ & $120$ & $1109$ \\ \hline
\ldots & $2000$ & $700$ &  $900$ & $1900$ & $200$ \\ \hline
\ldots & $-2000$ & $-147.4$ &  $-31.6$ & $-263$ & $-1421$ \\ \hline\hline
  &   &   &   &   &   \\\hline\hline
\ldots & $123.0$ D & $122.0$ S &  $123.7$ D & $48.9$ S & $69.1$ D \\ \hline
\ldots & $2012$ D & $123.4$ D &  $211.7$ S & $124.3$ D & $124.0$ D \\ \hline
\ldots & $2013$ D & $198$ S &  $123.6$ S & $244$ S & $61.7$ D \\ \hline
\ldots & $2014$ & $S_{13}^2\simeq76\%$ &  $P_{13}^2\simeq99.8\%$ & / & $BR(h_1\to2A_1)\sim0.7$ \\ \hline\hline
\ldots & $79.3$ D & $77.0$ &  $82.3$ & $82.8$ & $174.1$ \\ \hline
\end{tabular}

\caption{A few `best-fit points' emerging from the `global' scan.}
\label{gs_fit}
\end{table*}
Let us now investigate the outcome of this global scan in terms of the various scenarii that we considered. Representatives of all the types of spectra 
that we discussed are found: their best-fit points are summarized in Table~\ref{gs_fit}.
\begin{itemize}
\item Point A is the best-fit point of the whole scan. It exhibits a light CP-even singlet at $\sim101$~GeV. As for the doublet sector, it is
characteristic of the decoupling limit. That this best-fitting point involves a light CP-even singlet demonstrates clearly that this scenario
is currently the most interesting and motivated one arising in the context of the NMSSM. The doublet component of the light state is of the order of
$2\%$, so that the associated signals in direct production should be expected at the percent level of those of a SM Higgs boson.
\item Point B involves a light CP-even doublet state below $100$~GeV. The associated $\chi^2$ is quite high, illustrating the limited pertinence of
this interpretation of the LHC data.
\item Points C and D are two representatives of the NMSSM at low $\tan\beta$ and large $\lambda$: in the first case, the CP-even singlet state is
lighter than the doublet at $\sim125$~GeV, while it is heavier in the second case. Note that $(g-2)_{\mu}$ is a major pull in the present case, due to 
both low $\tan\beta$ and moderately light sleptons/neutralinos/charginos. In both cases, the mostly-singlet state has a doublet component at the percent
level only, leading to suppressed production rates.
\item Point E is a representative of the decoupling limit at `large' $\tan\beta$, although not in the MSSM limit. The large value of $\tan\beta\sim18$ 
is essentially driven by $(g-2)_{\mu}$. The heavy doublet states have large masses ($\sim2$~TeV), while a CP-even singlet with a $\sim2\%$ doublet 
component appears at $\sim200$~GeV.
\item Point F gives the best fit value for the MSSM limit: this is again a `Decoupling Limit'-like point with large $\tan\beta$, heavy $H^{\pm},\ H^0,\ 
A^0$ doublet states and even heavier singlets. The presence of relatively light sleptons and a bino at $\sim250$~GeV together with large $\tan\beta$ 
allow for a satisfactory fit to $(g-2)_{\mu}$, while, for all other aspects, this point could be regarded as belonging to the SM limit.
\item Two CP-even states, mixing singlet and light doublet almost at the level of $25\%$, intervene close to $\sim125$~GeV in Point G. While ATLAS and
CMS may be able to resolve their mass gap of about $1.5$~GeV, note that the degeneracy could be at the level of $100$~MeV. While points with 
smaller mass gaps were present in the global scan, none showed a mixing between singlet and doublet components as large as Point G: the latter would
allow for the observability of two separate peaks of comparable width in the spectrum.
\item Point H involves, in addition to a CP-even doublet, one singlet-like CP-odd state close to $\sim125$~GeV. As the corresponding state is singlet 
at $99.8\%$, the effect in direct production is completely negligible. Effects in pair production could develop however, due to the $h_1-A_1-A_1$ 
coupling.
\item For the Points I and J, light states are present below $m_{h_{SM}}/2$, hence potentially opening unconventional Higgs decays. For Point I, the 
light state is a CP-even singlet at $\sim50$~GeV: even though $\lambda$ and $\kappa$ are quite large, the unconventional Higgs decay is suppressed due 
to an accidental cancellation of the various terms entering the $h_2-h_1-h_1$ coupling: this avoids conflict with the LHC/TeVatron data. The doublet 
component of the light state is under $1\%$, explaining its compatibility with the LEP constraints. For Point J, a light 
doublet-like CP-odd state is present, leading to a large $h_1\to2A_1$ branching ratio $\sim0.7$: the quality of the fit is correspondingly quite poor, 
despite the presence of a CP-even doublet at $\sim124$~GeV. Note that the case of light CP-odd Higgs states is not really probed in this scan since the 
low scan density does not allow for many points in the low $|A_{\kappa}|$ region (which provides the light CP-odd singlet states).
\end{itemize}

\section{Conclusions}\label{Conc}
This discussion of the NMSSM Higgs scenarii in view of the Higgs-measurement data shows that, even though a conventional SM-like Higgs is expected at $\sim125$~GeV,
many scenarios that differ very significantly from the SM case remain possible and even give competitive fits to the LHC/TeVatron data. In most cases, 
these configurations involve light additional degrees of freedom, CP-even and/or CP-odd, singlet or doublet-like, hiding invisibly at lower masses or combining their signals with
those of the SM-like state. 

The most prominent NMSSM Higgs scenario which turned out to be favoured by the fits carried out in this paper involves a CP-even mostly singlet 
state with mass below that of the light doublet state (which is identified
with the signal observed at LHC) and includes a small mixing of the singlet and doublet components. From the point of view of the model, this setup 
has the desirable feature of uplifting somewhat the mass of the doublet-like state, arguably rendering a mass of $\sim125$~GeV more natural. Small 
deviations, at the percent level, from the SM rates, e.g.\ slightly suppressed decays into down-type fermions or slightly enhanced vector channels, can 
also be generated for the state at $\sim125$~GeV via a disturbed proportion in $H_u$ and $H_d$ components, which is associated to the appearance of a 
small singlet component (three-state mixing). Such features are shared with many other configurations however (e.g.\ perturbation of the couplings via 
heavier particles contributing at the loop level), and the crucial test of this scenario would lie in the detection of the light CP-even singlet itself. 
In the most favourable cases, the latter may acquire doublet components of the order of $10\%$ and could thus appear as a `miniature' Higgs boson, with 
a production cross-section reduced to $\sim10\%$ of that of a SM state at the same mass. Its dominant decays would likely be $b\bar{b}$ and 
$\tau^+\tau^-$, although unconventional rates, e.g.\ enhanced $\gamma\gamma$ or even $c\bar{c}$ channels, could also appear in certain configurations 
of the doublet composition of the lightest state. As an additional feature, the $2.3\,\sigma$ (local) excess noted at LEP in $e^+e^-\to Z h\to b\bar{b}$ 
could be accomodated with such a singlet carrying a sizable doublet component in the vicinity of $\sim100$~GeV. Direct searches, e.g.\ in the $\gamma\gamma$ 
or the $\tau^+\tau^-$ channel, may thus eventually
detect such a particle, provided that those searches are carried out in the low-mass region. On the other hand, reduced doublet components, in the percent range or below, are also compatible with the data and may
equally well account for an uplift of the mass of the second lightest CP-even state. Discovery of the light state could then prove difficult and,
while effects in Higgs pair production are possible, improved searches would be required to ensure that the singlet does not remain unnoticed.

In contrast, the popular NMSSM scenario where the SM-like Higgs boson decays unconventionally and sizably into lighter Higgs states 
(CP-even or CP-odd) has been essentially emptied of its phenomenological content, due to the success of the Higgs searches in the standard 
channels. Yet, light states below $\sim63$~GeV are still an open possibility provided the decay of the `observed' Higgs towards them is 
suppressed. From the point of view of the model, this amounts to a constraint on the Higgs sector. The main consequence for the hypothetical
light states is that their coupling to the observed state should remain small, which implies, in particular, that they should be almost purely
singlets. This further reduces the prospects for detecting the light Higgs state via channels alternative to the decays from
the doublet states, as all couplings to SM particles (hence the cross-section in production channels involving fermions or vector bosons) are 
correspondingly suppressed.

A more exotic possibility is a scenario with additional quasi-degenerate states at $\sim125$~GeV. The NMSSM indeed allows for CP-even singlet states
at this mass. While the off-diagonal singlet-doublet mass-entry should then remain small (so that the quasi-degeneracy is not lifted by this off-diagonal 
term), the mixing
of such a singlet with the SM-like state could be as large as $50\%$ (but also as small as $0\%$). Such a possible scenario motivates experimental
efforts to improve the resolution in 
the mass measurement of the state at $\sim125$~GeV, as two separate peaks may eventually become distinguishable. Other possible, but not guaranted,
effects would arise from modified apparent triple-Higgs couplings at $125$~GeV. The situation is somewhat different if the additional state is 
CP-odd. While current limits then favour a mostly singlet state, a sizable doublet component may also be present (hence allow the state
to be produced and contribute to the LHC rates). As the CP-odd Higgs does not couple to electroweak gauge bosons, its decay products, essentially
fermionic, would add to those of the SM-like state and displace the apparent proportion among the various channels. Note that this latter case
mimicks a scenario with CP-violation in the Higgs sector, where the observed state would have a CP-odd component.

A noteworthy aspect of the NMSSM, as compared to the MSSM, lies in its resilience at low values of $\tan\beta$, where specific tree-level 
contributions associated with a large $\lambda$ become important. Dependence on the supersymmetric spectrum to generate a Higgs mass close to
$\sim125$~GeV is of secondary importance in this case, so that the hypothetical presence of light stops and small trilinear couplings can be compatible 
in the NMSSM with the discovered signal. In addition to the most commonly considered case where the SM-like state is the lightest CP-even Higgs, the 
previous scenarios, involving e.g.\ a lighter CP-even singlet or quasi-degenerate states, can be accomodated in this context as well.

Finally, the prospect of a light doublet Higgs sector, with light neutral states below $100$~GeV, decoupling from electroweak gauge bosons, 
cannot be discarded altogether yet. While a significant adjustment of the parameters is necessary to avoid the relevant experimental limits, this 
scenario can serve as a motivation to cover remaining loopholes in the experimental searches. 
Further searches for light neutral but also charged Higgs states should soon provide more information about the validity of this scenario.

In summary, while the properties of the detected signal are so far compatible with the Higgs boson of the SM, the analyses of this paper clearly
demonstrate that the current results do not necessarily imply a minimalistic 
phenomenology but, on the contrary, call for a comprehensive investigation and precision tests of the Higgs properties.

\section*{Acknowledgements}
The authors acknowledge interesting discussions with U.~Ellwanger. They are also grateful to T.~Stefaniak and O.~St\aa{}l for their assistance in
the use of HiggsBounds and HiggsSignals, as well as to S.~Liebler for his
help with SusHi. 
The authors acknowledge support by
the DFG through the SFB~676 ``Particles, Strings and the Early Universe''.
This research was supported in part by the European Commission through the
``HiggsTools'' Initial Training Network PITN-GA-2012-316704.
\newpage
\appendix
\section{Best Fit points in Fig.~\ref{SMlim}-\ref{brtHnew}}
\begin{table*}[tbh!]
\null\vspace{2cm}\hspace{0cm}\begin{tabular}{|c||c|c|c|c|c|c}
\hline
NMSSMTools & Figure \ref{SMlim} & Figure \ref{MATBscan} & Figure \ref{MATBscan}  & Figure \ref{heavydoub} & Figure \ref{heavydoub} &  \ldots\\ 
Parameters & (SM limit) &  (Dec. Limit) &  (Dec. Limit) & (light sing.) & (2 CP-even) & \ldots\\\hline\hline
$\lambda$ & $1\cdot10^{-5}$ &  $0.2$ & $2\cdot10^{-4}$  & $0.45$ & $0.55$ & \ldots\\ \hline
$\kappa$ & $1\cdot10^{-5}$ &  $0.6$ & $2\cdot10^{-4}$ & $0.35$ & $0.3$ & \ldots\\ \hline
$\tan\beta$ & $10.8$ &  $23$ & $22$  & $8$ & $8$ &  \ldots\\ \hline
$\mu_{\mbox{\tiny eff}}$ (GeV) & $1000$ &  $200$ &  $200$  & $125$ & $125$ & \ldots\\ \hline
$M_A$ (GeV) & $2000$ &  $1000$ &  $1000$  & $1000$ & $1000$ & \ldots\\ \hline
$A_{\kappa}$ (GeV) & $-1000$ &  $-1500$ &  $-578$  & $-255$ & $-27$ &  \ldots\\ \hline
$M_1$ (GeV) & $500$ &  $250$ &  $250$  & $250$ & $250$ & \ldots\\ \hline
$M_2$ (GeV) & $1000$ &  $500$ & $500$  & $500$ & $500$ & \ldots\\ \hline
$M_3$ (TeV) & $3$ &  $1.5$ & $1.5$  & $1.5$ & $1.5$ & \ldots\\ \hline
$m_{\tilde{Q}_{1,2}}$ (TeV) & $2$ &  $1.5$ & $1.5$  & $1.5$ & $1.5$ & \ldots\\
$m_{\tilde{Q}_3}$ (TeV; if $\neq m_{\tilde{Q}_{1,2}}$) & / &  $1.1$ &  $1.1$  & $1$ & $1$ & \ldots\\ \hline
$m_{\tilde{L}}$ (GeV) & $1000$ &  $300$ & $300$  & $200$ & $200$ & \ldots\\ \hline
$A_{t}$ (TeV) & $-4$ &  $-2.3$ & $-2.3$  & $-2$ & $-2$ & \ldots\\
$A_{b,\tau}$ (TeV; if $\neq A_{t}$) & $-1.5$ &  $-1.5$ & $-1.5$  & $-1.5$ & $-1.5$ & \ldots\\ \hline\hline
Higgs Spectrum &   &   &  &  & \\\hline\hline
$m_{h_1}$ (GeV) & $125.7$ D &  $124.1$ D &  $125.0$ D  & $109.5$ S & $124.6$ D & \ldots\\ \hline
$m_{h_2}$ (GeV) & $1732$ S &  $732$ S &  $211$ S  & $124.5$ D & $124.9$ S & \ldots\\ \hline
$m_{A_1}$ (GeV) & $1732$ S &  $972$ S &  $589$ S  & $280$ S & $110$ S & \ldots\\ \hline
Other & / &  /  &  /  & $S_{13}^2=95\%$ & $S_{13}^2=20\%$ & \ldots\\ \hline\hline
$\chi^2$ (/89 obs.) & $86.9$ &  $80.5$ &  $81.4$  & $75.5$ & $80.1$ & \ldots\\ \hline

\end{tabular}

\caption{A few `best-fit points', with NMSSMTools input parameters and some data concerning the Higgs spectrum. The labels (SM limit), (Dec. limit),
(light doub.), (light sing.), (CP odd), (large $\lambda$), ($\lambda$+sing.), (2 CP-even), (close $A$) refer to the remarkable characteristics of the point,
belonging to the class of points respectively in SM-limit, in the decoupling limit, exhibiting a doublet CP-even Higgs much lighter than $\sim125$~GeV, exhibiting 
a singlet CP-even Higgs much lighter than $\sim125$~GeV, with a light CP-odd Higgs under $\sim125$~GeV, profiting from a large tree level $\lambda$ effect,
involving both a large $\lambda$ effect and a light CP-even singlet, involving several CP-even Higgs states close to $\sim125$~GeV, or a CP-odd Higgs
close to $\sim125$~GeV. The letters S and D coming together with the masses of the Higgs states indicate whether the
state is dominantly singlet or doublet. HB\_4.1.3 + HS\_1.2.0 indicate that the test for a particular point was performed with the versions 4.1.3 and 
1.2.0 of HiggsBounds and HiggsSignals instead of versions 4.2.0 and 1.3.1 as for the rest of the points.}
\label{BestFit}
\end{table*}
\begin{table*}[tbh!]

\null\hspace{0cm}\begin{tabular}{c|c|c|c|c|c|c|c}
\hline
\ldots & Figure \ref{heavydoub} &  Figure \ref{largemix} & Figure \ref{singrates} & Figure \ref{singrates} & Figure \ref{PQlim} &  Figure \ref{PQlim} & \ldots\\ 
\ldots & (CP-odd) & (light sing.) & (light sing.) & (light sing.) & ($\lambda$+sing.) &  (large $\lambda$) & \ldots\\\hline\hline
\ldots & $0.55$ & $0.6$ & $0.1$ & $0.1$ & $0.7$ & $0.7$&   \ldots\\ \hline
\ldots & $0.3$ & $0.35$ & $0.05$ & $0.05$ & $0.1$ & $0.1$ & \ldots\\ \hline
\ldots & $8$ &  $8$ & $12$ & $12$ & $2$ & $2$ &  \ldots\\ \hline
\ldots & $125$ & $125$ & $125$ & $125$ & $361$ & $324$ &   \ldots\\ \hline
\ldots & $1000$ & $1000$ & $1655$ & $1367$ & $870$ & $775$ &   \ldots\\ \hline
\ldots & $-46$ &  $-204$ & $-69$ & $-84$ & $-123$ & $31$ &  \ldots\\ \hline
\ldots & $250$ & $250$ & $250$ & $250$ & $75$ & $75$ &   \ldots\\ \hline
\ldots & $500$ & $500$ & $500$ & $500$ & $150$ & $150$ &  \ldots\\ \hline
\ldots & $1.5$ & $1.5$ & $1.5$ & $1.5$ & $1.5$ & $1.5$ &  \ldots\\ \hline
\ldots & $1.5$ & $1.5$ & $1.5$ & $1.5$ & $1.5$ & $1.5$ &  \ldots\\
\ldots & $1$ & $1$ & $1$ & $1$ & $0.5$ & $0.5$ &  \ldots\\ \hline
\ldots & $200$ & $200$ & $200$ & $200$ & $110$ & $110$ & \ldots\\ \hline
\ldots & $-2$ & $-1.5$ & $-2$ & $-2$ & $-0.1$ & $-0.1$ &  \ldots\\
\ldots & $-1.5$ &  /  & $1$ & $1$ & $-1.5$ & $-1.5$ &      \ldots\\ \hline\hline
\ldots &  &  &  &  &  &  & \\\hline\hline
\ldots & $119.8$ S & $61.8$ S & $105.3$ S & $101.6$ S & $105.3$ S & $123.2$ D &  \ldots\\ \hline
\ldots & $124.8$ D & $123.4$ D & $124.5$ D & $123.9$ D & $124.7$ D & $131.7$ S &  \ldots\\ \hline
\ldots & $124.8$ S & $-235$ S & $114.3$ S & $125.9$ S & $173$ S & $79$ S &  \ldots\\ \hline
\ldots & $P_{13}^2=99.4\%$ & $S_{13}^2=96\%$ & $S_{13}^2=94\%$ & $Br_{h_1\to c\bar{c}}\simeq31\%$ & $S_{13}^2=94\%$ & $S_{13}^2=4\%$ & \ldots\\ \hline\hline
\ldots & $79.4$ & $75.2$ & $79.8$ & $82.4$ & $76.1$ & $76.6$ &  \ldots\\ \hline

\end{tabular}\null\vspace{0.5cm}
\null\hspace{0cm}\begin{tabular}{c|c|c|c|c|c|c}
\hline
\ldots &  Figure \ref{PQlim} &  Figure \ref{PQlimmix} & Figure \ref{PQlimmix} & Figure \ref{lightA} & Figure \ref{lightA} & \ldots \\ 
\ldots & (2 CP-even) & ($\lambda$+sing.) & ($\lambda$+sing.) &  (CP odd) & (CP odd) & \ldots \\\hline\hline
\ldots & $0.7$ & $0.7$  & $0.7$ & $0.4$ &  $0.25$ & \ldots \\ \hline
\ldots & $0.1$ & $0.1$ & $0.1$ & $0.25$ &  $0.15$ & \ldots \\ \hline
\ldots & $2$ & $2.25$ & $2.25$ & $8$ & $11$ & \ldots \\ \hline
\ldots & $398$ & $315$ & $343$ & $120$ &  $120$ & \ldots \\ \hline
\ldots & $951$ & $826$ & $898$ & $900$ &  $1200$ & \ldots \\ \hline
\ldots & $-68$ & $-83$ & $-51$ & $-31$ &  $-1$ & \ldots \\ \hline
\ldots & $75$ & $75$ & $75$ & $250$ & $250$ & \ldots \\ \hline
\ldots & $150$ & $150$ & $150$ & $500$ & $500$ & \ldots \\ \hline
\ldots & $1.5$ & $1.5$ & $1.5$ & $1.5$ & $1.5$ & \ldots \\ \hline
\ldots & $1.5$ & $1.5$ & $1.5$ & $1.5$ & $1.5$ & \ldots \\
\ldots & $0.5$ & $0.5$ & $0.5$ & $1.2$ & $1.2$ & \ldots \\ \hline
\ldots & $110$ & $150$ & $150$ & $300$ & $300$ & \ldots \\ \hline
\ldots & $-0.1$ & $-0.1$ & $-0.1$ & $-2.5$ & $-2.5$ & \ldots \\
\ldots &  /  &  /  &  /  &  /  &  /  & \ldots\\ \hline\hline
   &  &  &  &  &  &   \\\hline\hline
\ldots & $123.4$ S/D & $100.5$ S & $108.9$ S & $123.0$ D & $123.2$ D & \ldots \\ \hline
\ldots & $123.8$ D/S & $123.4$ D & $125.8$ D & $141$ S & $144$ S & \ldots \\ \hline
\ldots & $154$ S & $145$ S & $135$ S & $103.1$ S & $41.3$ ($P_{13}^2\simeq100\%$) & \ldots \\ \hline
\ldots & $S_{13}^2=63\%$& $S_{13}^2=81\%$ & $S_{13}^2=73\%$ & $P_{13}^2=99.7\%$ & $Br_{h_1\to2A_1}\simeq1\cdot10^{-3}\%$ & \ldots \\ \hline\hline
\ldots & $77.1$ & $77.0$ & $79.2$ & $75.9$ & $78.7$ & \ldots \\ \hline

\end{tabular}

\end{table*}

\begin{table*}[tbh!]
\null\hspace{0cm}\begin{tabular}{c|c|c|c|c|c|}
\hline
\ldots & Figure \ref{lightA} &  Figure \ref{lightA} & Figure \ref{2CPeven} & Figure \ref{deg2} & Figure \ref{lightdoublk} \\ 
\ldots & (CP odd) &  (close CP odd) & (2 CP-even) & (2 CP-even) & (light doub.)  \\\hline\hline
\ldots & $0.2$ & $0.5$ & $1\cdot10^{-3}$ & $0.7$ & $0.2$ \\ \hline
\ldots & $0.1$ & $0.4$ & $1\cdot10^{-3}$ & $0.1$ & $0.6$ \\ \hline
\ldots & $10$ & $11$ & $18$ & $2$ & $9.25$ \\ \hline
\ldots & $120$ & $120$ & $120$ & $403$ & $200$  \\ \hline
\ldots & $1200$ & $1200$ & $1903$ & $963$ & $130$  \\ \hline
\ldots & $4$ & $-31$ & $-350$ & $-75$ & $-1392$ \\ \hline
\ldots & $250$ & $250$ & $250$ & $75$ & $250$  \\ \hline
\ldots & $500$ & $500$ & $500$ & $150$ &  $500$ \\ \hline
\ldots & $1.5$ & $1.5$ & $1.5$ & $1.5$ & $1.5$ \\ \hline
\ldots & $1.5$ & $1.5$ & $1.5$ & $1.5$ &  $1.5$ \\
\ldots & $1.2$ & $1.2$ & $1.2$ & $0.5$ & $1.1$ \\ \hline
\ldots & $300$ & $300$ & $300$ & $150$ & $300$ \\ \hline
\ldots & $-2.5$ & $-2.5$ & $-2.5$ & $-0.1$ &  $-2.3$ \\
\ldots & $-1.5$ & $-1.5$ & $-1.5$ & / &  $-1.5$ \ldots \\ \hline\hline
  &   &   &   &   &  \\\hline\hline
\ldots & $120.6$ S & $123.2$ D & $125.0$ S/D & $122.9$ S/D & $75.9$ D \\ \hline
\ldots & $124.7$ D & $181$ S & $125.1$ D/S & $123.9$ D/S & $125.1$ D \\ \hline
\ldots & $6.4$ ($P_{13}^2\simeq100\%$)& $124.2$ S & $355$ S & $160$ S & $70.3$ D \\ \hline
\ldots & $Br_{h_1\to2A_1}\simeq98\%$ & $P_{13}^2\simeq99.8\%$ & $S_{13}^2=53\%$ & $S_{13}^2=91\%$ & $m_{H^{\pm}}=107.3$ \\ \hline\hline
\ldots & $80.6$ &  $76.5$ & $79.8$ & $76.2$ & HB\_4.1.3 + HS\_1.2.0: $88.0$ \\ \hline

\end{tabular}
\end{table*}

\end{document}